\documentclass[%
 aip,
 amsmath,amssymb,
 reprint,%
]{revtex4-2}

\usepackage{graphicx}
\usepackage{dcolumn}
\usepackage{bm}
\usepackage[linktoc=page]{hyperref}

\usepackage[utf8]{inputenc}
\usepackage[T1]{fontenc}
\usepackage{mathptmx}
\usepackage{etoolbox}

\usepackage{verbatim}
\usepackage{xcolor}
\usepackage{units}
\usepackage{tensor}

\usepackage{tikz}

\makeatletter
\def\@email#1#2{%
 \endgroup
 \patchcmd{\titleblock@produce}
  {\frontmatter@RRAPformat}
  {\frontmatter@RRAPformat{\produce@RRAP{*#1\href{mailto:#2}{#2}}}\frontmatter@RRAPformat}
  {}{}
}%
\makeatother

\newcommand\avg[1]{\left< #1 \right>}
\newcommand\abs[1]{\left| #1 \right|}
\newcommand\norm[1]{\left\| #1 \right\|}

\newcommand\ket[1]{\left| #1 \right>} 
\newcommand\bra[1]{\left< #1\right|} 
\newcommand\braket[2]{\left< #1 \left| \right. #2 \right>} 
\newcommand\braOpket[3]{\left< #1 \left| #2 \right| #3 \right>} 
\newcommand\ketbra[2]{\left| #1 \left> \right< #2 \right|} 

\newcommand\Operator[1]{\mathbf{\boldsymbol{#1}}}
\newcommand\Diag[1]{{\rm Diag}\;#1} 
\newcommand\Tr[1]{\mathrm{Tr}\;#1}
\newcommand\poly[1]{{\mathrm{poly}\:#1}}
\newcommand\PoissonBracket[2]{\left\{#1,#2\right\}_{PB}}
\newcommand\Commutator[2]{\left[#1,#2\right]}
\newcommand\AntiCommutator[2]{\left\{#1,#2\right\}}

\def\Probabilty{{\rm  Prob}}
\def\CO{{\mathcal O}}

\def\Manifold{\mathcal M} 
\def\Reals{{\mathbb R}}
\def\Complex{{\mathbb C}}

\def\pdf{p} 
\def\waveFunc{\psi}
\def\basisFunc{\phi} 
\def\densityMatrix{\rho} 
\def\densityOp{\Operator{\rho} }
\def\UnitOp{\Operator{1}}
\def\time{t} 

\def\kernelFunc{K}

\def\Observable{O} 
\def\ObservableOp{\Operator{\Observable}}
\def\Lag{L} 
\def\CL{{\mathcal L}} 

\def\Ham{H}
\def\pcoord{p} 
\def\HamiltonianOp{\Operator{\Ham}}

\def\conditionNumber{\kappa}
\def\generalOp{\Operator{A}} 
\def\TimeOrder{\mathcal{T}}

\def\UnitaryOp{\Operator{U}}

\def\rateMatrix{\nu} 
 
\def\processMatrix{\chi}

\def\CollapseMatrix{L}
\def\CollapseOp{\Operator{\CollapseMatrix}}

\def\hPlanck{h}
\def\Energy{E}
\def\nqubits{n}
\def\Nstates{N} 

\def\xcoord{x} 
\def\ycoord{y} 
\def\zcoord{z} 
\def\xhat{\mathbf{\hat x}}
\def\yhat{\mathbf{\hat y}}
\def\zhat{\mathbf{ \hat z}}

\def\Velocity{V} 

\def\action{J}
\def\actionOp{\Operator{J}}
\def\angle{\theta} 

\def\angularMomOp{\Operator{L}}

\def\qcoord{q}
\def\pcoord{p} 
\def\qOp{\Operator{\qcoord}}
\def\pOp{\Operator{\pcoord}}

\def\QFTOp{\Operator{F}}
\def\PhaseShiftOp{\UnitaryOp}
\def\destroyOp{\Operator{a}} 
\def\createOp{\Operator{a}^\dagger} 
\def\numberOp{\Operator{\Nstates} } 

\def\generalOp{\Operator{A}} 

\def\RotOp{\Operator{R}} 
\def\RX{\Operator{RX}}
\def\RY{\Operator{RY}}
\def\RZ{\Operator{RZ}}
\def\CNOT{\Operator{CNOT}}
\def\CZ{\Operator{CZ}}
\def\SWAP{\Operator{SWAP}}

\def\Nsamples{S}
\def\accuracy{\epsilon}

\def\Kick{K}

\def\chaotic{{c}}
\def\Nchaotic{\Nstates_\chaotic}
\def\localizationLength{\ell_\chaotic}
\def\diffusion{D_\chaotic}
\def\LyapunovRate{\lambda} 

\def\wOp{\Operator{w}}
\def\zOp{\Operator{z}}
\def\weightFunc{W}
\def\weightCoef{w}
\def\jindex{j}

\newcommand\set[1]{\left\{ #1 \right\}}

\def\RotationOp{\Operator{R}}
\def\PauliOp{\Operator{\sigma}}
\def\PauliOpVec{{\vec \PauliOp}}
\def\nhat{{\hat n}}
\def\SU{SU}
\def\SO{SO}

\def\prob{\pdf}
\def\Interval{{\mathcal I}}
\def\pdfOp{\Operator{\pdf}}

\def\XOp{\Operator{X}}
\def\YOp{\Operator{Y}}
\def\ZOp{\Operator{Z}}
\def\HadamardOp{\Operator{H}}
\def\kindex{k}
\def\CROp{\Operator{CR}}

\def\CUOp{\Operator{CU}}

\def\Integers{\mathbb{Z}}
\def\eigenphase{\alpha}
\def\eigenvector{\waveFunc_\eigenphase}

\def\Naturals{\mathbb{N}}
\def\zCoherent{y}
\def\ndim{d} 
\def\wcoord{w}
\def\weightFunc{\rho}

\def\QFTOp{\Operator{QFT}}
\def\superpositionOp{\Operator{S}} 
\def\OracleOp{\Operator{O}}
\def\OracleFunc{f}
\def\superpositionState{\ket{s}}

\def\GroverOp{\Operator{G}}
\def\good{{\rm good}}
\def\bad{{\rm bad}}
\def\Mstates{M}
\newcommand\round[1]{{\rm round}\,\,#1}

\def\sparsity{s}
\def\phase{\theta}
\def\superOp{{\mathcal E}}
\def\superOpGenerator{{\mathcal L}}

\def\conditionNumber{\kappa}

\def\HSmetric{g} 
\def\HSmetricOp{\Operator{\HSmetric}}

\def\conj{*}
\newcommand\conjugate[1]{ { {#1}^\conj } }
\def\zconj{\conjugate{\zcoord}}
\def\Vconj{\conjugate{\Velocity}}

\def\scalar{\theta}
\def\Jac{{\mathcal J}}

\def\weightFunc{{\mathcal G}}

\newcommand\wket[1]{\ket{#1}}
\newcommand\wbra[1]{\bra{#1}}
\newcommand\wbraket[2]{\braket{#1}{#2}}
\newcommand\braketw[2]{\braket{#1}{#2}}

\newcommand\nket[1]{ \ket{\widehat{#1}} }
\newcommand\nbra[1]{ \bra{\widehat{#1}} }

\newcommand\dbra[1]{\bra{\overline{#1}}}
\newcommand\dbraket[2]{ \braket{\overline{#1}}{#2} }

\def\yCoherent{y} 
\def\zCoherent{z} 

\def\mqubits{m}
\def\numOp{\Operator{n}}
\def\ROp{\Operator{R}}

\begin{document}

\preprint{LLNL-JRNL-839545}

\title[Quantum Computing for Fusion Energy Sciences]{Quantum Computing for Fusion Energy Science Applications }
\author{I. Joseph, Y. Shi,  M. D. Porter,   
A. R. Castelli,  V. I. Geyko, \\
 F. R. Graziani, S. B. Libby, J. L. DuBois}
 \email{joseph5@llnl.gov}
\affiliation{ 
Lawrence Livermore National Laboratory, P.O. Box 808, Livermore, CA  94551, USA
}%


\date{\today}

\begin{abstract}
This is a review of recent research exploring and extending present-day quantum computing capabilities for fusion energy science applications.
We begin with a brief tutorial on both ideal and open quantum dynamics, universal quantum computation, and quantum algorithms.
Then, we explore the topic of using quantum computers to simulate both linear and nonlinear dynamics in greater detail.
Because quantum computers can only efficiently perform linear operations on the quantum state, it is challenging to perform nonlinear operations that are generically required to describe the  nonlinear differential equations of interest.
In this work, we extend previous results on embedding nonlinear systems within linear systems by explicitly deriving the connection between the Koopman evolution operator, the Perron-Frobenius evolution operator, and the Koopman-von Neumann evolution (KvN) operator.
We also explicitly derive the connection between the Koopman and Carleman approaches to embedding.
Extension of the KvN framework to the complex-analytic setting relevant to Carleman embedding, and the proof that different choices of complex analytic reproducing kernel Hilbert spaces depend on the choice of Hilbert space metric are covered in the appendices.
Finally, we conclude with a review of recent quantum hardware implementations of algorithms on present-day quantum hardware platforms that may one day be accelerated through Hamiltonian simulation.
We discuss the simulation of toy models of wave-particle interactions through the simulation of quantum maps and of wave-wave interactions important in nonlinear plasma dynamics. 
\end{abstract}


\maketitle

\tableofcontents

\section{ Introduction \label{sec:intro}}

\subsection{Motivation \label{sec:intro:motivation} }
The three pillars of Quantum information science (QIS): quantum sensing, quantum communications, and quantum computing,  promise to have transformative impact on science, engineering and technology as we know it.\cite{Raymer19qst}
This article presents a pedagogical introduction to quantum computing and reviews recent research to develop and apply quantum algorithms and utilize quantum computing hardware platforms for fusion energy science (FES) applications. 
The interesting results obtained so far make it hopeful that QIS may one day lead to game-changing capabilities for FES.

Before diving into quantum computing (QC), let us briefly mention the other two pillars of QIS. 
First, \emph{quantum sensing} techniques are already being used today to improve measurement sensitivity by orders of magnitude. 
Instead of being limited by the central limit theorem to yield a noise-to-signal ratio of $1/\sqrt{\Nsamples}$, were $\Nsamples$ is the number of samples, using intrinsically quantum entangled states, such as squeezed states, as sensitive probes can reduce the detection threshold to the Heisenberg limit $1/\Nsamples$.
Such techniques have been used to improve the sensitivity of the LIGO gravitational wave detector by a factor of 2 for nearly a decade. \cite{Tse19prl}
New detectors based on nitrogen vacancy centers (NV-centers) in diamond have provided unprecedented sensitivity for measurements of magnetic and electric fields as well as  temperature and pressure.
Advances in atom interferometry \cite{Narducci2022aipx} have led to a revolution in gravitational and inertial (gyroscopic) sensing, now transitioning to real world applications such as non-GPS navigation, rapid passive sensing of mass distributions, and underground structure discovery, as well as basic science applications such as gravitational wave detection in the frequency regime between LIGO and LISA.

Second, \emph{quantum communications} offer the possibility of secure and intrinsically parallel information transfer. \cite{Nordrum17ieee}
The goal is to transform and  transport quantum information over long distances and to transduce information between different quantum hardware platforms.
This field generated some of the earliest rigorous proofs that combining entangled states with quantum communication protocols could surpass their classical counterparts in their ability to carry and manipulate information. 
The application to network security is considered to be so important that many researchers are already working on the \emph{quantum internet.}

Finally, \emph{quantum computing}, which is the focus of this review article, promises to efficiently perform quantum algorithms that have the power to achieve polynomial to exponential improvements in complexity while manipulating exponentially large quantum memory resources. \cite{KitaevBook, NielsenBook,  KayeBook} 
The recent growth in the capabilities of today's quantum computing devices has spurred great interest in quantum simulation and they have been used to perform key demonstrations of quantum calculations. \cite{Martinez16nat,Roushan17sci,Arute19nat}
The potential impact of this so-called \emph{quantum advantage} is so exciting that many government and private industry laboratories around the world are working to perfect the technology.
In fact, in recent years, there have been claims that some demonstrations have already passed the point of \emph{quantum supremacy,} \cite{Arute19nat, Wu21prl, Zhong21prl}
the point at which quantum computers can surpass even the world's best supercomputers at performing certain computational tasks.

While there has certainly been enormous progress over the last two decades, we shall see that the practical use of today's hardware platforms is still quite limited due to the lack of fault-tolerant error correction.
Because present-day and near-term quantum devices offer large quantum memory registers but lack error correction, the present era has been dubbed the \emph{Noisy Intermediate-Scale Quantum} (NISQ) era. \cite{Preskill18quantum}
The search for the  ultimate physical platform for realizing a quantum computer spurs research  onward to invent new quantum materials, new hardware platforms, and new algorithms that can surpass today's limitations.

\subsection{Fusion Energy Science Needs  \label{sec:intro:fes} }
Fusion energy science (FES) is defined by the mission of: \emph{Achieving a safe sustainable fusion energy source for the foreseeable future.}
Hence, it encompasses all fields of science required to achieve this goal, notably including plasma physics, nuclear physics, materials science, and chemistry.
Thus, the FES mission requires developing  accurate predictive models within all of these disparate fields, to form combined whole-device models that ultimately span a wide range of physical regimes from  classical to quantum. 

To stress the difficulty of this endeavor, FES requires calculations of strongly correlated quantum materials such as  superconductors at temperatures below 1 $\unit{meV}$ to the collective motions of plasma at 1-10 $\unit{keV}$ as well as nuclear reactions at 10's of $\unit{MeV}$.  
Whereas magnetic confinement fusion uses low-density plasmas on the order of $10^{20}$ $\unit{m}^{-3}$, inertial confinement fusion seeks to obtain a compression of particle density on the order of 1000 times that of solid matter, for a particle densities beyond $10^{30}$ $\unit{m^{-3}}$.
Thus, this represents a range of 9 orders of magnitude in energy and 10 orders of magnitude in density (without considering the density of nuclear matter).

Note, however, that, today, the vast majority of scientific computations in the fusion energy sciences are classical in nature.
The meaning of this statement is that such calculations seek to predict the time-dependent evolution of \emph{nonlinear} sets of both ordinary differential equations (ODEs) and  partial differential equations (PDEs).
Ensembles of systems are typically evolved using classical \emph{probability distribution functions} (PDFs), which are simply advected along the trajectories.
While this is certainly true in the traditionally classical realm of plasma physics, it is also quite true in the traditional quantum realm of chemistry and materials science.
The computational workhorses of density functional theory (DFT) and molecular dynamics crucially rely on the ability to efficiently evaluate nonlinear functions and to evolve nonlinear differential equations.
The same is true for evolving classical kinetic equations that describe the evolution of nuclear and chemical species with reaction rate theory.

An essential limitation of the standard quantum computing paradigm is that, aside from measurements, only linear unitary operations can be performed efficiently in the quantum Hilbert space.
If one believes that quantum mechanics is the correct physical theory of nature, then linear dynamics is all that is needed to describe the evolution of the universe.
(A notable exception to this rule is that some physicists believe that the measurement process might require nonlinear dynamics to be described completely.)
However, according to the \emph{exact factorization theorem}\cite{Abedi10prl, Abedi12jcp} the process of generating reduced order models typically generates nonlinear interactions between subsystems.
This is the reason that essentially all useful physical models that achieve significant complexity reduction, e.g. fluid dynamics, density function theory (DFT), and kinetic theory, are nonlinear.

\subsection{ Quantum Computing Capabilities  \label{sec:intro:qc} }
Many scientific computations of interest to FES can be solved more efficiently using quantum algorithms (see the excellent overviews given in Refs.~\onlinecite{Schenkel18fesqis} and \onlinecite{Dodin21pop}).
Quantum algorithms promise to accelerate many of the algorithms that are central to scientific computing, including the Fourier transform, sparse linear solvers, and sparse Hamiltonian simulation.
Quantum algorithms also exist for linear ODEs \cite{Berry17cmp} and linear PDEs \cite{Costa19pra, Childs21quantum} that can be cast in the form of initial value problems (IVPs),  boundary value problems (BVPs), and eigenvalue problems (EVPs).

For example, for plasma and materials science applications, some IVPs are naturally posed as being Hamiltonian, and, hence, can naturally be presented as a unitary evolution resulting from a Hermitian Hamiltonian.
If the Hamiltonians of interest have a sparse structure, then they can be solved using efficient quantum algorithms for Hamiltonian simulation,\cite{Berry16qic, Low17prl} which we will discuss in detail in Sec.~\ref{sec:qsim}.
Notably, Ref.~\onlinecite{Engel19pra} developed a quantum algorithm to solve the linearized Vlasov-Poisson system and Ref.~\onlinecite{Novikau22pra} developed a quantum algorithm to solve the problem of cold plasma wave propagation and linear mode conversion. Yet, the majority of IVPs for reduced models include dissipation. Since the dynamics is not Hamiltonian, they need to be treated with specialized methods, for example, the general purpose quantum linear differential equation solver, \cite{Berry17cmp} discussed in Sec.~\ref{sec:linear}.

For many fields of science and engineering, including biology, chemistry, and physics, a large share of today's computational resources are used for the simulation of classical nonlinear dynamics. Hence, it is important to understand whether quantum computers can provide similar gains in efficiency for the simulation of nonlinear problems. 
Nonlinear operations can be performed by forming the tensor product of identical states. 
However, due to the \emph{no-cloning theorem}, \cite{ Park70fp, Dieks82pla,Wooters82nat} it is not possible to make copies of a quantum state; rather, one must prepare multiple identical states from scratch. 
Thus, if a nonlinear operation is to be iterated, e.g. in time, then this will require preparing a number of identical states that is exponentially large in the number of iteration steps.\cite{Leyton08arxiv}

Efficient quantum algorithms for simulating nonlinear physical processes have only recently begun to be invented and improved \cite{Joseph20prr, Dodin21pop, Dodin21arxiv, Liu21pnas, Engel2021pop, Lin22arxiv, Krovi22arxiv} and some of these efforts were initiated for FES applications. \cite{Schenkel18fesqis}
The new algorithms all use the strategy of embedding the nonlinear system within an infinite-dimensional  linear system that is then truncated to finite dimension. 
For systems of nonlinear ODEs, the general method of using a quantum computer to solve for the evolution of the wavefunction corresponding to a classical PDF was first clearly described in Refs.~\onlinecite{Joseph20prr, Dodin21pop}, using what is known as the Koopman-von Neumann embedding.
Using an alternative approach, Ref.~\onlinecite{Liu21pnas} developed concrete quantum algorithms for dissipative differential equations with quadratic nonlinearity based on a approach known as Carleman embedding, which is applied to simulate the Burgers equation. 
Connections between these embedding techniques are discussed in Ref.~\onlinecite{Engel2021pop}, and comparisons of these approaches are elucidated using numerical examples in Ref.~\onlinecite{Lin22arxiv}. 
Building upon ODE solvers, quantum algorithms for solving PDEs, including the Navier-Stokes equations, have also been explored. \cite{Gaitan2021finding, Jin2022quantum}

Apart from deterministic methods, algorithms based on random processes have also been developed. For example, quantum versions of Monte Carlo algorithms can generally be performed with a quadratic speedup over their classical counterparts.\cite{Montanaro15prsa}
Similar concepts can then be used to speed up calculations of stochastic differential equations (SDEs). \cite{Kacewicz06jc}
General multi-level Monte-Carlo methods for stochastic differential equations can also been accelerated. \cite{An2021quantum}
In fact, quantum algorithms based on these concepts have been proposed for computing turbulent mixing and the reaction rate within turbulent flows. \cite{Xu2018aiaa, Xu2019ctm}

The algorithms mentioned above requires fault-tolerant quantum computers to implement various oracles that are required by quantum subroutines. 
Alternatively, on NISQ devices, which have been shown to be capable of performing classical-quantum hybrid optimization problems, \cite{Farhi2014quantum,Kandala2017hardware} 
advancing nonlinear equations may be treated as an optimization problem using variational algorithms. \cite{Lubasch20pra}
In this case, evaluating the nonlinear functional only requires a fixed number of copies per time step.
Hence, there have been recent proposals for simulating the Navier-Stokes equations using the variational technique. \cite{Steijl2020aqci,Steijl2020qcc,Griffin2019turb}

Even though simulating nonlinear problems is perhaps one of the most useful applications of scientific computation, relevant quantum algorithms are still in their infancy and there is much work to be done in terms of improving their speed and accuracy and verifying their performance on quantum computing hardware platforms.
Once fully vetted and understood, they can be used as subroutines to provide a foundation for quantum programs that  perform more useful and more complex tasks.

\subsection{ Overview of Contents  \label{sec:intro:outline} }
The next section will provide a brief introduction to the principles of quantum information and explain the reason that quantum memory registers can be considered to be much larger than their classical counterparts.
Then, Sec.~\ref{sec:qcomp} will discuss the digital circuit model of universal quantum computation and review the key quantum subroutines that are useful for accelerating classical computations and for building quantum programs. 
Next, Sec.~\ref{sec:qsim} will discuss  quantum approaches to simulating differential equations, the type of application that is used most by FES researchers today.
Then, Sec.~\ref{sec:nonlinear} will unify several recent proposals for using quantum algorithms to  simulate nonlinear dynamics, including generalized approaches to Koopman-von Neumann (KvN) and Carleman embedding, as well as explain their differences.
The penultimate section, Sec.~\ref{sec:apps} will discuss recent findings in applying several leading quantum computing hardware platforms to quantum simulation problems of interest to FES.    
Appendix~\ref{sec:complex-KvN} extends the KvN framework to the complex-analytic setting relevant to Carleman embedding.
Appendix~\ref{sec:RKHS} proves that different choices of the relevant Hilbert space bases, i.e. reproducing kernel Hilbert spaces, depend on the choice of Hilbert space metric.
The concluding section summarizes our findings and offers perspectives on the future outlook for quantum computing for FES applications.

\section{ Quantum Information \label{sec:qinfo}}
Paul Benioff, \cite{Benioff80jsp}  Yuri Manin, \cite{ManinBook} and Richard Feynman, \cite{Feynman82ijtp, Feynman86fp} were some of the first scientists to think seriously about the potential of using machines that satisfy the laws of quantum mechanics for performing scientific calculations. 
Feynman also left us with one of the great quotes in the field: \cite{Feynman82ijtp} 
\emph{Nature isn't classical, dammit, and if you want to make a simulation of nature, you'd better make it quantum mechanical, and by golly it's a wonderful problem, because it doesn't look so easy.}
Understanding this statement is the key to understanding why scientists believe that quantum computers can, in principle, be so powerful.

\subsection{The Postulates of Mechanics \label{sec:qinfo:postulates} }
To see why quantum computers are fundamentally different from classical computers, let us begin by recalling the postulates of quantum mechanics \cite{NielsenBook, KayeBook, PreskillNotes} and compare them with postulates of classical mechanics. 
A primer on Hilbert spaces is given in Appendix \ref{sec:HilbertSpace}.
In the following, we assume that the reader is familiar with Dirac's bra-ket notation.

\vspace{6pt}
\paragraph*{\bf Postulates of Quantum Mechanics (QM).} 
\emph{In the following formulae, the Hilbert space will be treated as if it were discrete (which is typically true for digital models of quantum computation), but if it it is continuous, then the sums should be replaced by  integrals.}
\begin{enumerate}
\item {\bf Physical States
\label{QM:states}
}
\begin{enumerate}
	\item { \label{QM:pure-states}
	\emph{Pure physical states}, denoted by the\emph{ wavefunction}  $\waveFunc$, are rays in \emph{Hilbert space}.
	}
	\begin{itemize}
		\item Pure states are vectors in Hilbert space, generically, a linear \emph{superposition} of basis elements that span the Hilbert space.
		\item Pure states are elements of \emph{projective Hilbert space}; i.e. the overall complex constant   is unimportant, and,  the normalization can be set to unity, $\norm{\waveFunc}=1$.
		\item The Hilbert space is typically a space of functions that is associated with a configuration manifold, $\Manifold$, but it can also be a discrete set. 
	\end{itemize}
	
	\item { 	\label{QM:mixed-states}
	\emph{Mixed physical states} are probability distributions over pure states and, hence, are represented by a positive self-adjoint operator on Hilbert space, $\densityOp^\dagger=\densityOp$, called the probability \emph{density matrix}.
	}
	\begin{itemize}
		\item Due to Hermiticity, mixed states can be diagonalized
	\begin{align} 
	\densityOp=\sum_\jindex \pdf_\jindex \ketbra{\waveFunc_\jindex} {\waveFunc_\jindex} 
	.
	\end{align}
	 Because the eigenvalues, $\prob_\jindex \geq 0$, must be non-negative
	  this represents a probability distribution  over a set of pure states, $\ket{\waveFunc_\jindex}$, with probability $\prob_\jindex$. 
		\item Probability is normalized to unity, \\
		$\Tr{(\densityOp)}=\sum_\jindex\prob_\jindex=1$.  
		\item Proper mixed states must satisty $\Tr{(\densityOp^a)}<1$ for $a>1$ and $\Tr{(\densityOp^a)}>1$ for $0<a<1$.
		\item Mixed stated are required to represent the outcome of measurements. 
	\end{itemize}
	\item { The Hilbert space of a \emph{composite system} is the tensor product of the Hilbert spaces of the subsystems. 	
	\begin{itemize}
		\item The Hilbert space dimension of the composite system is the product of the dimensions of the subsystems.
		\item \emph{Entangled states} are states that cannot be represented as tensor products of the states of the individual subsystems; i.e. states that are a nontrivial superposition of the states of the subsystems.
	\end{itemize}
	\label{QM:Composite_Hilbert-space} 
	}
	\end{enumerate}
\item {\bf Measurement of Physical Observables
\label{QM:measurement}
}
	\begin{enumerate}
	\item { \emph{Physical observables} are \emph{self-adjoint  operators} $\ObservableOp^\dagger=\ObservableOp$ on Hilbert space.
	\label{QM:observables}
	}
	\item { \emph{Measurements} yield eigenvalues of the operator corresponding to the observable  
	\label{QM:measuring-eigenvalues} 
	}
	\begin{align}
	  \ObservableOp\ket{\waveFunc_\alpha} = \alpha\ket{\waveFunc_\alpha} .
	\end{align} 
	\item { The probability of observing a particular eigenvalue, $\alpha$, is given by the overlap between the eigenstate and the physical state  }
	\begin{align}
	 \Probabilty( \alpha)
	 =  \braOpket{\waveFunc_\alpha}{\densityOp}{\waveFunc_\alpha}    
	 =\sum_\jindex \prob_\jindex \norm{\braket{\waveFunc_\alpha}{\waveFunc_\jindex}}^2 
	 . 
	\end{align} 
 
		 { Therefore, the expectation value of measuring an observable is given by}
		\begin{align}
		 \avg{\ObservableOp}  &=\Tr{(\densityOp\ObservableOp)}=   \sum_\jindex \prob_\jindex  \braOpket{\waveFunc_\jindex}{\ObservableOp}{\waveFunc_\jindex}
	 .
		\end{align} 
 
	\item { Immediately after a measurement of eigenvalue $\alpha$, the wavefunction \emph{collapses} to the state $\ket{\psi_\alpha}$. 
	}
	\end{enumerate}

\item{\bf Temporal Evolution}
	\begin{enumerate}
	\item \emph{General open systems  \label{QM:open-systems}}
		\begin{itemize}
		\item Physical states evolve via linear \emph{superoperations} that act on the density matrix
		\begin{align}
		\densityOp(\time) &= \superOp(\time,\time_0)  \densityOp(\time_0).
		\end{align}
	The  evolution must preserve the Hermiticity, positivity, and trace of any density matrix.
		\item Continuous time evolution is generated by the generator of a linear superoperator
		\begin{align}
		d\densityOp/d\time &= \superOpGenerator(\time) \densityOp (\time) .
		\end{align}
		The  evolution must preserve the Hermiticity, positivity, and trace of any density matrix.
		\end{itemize}
	
	\item   \emph{ Ideal closed systems \label{QM:closed-systems}}
		\begin{itemize}	
		\item {Ideal closed systems evolve via a  linear \emph{unitary}, 
		$\UnitaryOp^{-1}=\UnitaryOp^\dagger$, transformation of Hilbert space, 
		\label{QM:UnitaryEvolution}
		}
		\begin{align}
		\waveFunc(\time)  = \UnitaryOp   \waveFunc(\time_0)
		& &
		\densityOp(\time) = \UnitaryOp   \densityOp(\time_0)\UnitaryOp^{\dagger} 
		.
		\end{align} 

		\item {  Continuous time evolution is generated by a Hermitian \emph{Hamiltonian}  operator, $\HamiltonianOp$
		\label{QM:HamiltonianEvolution}
		}
		\begin{align}
		i\hbar\partial_\time \waveFunc=\HamiltonianOp \waveFunc &&
		i\hbar\partial_\time \densityOp=\Commutator{\HamiltonianOp}{ \densityOp}
		\end{align} 
		where the \emph{commutator} is defined via 
		\begin{align}
		\Commutator{\Operator{A}}{\Operator{B}}:=\Operator{A}\Operator{B}-\Operator{B}\Operator{A}.
		\end{align}
		
		\end{itemize}

	\end{enumerate}
\end{enumerate}

The usual statement of Newton's laws\cite{GoldsteinBook, SudarshanBook}  discusses the kind of equations of motion that are allowed rather than the underlying mathematical assumptions. 
This does not make it easy to compare the underlying mathematical assumptions of quantum and classical mechanics.
Perhaps this is one of the reasons that quantum mechanics took some time to be developed and understood.
If one were to attempt to write the postulates of classical mechanics in a form that parallels that of quantum mechanics, then one might arrive at the following.

\vspace{6pt}
\paragraph*{\bf Postulates of Classical Mechanics (CM).} 
\emph{In the following formulae, the classical phase space will be treated as if it were continuous, but if it is discrete, then the  integrals should be replaced by sums.}
\begin{enumerate}
\item {\bf Physical States}
	\begin{enumerate}
	\item { \emph{Pure physical states} are points in a set, $\zcoord\in  \Manifold$,  called the \emph{phase space}. }
	\begin{itemize} 
		\item Phase space is typically a continuous and smooth space such as a differentiable manifold, but it can also be a discrete set. 
	\end{itemize}
	\item { \emph{Mixed physical states} are represented by a probability distribution function, $\pdf(\zcoord)$, over phase space. } 
	\begin{itemize}
		\item  Probability is normalized to unity $\int \pdf d\zcoord=1$.
	\end{itemize}
	\item { The phase space of a  \emph{composite system} is the tensor product of the phase spaces of the subsystems.}
	\begin{itemize}
		\item The dimension of the composite phase space is the product of the dimensions of the subsystems.
	\end{itemize}
	\end{enumerate}
\item {\bf Measurement of Physical Observables}
	\begin{enumerate}
	\item { \emph{Physical observables} are real functions of coordinates in phase space, $\Observable:\Manifold\rightarrow\Reals$.}
	\item {In principle, \emph{measurements} of the coordinates, $\zcoord$, can be made with arbitrary precision, and, so, a measurement of an observable at the point $\zcoord$ yields its value, $\Observable(\zcoord)$. }
	\item { The probability of measuring a particular value of the observable, $\Observable(\zcoord)$, is given by the probability, $\pdf(\zcoord)$, of observing the pure state at the point $\zcoord$. }
	Therefore, the expectation value of an observable is given by 
	\begin{align}
	\avg{\Observable}=\int\Observable\pdf d\zcoord
	.
	\end{align}
	\end{enumerate}
\item{\bf Temporal Evolution}
	\begin{enumerate}
	\item  \emph{General open systems \label{CM:open-systems}}
	\begin{itemize}
		\item { Physical states evolve via (generically nonlinear) coordinate transformations, $\zeta:\Manifold \times \Interval \rightarrow\Manifold$, }
		\begin{align}
		\zcoord(\time) = \zeta(\zcoord(\time_0) ,\time).
		\end{align} 
		\item { Continuous time evolution is generated by (generically nonlinear)  differential equations }
		\begin{align}
		 d\zcoord/d\time=\Velocity(\zcoord, \time).
		\end{align} 
	\end{itemize}
	\item  \emph{Ideal closed systems \label{CM:closed-systems}}
	\begin{itemize}
	\item{ 
	Ideal closed systems are Hamiltonian systems that evolve via Hamilton's equations of motion, which yield \emph{infinitesimal symplectic transformations} of phase space.
	}
		\item In canonical coordinates, $\zcoord=\{\qcoord,\pcoord\}$, composed of equal numbers of configuration space coordinates, $\qcoord$, and momentum space coordinates, $\pcoord$, Hamilton's equations of motion for Hamiltonian $\Ham(\zcoord)$ are
		\begin{align}
			d\qcoord/d\time = \partial_\pcoord\Ham &&d\pcoord/d\time =- \partial_\qcoord\Ham 
			.
		\end{align}
		\item Evolution over a finite time interval yields a (generically nonlinear) \emph{symplectic transformation} of phase space, $\zeta:\Manifold \times \Interval \rightarrow \Manifold$.
	\end{itemize}
	\end{enumerate}
\end{enumerate}

\subsection{ Open Quantum Dynamics \label{sec:qinfo:open-quantum} } 

\subsubsection{ Open Systems and Decoherence \label{sec:qinfo:decoherence} } 
Many physicists and mathematicians are familiar with the temporal evolution of ideal closed systems, e.g. in the classical CM Postulate \ref{CM:closed-systems} and quantum QM Postulate \ref{QM:closed-systems} contexts.
The more general formulation for open classical systems in CM Postulate \ref{CM:open-systems} is also quite natural. 
This form applies to non-ideal open classical systems, which typically experience various forms of dissipation such as friction, diffusion, collisions, etc.

In contrast, the discussion of quantum dynamics in QM Postulate \ref{QM:open-systems} was far too brief to do justice to this important topical area. 
While the mathematical justification is on solid footing today, in terms of applications to real-world quantum hardware platforms, this field is still under active development.

For closed ideal quantum systems, QM Postulate \ref{QM:UnitaryEvolution} requires time evolution to be unitary and, hence, reversible.
However, similar to the way that classical chaos can scramble classical information, quantum dynamical processes can scramble quantum information.
Classically, the combination of the butterfly effect (positive Lyapunov exponents)
and \emph{coarse-graining} due to the finite precision of measurements leads to irreversibility.
However, \emph{quantumly}, the unitary dynamics only has vanishing Lyapunov exponents, and, so, while the evolution may track the semiclassical dynamics for long periods of time, any apparent exponential instability associated with the Lyapunov exponent must always eventually cease.

The paradox is resolved by the fact that most quantum systems are actually open; i.e. the observer does not  have access to information to all parts of the system.
The parts of the system that either are not
or cannot be measured are usually referred to as the \emph{environment}. 
The existence of this  unmeasured or \emph{hidden} information has effects that are similar to coarse-graining.
Thus, open quantum systems can display many different types of non-ideal and non-unitary behavior.

In fact, the act of measurement is the primary example of a non-ideal but linear evolution process.
Because the measurement apparatus in QM Postulate \ref{QM:measurement} is assumed to be classical, the probability of each eigenvalue is described by classical probability theory.
Thus, even if one begins with a pure state, after the measurement is performed, the resulting knowledge about the system must become a mixed state: a probability distribution over possible eigenfunctions of the observable operator that was measured.

\emph{Decoherence} refers to the way in which small but persistent interactions with the environment cause an almost irretrievable loss of information for open quantum systems.
The phrase decoherence was introduced by H. Dieter Zeh\cite{Zeh70fop} and has been explored in great detail by Wocjeck Zurek \cite{Zurek03rmp, Zurek03arxiv} and collaborators.
Physicists and mathematicians such as Choi, Gorini, Krauss, Kossakowski, Lindblad, Redfield, Stinespring, and Sudarshan explored possible types of evolution laws, called \emph{master equations}, that an open quantum system can undergo.\cite{PreskillNotes}
The goal is to develop a master equation that describes the loss of information in a manner similar to the way the Fokker-Planck equation describes the loss of information due to diffusive processes for classical probability theory.

If one were to use an accurate quantum model of the environment, then the state of the system can interact in many different ways with the environmental degrees of freedom.
For simplicity, it is customary to attempt to determine a reduced order model that has the expected qualitative behavior. 
For example, it is customary to model the interactions with the environment as a \emph{quantum process}: a linear transformation of the density matrix that has no memory, i.e. a Markovian process.
Note, however, that using a more realistic reduced order model of the environment that has
nontrivial dynamics will typically generate a nonlinear model that is not Markovian.
 
\subsubsection{CPTP Quantum Processes}
Linear operators that act on the density matrix, rather than on pure states, are referred to as \emph{quantum processes}, \emph{quantum channels}, or \emph{superoperators}.
If the dimension of the Hilbert space is $\Nstates$, then there are $\Nstates^2$  unitary operations that describe ideal evolution. 
However, there are many more, $\Nstates^4-\Nstates^2$, quantum processes that describe decoherence.

If the density matrix is to remain a positive Hermitian operator, then the superoperator must itself be Hermitian and positive.
In fact, it is common to demand a stronger criterion called \emph{complete positivity} (CP) \cite{Stinespring55pams, Choi75laa} that results from considering the evolution of the system when coupled to a reservoir that represents the environment with a Hilbert space of arbitrary dimension, $\ndim$.
In this case, the state is $\densityOp\otimes \UnitOp_\ndim$, and evolution of the subsystem with superoperator, $\superOp$, while the reservoir is unchanged results in the state $\superOp(\densityOp)\otimes \UnitOp_\ndim$.
The map, $\superOp(\densityOp)$, must be completely positive in order for the coupled map, $\superOp(\densityOp)\otimes \UnitOp_\ndim$, to be positive for the total system $\otimes$ reservoir Hilbert space.
It turns out that complete positivity is a stronger condition than positivity alone as there are examples of maps, e.g. the transpose operation, that are positive but not completely positive.
In order to conserve probability, the superoperator must also be trace-preserving (TP).
 Such superoperators are called CPTP operators.

A Hermiticity preserving superoperator must have the form \cite{Choi75laa, PreskillNotes}
\begin{align}
 \superOp \densityOp =\sum_{\jindex, \kindex=1}^{\Nstates^2}  \processMatrix_{\jindex \kindex} \CollapseOp_\jindex \densityOp \CollapseOp^\dagger_\kindex 
\end{align}
where $\set{\CollapseOp_\jindex}$ defines a complete basis of  operators of size $\Nstates\times \Nstates$ and the matrix $\processMatrix_{ij}$ is Hermitian. 
In order for the map to be CP, the matrix $\processMatrix_{ij}$ must also be positive semi-definite. 
(Note that \emph{completely} positive refers to the action on the total Hilbert space rather than the definiteness of the eigenvalues.)
In order to be trace preserving (TP), the condition
\begin{align}
 \UnitOp =\sum_{\jindex, \kindex=1}^{\Nstates^2}   \processMatrix_{\jindex \kindex} \CollapseOp^\dagger_\kindex  \CollapseOp_\jindex 
\end{align}
must hold.
The trace preservation criterion represents $\Nstates^2$ constraints. 
Since there are $\Nstates^2$ unitary operations, this implies that there are $\Nstates^4-2\Nstates^2$ non-unitary decoherence operations.

 Because the $\processMatrix_{\jindex \kindex}$ matrix is Hermitian, this form can be simplified by diagonalizing this matrix with a unitary transformation. 
 Applying this transformation to the operator basis allows one to find a new basis, $\set{\hat \CollapseOp_\alpha}$,
that simplifies the equation to a sum over the eigenvalues of the $ \processMatrix_{\jindex \kindex}$ matrix, which we denote as $\processMatrix_\alpha$:
\begin{align}
 \superOp \densityOp =  \sum_{\alpha=1}^{\Nstates^2} \processMatrix_\alpha \hat\CollapseOp_\alpha \densityOp \hat\CollapseOp^\dagger_\alpha 
 .
\end{align}
Because the eigenvalues $\processMatrix_\alpha$ are positive, the so-called \emph{Krauss form}  \cite{KraussBook} can be achieved by absorbing the square root of the eigenvalue into the definition of the operators $\hat \CollapseOp_\alpha\rightarrow \processMatrix_\alpha^{1/2} \hat \CollapseOp_\alpha$. 

Note that complete positivity only requires the eigenvalues to be positive semi-definite $\chi_\alpha\geq 0$. 
If the map is positive but not completely positive, then it must have at least one negative eigenvalue.  

\subsubsection{CPTP Evolution}
The Gorini-Kossakowsi-Lindblad-Sudarshan (GKLS) master equation \cite{Lindblad76cmp, Gorini76jmp} represents the most general form of a continuous CPTP evolution of the density matrix when the system is separable from the environment and when their interactions are Markovian. 
Just as the Fokker-Planck equation describes processes such as diffusion due to collisions for the Liouville evolution of the classical PDF, the Lindblad equation adds nonunitary CPTP  processes to the von Neumann equation for the density matrix.
The {\bf GKLS equation} is given by
\begin{align}
 \partial_\time \densityOp = \Commutator{\HamiltonianOp}{\densityOp}/i\hbar  
 -\AntiCommutator{\generalOp}{\densityOp} 
 + \sum_{\jindex, \kindex=1}^{\Nstates^2-1} \rateMatrix_{\jindex \kindex}  \CollapseOp_\jindex \densityOp \CollapseOp^\dagger_\kindex 
\end{align}
where the anticommutator is defined via $\AntiCommutator{\Operator{A}}{\Operator{B}}:=\Operator{A}\Operator{B}+ \Operator{B}\Operator{A}$.
Thus, in addition to unitary evolution determined by the Hamiltonian $\HamiltonianOp$, one adds a general CP evolution as well as the TP damping term 
\begin{align}
\generalOp =  \sum_{\jindex,\kindex=1}^{\Nstates^2-1} \rateMatrix_{\jindex \kindex}\CollapseOp^\dagger_\kindex \CollapseOp_\jindex /2 
\end{align}
required to preserve the trace. 
 In this case, the \emph{collapse operators}, $\set{\CollapseOp_\jindex}$, define a basis over trace-free matrices of size $\Nstates\times\Nstates$.
(The freedom of including the unit matrix, which possesses a trace, was already used in the construction of the damping operator $\generalOp$.)
This evolution is CPTP as long as the rate matrix, $\rateMatrix_{ij}$, is Hermitian and positive semidefinite.
Information loss occurs whenever the rate matrix has positive eigenvalues.

Again, because the rate matrix is Hermitian, this form can be simplified by diagonalizing $\rateMatrix_{\jindex \kindex}$ with a unitary transformation. 
Applying this transformation to the  collapse operators allows one to find a new basis, $\set{\hat \CollapseOp_\alpha}$, that simplifies the equation to a sum over the eigenvalues of the $\chi_{ij}$ matrix, which we denote as $\nu_\alpha$:
\begin{align}
 \partial_\time \densityOp = \Commutator{\HamiltonianOp}{\densityOp}/i\hbar +\sum_\alpha \nu_\alpha \left(\hat\CollapseOp_\alpha \densityOp \hat\CollapseOp^\dagger_\alpha- \AntiCommutator{ \hat \CollapseOp^\dagger_\alpha\hat \CollapseOp_\alpha}{\densityOp}/2\right)
 .
\end{align}
Since the eigenrates $\nu_\alpha$ are positive, the square root of the rates can be absorbed into the definition of the collapse operators, as in the Krauss form above.
While the diagonal form offers simplicity, it hides the complexity associated with the myriad ways in which the environment can potentially interact with a quantum system to form an arbitrary rate matrix. A pedagogical introduction to the GKLS master equation is given in Ref.~\onlinecite{Manzano20aipa}.

There are a few important environmentally-induced decoherence processes that affect almost all  quantum systems: (1) \emph{relaxation:}   of higher energy levels to lower energy levels,  (2) \emph{excitation:} of lower energy levels to higher energy levels, and (3) \emph{dephasing:} the diffusion of the complex phases within a linear superposition.
For a harmonic oscillator,
the relaxation collapse operator is the destruction operator, $\destroyOp$, the excitation collapse operator is the creation operator, $\createOp$, and the dephasing collapse operator is the number operator, 
$\numberOp=\createOp\destroyOp$.
While excitation is  important in general contexts, it is often much slower than the  other two processes for typical quantum hardware platforms.

\subsection{The Qubit \label{sec:qinfo:qubit} } 
\begin{figure}
\centering
\includegraphics[width=3in]{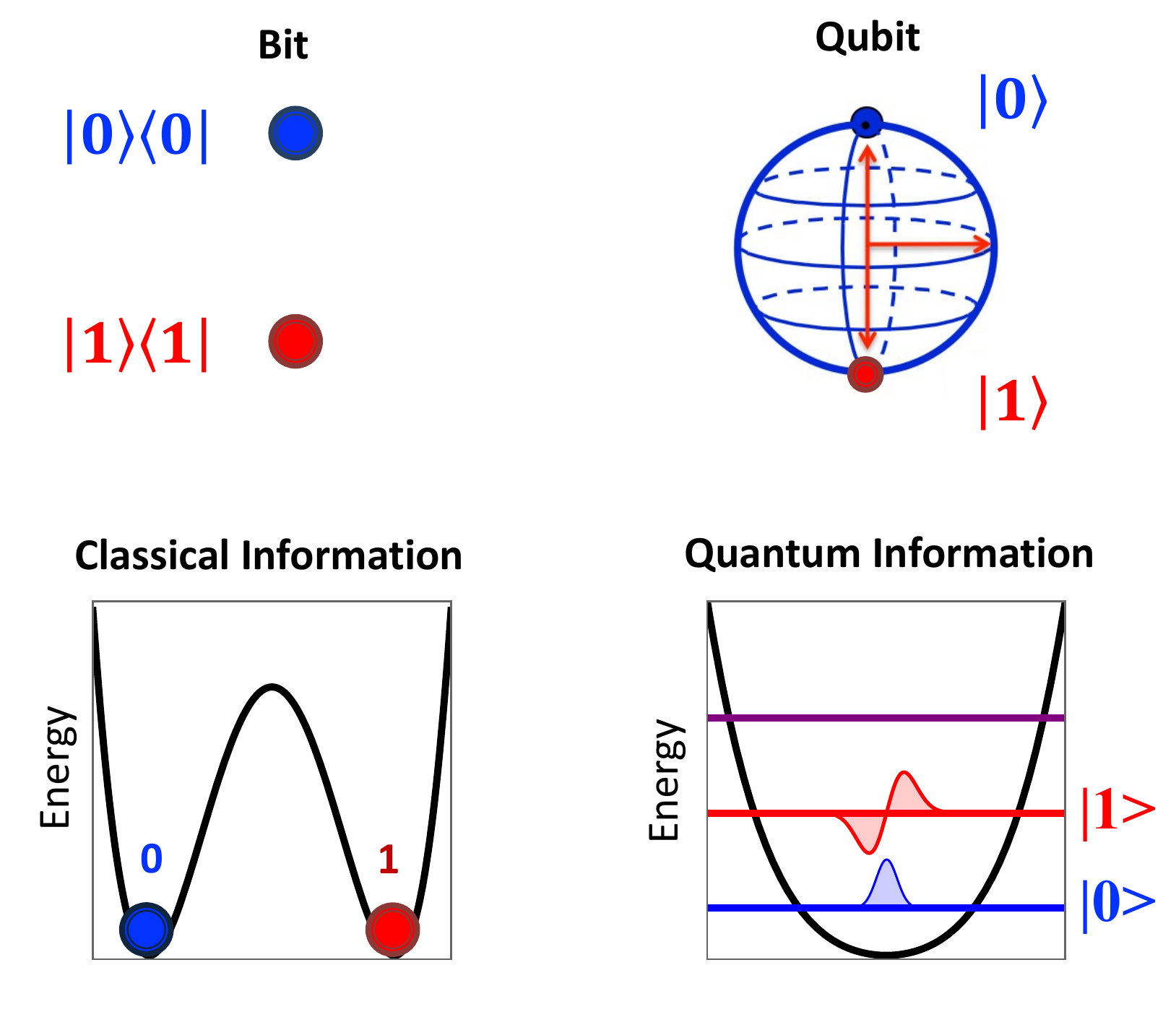}
\caption{A classical bit (top left) can only ever be found in the states $\ketbra{0}{0}$ or $\ketbra{1}{1}$
(bottom left). 
Physically, this is created with two deep potential wells that trap the physical state in one region or the other.
A quantum qubit (top right) can be in any superposition of states $\ket{0}$ and $\ket{1}$ (bottom right).  
Physically, this is created using the ground and first excited states of a quantum system.
}
\label{fig:qubit-bit}
\end{figure}

Just as the  \emph{bit} is the smallest unit of classical information, the  \emph{qubit} is the smallest unit of quantum information. 
Both are shown schematically in  Fig.~\ref{fig:qubit-bit}.
The bit can only ever be in one of two classical states: 0 or 1. 
In the language of quantum information, we might call these classical states $\ket{0}$ and $\ket{1}$.
However, it would be clearer to label these states as density matrices $\ketbra{0}{0}$ and $\ketbra{1}{1}$ to signify the fact that, in the classical world,  superpositions are not allowed and, hence, these two states are mutually exclusive.
In other words, in the classical world,  only one of the two possible pure states $\ketbra{0}{0}$ and $\ketbra{1}{1}$ is allowed to exist at any given moment in time.

To make  a  \emph{physical bit} from a physical device, one must devise a physical system that is subject to a potential that has two deep wells, as shown in Fig.~\ref{fig:qubit-bit}. 
Then, one must ensure that the temperature of the system and, more generally, the possible range of energy exchanges with the environment  remains far below the energy barrier separating the two wells.

If one constructs a similar potential with a quantum mechanical device, then the wavefunction can potentially be in a superposition of all accessible states, including the lowest energy level,  called the  \emph{ground state} or  \emph{vacuum state}, as well as the higher energy levels, called the  \emph{excited states}, also shown in Fig.~\ref{fig:qubit-bit}.
To make a  \emph{physical qubit} from a physical device, one must confine the wave function to two states.
For this two-state system, $\ket{0}$ is the ground state and $\ket{1}$ is the first excited state. 
Again, one must control the temperature and the interactions with the environment so that the probable range of energy exchanges with the environment is much less than the energy differences between these two states as well as the energy differences between these two states and all other states.
Moreover, one must also be careful to control environmental interactions that cause decoherence and destroy superposition  and entanglement.

The wavefunction of a qubit lives in a 2-dimensional complex Hilbert space 
\begin{align}
 \ket{\waveFunc} = \sin{(\theta/2)} e^{i\phi/2}  \ket{0} + \cos{(\theta/2)} e^{-i\phi/2} \ket{1}
 .
\end{align} 
Due to the fact that the wavefunction is a ray in Hilbert space, i.e. it actually lives in projective Hilbert space, the overall complex constant is unimportant and is normalized to unity.  
Note, however, that for describing a normalized vector in Hilbert space, and for describing unitary transformations of physical states, one must include the overall complex phase factor, $ e^{i\chi} \ket{\waveFunc}$.

Thus, the qubit is described by two real parameters that form the surface of a sphere, called the \emph{Bloch sphere}.
By convention, the $+\zhat$ direction that points towards the top of the sphere corresponds to the $\ket{0}$ state while the $-\zhat$ direction that points towards the bottom of the sphere corresponds to the $\ket{1}$ state.
Thus, an arbitrary qubit state can also be defined by the direction that it points in; if the the unit direction vector is $\nhat$, then the state is denoted  $ \ket{\nhat}$.
 For example, $\ket{\zhat}:=\ket{0}$ and $\ket{-\zhat}:=\ket{1}$, but remember that these states are orthogonal.
 Similarly, one can define the states
 \begin{align}
 \ket{\pm\xhat} &= \left(\ket{0} \pm \ket{1}\right)/2^{1/2} 
 \\
 \ket{\pm\yhat} &= \left(\ket{0} \pm i \ket{1}\right)/2^{1/2}.
\end{align}

Given the definition above, when measurements are made along the $\zhat$ axis, the probability of  $\ketbra{0}{0}$ is $\prob_0=\cos^{2}(\theta/2)=(1+\cos{\theta})/2$ and the probability of $\ketbra{1}{1}$ is $\prob_1=\sin^{2}(\theta/2)=(1-\cos{\theta})/2$. 
In fact, the probability of measuring a given outcome is simply given by the \emph{binomial probability distribution} with probability $\prob_{1}$ of obtaining the value 1 on each trial.
For $\Nsamples$ trials, the binomial distribution has mean $ \prob_1\Nsamples$ and  variance $ \prob_0\prob_1\Nsamples $, so the variance of the mean decreases as the central limit theorem would predict, $  \prob_0 \prob_1/\Nsamples$.
In agreement with QM Postulate \ref{QM:measurement}: immediately after a measurement, the state is known with certainty, so the variance must vanish.
As $\Nsamples\rightarrow \infty $ for fixed probability, the binomial distribution is well-approximated by a  normal Gaussian distribution, but this approximation breaks down for small sample sizes or  if one of the probabilities becomes vanishingly small.
For example, if $\Nsamples\rightarrow \infty$, but the product $\lambda =\prob_1 \Nsamples $ is held fixed, then the binomial distribution tends toward the Poisson distribution with mean and variance $\lambda$.

There is also the possibility of rotating the measurement basis to point along the $\xhat$ or $\yhat$ directions. 
In this case the probability is controlled by the angle $\phi$ rather than $\theta$.
In practice, it usually easier to keep the measurement apparatus fixed and, instead, to rotate the state.
For example,  to measure the probabilities along the $\pm\xhat$ axis, simply 
rotate the state counter-clockwise around the $\yhat$ axis by $\mp90^\circ$.

It is useful to explicitly define the effect of a rotation as a unitary operation on the qubit.
Define the Pauli vector as the vector of Pauli matrices $\vec \PauliOp  = \left( \PauliOp _\xcoord,  \PauliOp _\ycoord,  \PauliOp _\zcoord\right)^T$.
A rotation around the unit vector $\nhat$ by the angle $\theta$ is defined by 
\begin{align}
\RotationOp_\nhat(\theta)=e^{-i\theta \nhat\cdot \PauliOpVec /2}=\cos{(\theta/2)}\UnitOp -i \sin{(\theta/2)}\nhat\cdot \PauliOpVec.
\end{align}
Note that a rotation by $\theta=2\pi$ multiplies the qubit by the overall phase $e^{-i\pi}=-1$. 
This explicitly demonstrates the fact that the special unitary group, $\SU(2)$, is actually a double cover of the special orthogonal group of rotations $\SO(3)$.

A classical probability distribution over qubit wavefunctions is equivalent to a positive Hermitian matrix called the density matrix
\begin{align}
\densityOp = \begin{array}{cccc}
 &\densityMatrix_{00} \ketbra{0}{0} &+& \densityMatrix_{01} \ketbra{0}{1} 
 \\
+& \densityMatrix_{10} \ketbra{1}{0} &+& \densityMatrix_{11} \ketbra{1}{1}
\end{array}
.
\end{align} 
Due to Hermiticity, the off-diagonal elements must be complex conjugates,  $\densityMatrix_{10}^*=\densityMatrix_{01}$ and the diagonal elements must be real.
These diagonal elements represent the probability of measuring the two states, $\ketbra{0}{0} $ and $\ketbra{1}{1}$, and are  guaranteed to be positive due to the positivity of the density matrix. 
In order for the probabilities to sum to one, there is the constraint $\Tr{\densityOp}=\densityMatrix_{00}  + \densityMatrix_{11} =1$. 
The off-diagonal elements represent non-classical superposition states that have no analogue in the classical world.

Again, due to Hermiticity, the density matrix can be diagonalized, so that it has the form
\begin{align}
\densityOp =  
 \prob_+ \ketbra{\nhat}{\nhat}  + \prob_{-}  \ketbra{-\nhat}{-\nhat}
\end{align} 
 where $\ket{\pm \nhat} $ are orthogonal states and $\prob_\pm$ are the classical probabilities of each of these states occurring.
Any 2$\times$2 density matrix can also be written in the form
\begin{align}
\densityOp =  \left[ \UnitOp  +  (\prob_+-\prob_-) \nhat \cdot \PauliOpVec\right] /2.
\end{align} 
Thus, the qubit density matrix can be visualized as an arrow of length $\abs{\prob_+-\prob_-}$ pointing in the $\nhat$ direction, called the \emph{Bloch vector}. 
A proper mixed state must have $\abs{\prob_+-\prob_- }<1$.
 A pure state can only have eigenvalue 1 for one entry and eigenvalue 0 for the other, so that $\Tr\densityOp^a=1$ for any $a>0$. 
In contrast, for a \emph{proper mixed state}, both eigenvalues must be less than 1, so that $\Tr(\rho^a)<1$  for any power $a>1$ (and $\Tr(\rho^a)>1$  for any power $0<a<1$).

Given the discussion above,  a classical probability distribution function (PDF) over classical bits is simply a diagonal density matrix
\begin{align}
\pdfOp = \Diag{(\densityOp)} = \densityMatrix_{00} \ketbra{0}{0} + \densityMatrix_{11} \ketbra{1}{1}
.
\end{align} 
Because it is diagonal, it can also simply be considered to be a function of the classical states alone, $\pdf: \set{0,1}\rightarrow \Reals$, with the two values $\pdf(0)=\prob_0= \densityMatrix_{00}$ and $\pdf(1) =\prob_1= \densityMatrix_{11}$. 
Again the normalization condition, $\Tr{(\pdfOp)}=1$, ensures that the probabilities sum to unity.

\subsection{Multiple Qubits \label{sec:qinfo:qubits}} 

\begin{figure}
\centering
\includegraphics[width=3in]{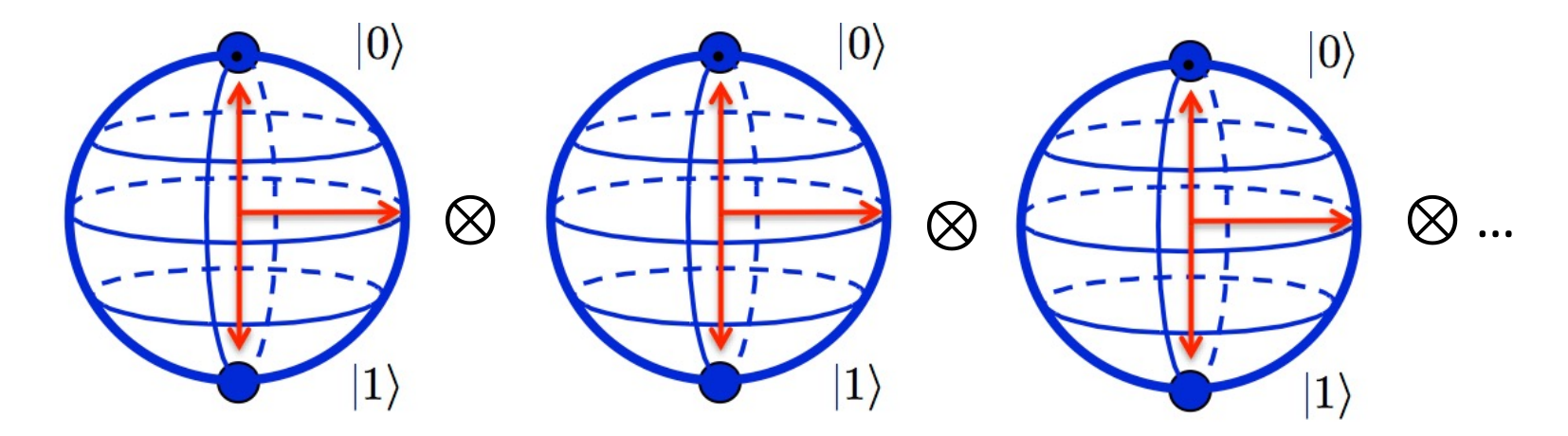}
\caption{The Hilbert space of multiple qubits is the tensor product of the Hilbert spaces of each qubit in the system.
The total composite Hilbert space dimension is $2^\nqubits$
where $\nqubits$ is the number of qubits.}
\label{fig:qubit-many}
\end{figure}

In the following, we assume that individual qubit states can be physically prepared and measured in the same basis each time, which is referred to as the \emph{computational basis}.
Two qubits live within a four-dimensional Hilbert space, which is the tensor product of the two individual qubit Hilbert spaces. Hence, basis states that span the two qubit Hilbert space can be denoted $\ket{q_1 q_2}=\ket{q_1}\otimes \ket{q_2} $, where $q_{1,2}$ refer to the state of  qubit 1 and 2 respectively.
There are 4 possible measurement outcomes, which correspond to 4 possible classical bit strings.
However, there are many more quantum states that are not classical, corresponding to 3 complex  degrees of freedom.

In addition to single-qubit superpositions, the key resource to be exploited is the ability to create \emph{entangled states}.
Entangled states are defined as states that cannot be represented as tensor products of the states of the individual subsystems.
Thus, a state such as $\ket{\xhat \xhat}=(\ket{00}+\ket{01}+\ket{10}+\ket{11})/2$ is not entangled, even though each qubit is in a superposition state itself.
In contrast, a two-qubit state such as the \emph{Bell state} $(\ket{00}+\ket{11})/2^{1/2}$ is truly entangled.
When quantum information  is entangled, but only part of the system can be measured,  this leads to kind of information hiding that causes decoherence. 
Note that the concept of entanglement explicitly refers to consideration of the tensor product structure of the Hilbert space, and, hence, the basis that is being used.

In order to find the size of a multi-qubit wavefunction, simply apply QM Postulate \ref{QM:Composite_Hilbert-space}:  multiply the dimensions of the Hilbert spaces of the subsystems and apply the ray condition.
As illustrated in Fig. \ref{fig:qubit-many}, for $n$ qubits, there are $\Nstates=2^\nqubits$ computational basis states, which is equivalent to the number of possible measurement outcomes, and, hence, to the number of possible classical bit strings. 
For $\nqubits$ qubits, the wavefunction is specified by $\Nstates = 2^{\nqubits}-1$ complex numbers. 
Although classically, the state is only ever in one of the possibilities, a classical probability distribution (PDF) of $\nqubits$ bits  is specified by $\Nstates=2^\nqubits-1$ real numbers.
Hence, the quantum wavefunction holds exactly twice the information of the classical PDF due to the fact it has a complex phase associated with each probability amplitude.
Yet, the the classical PDF should really be compared to the quantum density matrix, which is specified by $\Nstates^2-1=2^{2\nqubits}-1$ real numbers, and, thus, has quadratically more information.

In many ways, these simple facts are the reason for all of the excitement about quantum computing.  
This capacity of a quantum memory register for holding  an exponentially large amount of information and the potential ability of a quantum computer to efficiently manipulate this exponentially large amount of information is amazing from the standpoint of classical computation.

This also implies that performing a direct classical simulation of a multi-qubit quantum system is exponentially hard!
Although many ingenious algorithms have been invented for approximately solving these kinds of problems, progress with direct solvers is limited due to the exponentially large size of the required Hilbert space.
Classical computations much use drastically reduced models to even attempt the simulation of large systems. 
Now we can understand the first part of Feynman's statement: \emph{If you want make a [quantum-mechanical] simulation, you'd better make [the computational engine] quantum mechanical ...}  so that it has a sufficiently large Hilbert space and so that this Hilbert space can be manipulated efficiently.

Note that classical algorithms that can efficiently approximate the classical PDF are also quite powerful.
From the simple scaling of the amount of information in the classical PDF vs. the quantum density matrix, one might expect that  quantum algorithms can generically outperform classical probabilisitic algorithms by this same quadratic factor.
As will be discussed in detail later, this is indeed the case.\cite{HeinrichNovak01arxiv, Montanaro15prsa}

\subsection{Qudits \label{sec:qinfo:qudits}} 
\begin{figure}
\centering
\includegraphics[width=2in]{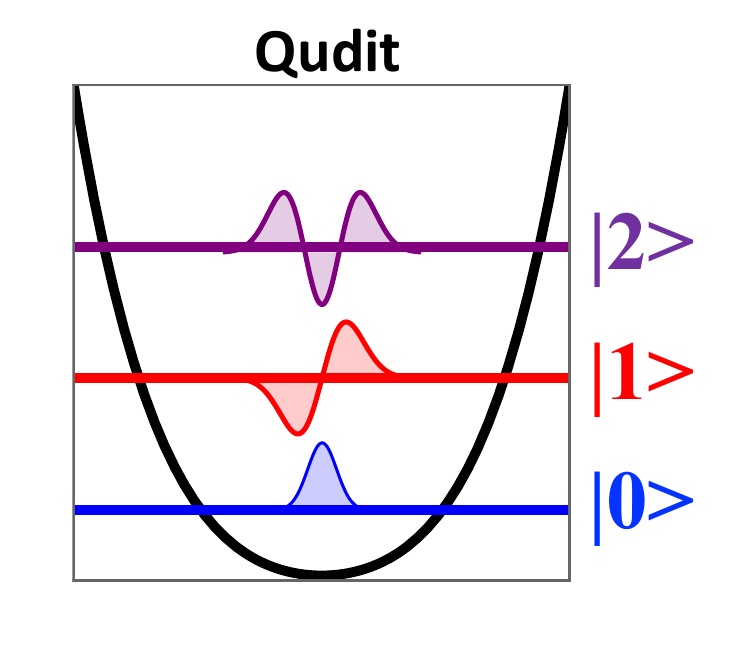}
\caption{Qudits use more than 2 states, e.g. $\ket{0}$, $\ket{1}$, $\ket{2}$, \dots.
}
\label{fig:qudit}
\end{figure}

Just as classical information with $d$ possible values is referred to as a \emph{dit},
the quantum analogue of a wavefunction with $d$ possible states is called a \emph{qudit}.
For both the classical and quantum cases, systems with multiple bits and dits can be used to simulate each other with polynomial efficiency, so there is no change in computational complexity class.
In the classical world, transistors are so cheap that essentially all modern computing technology is based on bits.
In the quantum world, experimentalists are still striving to reduce the  error budget to the point at which error correction can be applied.
For this reason, a number of researchers have proposed using the higher energy levels of a quantum system to encode information, as illustrated in Fig.~\ref{fig:qudit}.
For certain quantum device platforms, these higher energy states are naturally present, so it could be a good idea to effectively pack more qubits into a single device, potentially allowing less error-prone communication between these states. \cite{Wu20prl}
However, since decoherence rates are often faster for higher energy levels, there is clearly a tradeoff in performance that must be investigated for each experimental hardware platform.

\section{Quantum Computation \label{sec:qcomp}}

\subsection{Models of Universal Quantum Computation  \label{sec:qcomp:model}}
\emph{Universal quantum computation} \cite{KitaevBook, NielsenBook, KayeBook, PreskillNotes} refers to models of quantum computing that can efficiently simulate a universal classical computer, i.e. a Turing machine.
There are a number of different models of universal quantum computation that have been proven to be equivalent to one another in computational capabilities; i.e. they can simulate each other with polynomial complexity. 
Popular models include the \emph{digital circuit model}, \cite{KitaevBook, NielsenBook, KayeBook}
the \emph{quantum walk model}, \cite{Childs09prl, Childs13sci}
the \emph{adiabatic model},  \cite{Farhi00arxiv}
the \emph{quantum annealing model}, \cite{Farhi01science, Morita08jmp}
and the \emph{measurement-based model}. \cite{Raussendorf03pra}
The digital circuit model, also known as the \emph{digital quantum computational model}, has power and simplicity and is perhaps closest to digital classical computation based on classical circuits, which is familiar to most scientists and engineers today.

The basic idea of the digital circuit model is that quantum states can be transformed efficiently via linear unitary operations: 
$\waveFunc=\UnitaryOp\waveFunc_0$.
In fact, according to QM Postulate \ref{QM:UnitaryEvolution}, this is the only operation that can be performed without making a measurement. 
After making a measurement, the state generically becomes a density matrix $\densityOp$, but in between measurements, linear unitary operations can be performed: $\densityOp=\UnitaryOp\densityOp_0\UnitaryOp^\dagger$.
The fact that a quantum computer can do this efficiently is unbelievable from the classical standpoint, because it allows one to efficiently perform matrix-vector operations with exponentially large matrices and exponentially large vectors.

While there are a large number, $\Nstates^2$, of independent unitary operations, they are generated by a small, $\sim \CO(\nqubits)-\CO(\nqubits^2),$ number of basic operations called a \emph{gate set.}
Single qubit operations can be performed efficiently with just two standard gates  that perform rotations of the Bloch sphere.
Rotations by angle $\theta$ around the $\xhat$, $\yhat$, and $\zhat$ axes of the Bloch sphere, are denoted $\RX(\theta)$, $\RY(\theta)$, and $\RZ(\theta)$, respectively.
A rotation by angle $\theta=\pi$, simply generates the Pauli operations, denoted $\XOp  = \PauliOp_\xcoord=\Operator{NOT}$,  $\YOp =\PauliOp_\ycoord$,
and $\ZOp=\PauliOp_\zcoord$. 
A standard result in Lie group theory says that only two such operators are required to generate the whole group.
Another important example is the Hadamard gate:
\begin{align}
\HadamardOp:=\ZOp\circ \RY(-\pi/2)=\frac{1}{2^{1/2}}
\left(
\begin{array}{cc}
1 & 1\\
1 & -1
\end{array}
\right).
\end{align}
 Thus, starting from the $\ket{\zhat}:=\ket{0}$  
state pointing along the $\zhat$ axis, this generates the uniform superposition state $\ket{\xhat}$.
Similarly, starting from the state $\ket{-\zhat}=\ket{1}$, this generates the superposition state $\ket{-\xhat}$.

Adding just one nontrivial 2-qubit entangling gate between nearest neighbor qubits is enough to generate all other operations.
The 2-qubit Hilbert space is a 4-state system and these operations have the form of a $4\times4$ matrix.
Common choices are $\CNOT=\Operator{CX} $ and $\CZ$
\begin{align}
\CNOT := \left(
\begin{array}{cccc} 
1 &0 && \\
 0 & 1 && \\
& &0 & 1\\
& & 1 & 0
\end{array}
\right) 
&& 
\CZ := \left( 
\begin{array}{cccc} 
1 & && \\
& 1 && \\
& & 1&  \\
& & & -1  
\end{array}
\right) .
\end{align}
Another useful operation is the $\SWAP$ operation which simply exchanges information between qubits
\begin{align}
\SWAP :=\left( 
\begin{array}{cccc} 
1&\\
& 0 & 1 \\
& 1 & 0 \\
& & & 1
\end{array}
\right) .
\end{align}

If $\UnitaryOp_{2 \times 2}$ is a single-qubit unitary and $\UnitOp_{2\times 2}$ is the single-qubit unit matrix,  then a controlled $\UnitaryOp$ operation has the form
\begin{align}
\CUOp:=\left( 
\begin{array}{cc} 
\UnitOp_{2\times 2} &\\
& \UnitaryOp_{2\times 2}
\end{array}
\right) .
\end{align}
This shows that the operation $\UnitaryOp$ is performed if the first qubit is in the $\ket{1}$ state.
To make this clearer, use the notation $\CUOp_{jk}$ to indicate the fact that the state of the $\kindex$'th qubit is controlled on the state of the $\jindex$'th qubit. 

\begin{figure}
\centering
\includegraphics[width=2.5in]{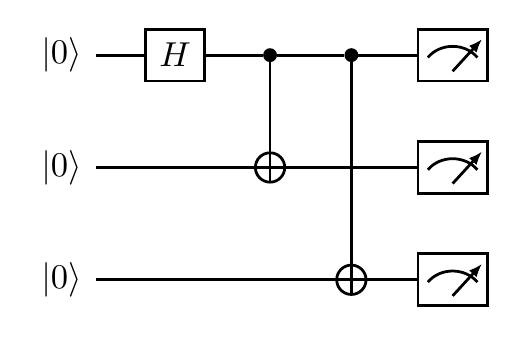}
\caption{An example of a circuit diagram for the creation of the entangled GHZ state $\left(\ket{000} + \ket{111}\right)/2^{1/2}$.
Apply $\HadamardOp$ to the first qubit and $\CNOT$ gates between the second and third qubits and the first qubit.
Finally, measure the result.
}
\label{fig:ghz}
\end{figure}

A sequence of unitary operations can be visualized through a \emph{circuit diagram}.
\emph{Operations are applied from left to right} in a circuit diagram -- which is opposite to the usual right to left operator ordering convention in mathematical expressions.
A circuit diagram has a number of lines going from left to right that each indicate individual qubits.
Blocks in the middle of the diagram represent single-qubit operations.
Blocks that are controlled on another qubit have a line from the block to the control qubit.
For example, consider the creation of the entangled  \emph{Greenberger-Horne-Zeilinger (GHZ) state}\cite{Greenberger07arxiv} on three qubits defined by $(\ket{000}+\ket{111})/2^{1/2}$ (an analogue of the Bell state for two qubits). 
This state can be created with the circuit diagram shown in Fig.~\ref{fig:ghz}.
The instructions are to: 
(i) apply $\HadamardOp$ to the first qubit; 
(ii) perform a $\CNOT_{1,2}$ operation, where the state of the second qubit is controlled by the state of the first qubit; 
(iii) perform a $\CNOT_{1,3}$ operation, where the state of the third qubit is controlled by the state of the first qubit; and
(iv) finally, the quantum state can be verified by measuring all qubits.
If one rotates the states before measurement, then one can test the consequences of \emph{Bell's theorem}.\cite{Bell1964physics}

The fundamental discovery of early research on quantum algorithms is that many useful computations can be performed using relatively few $\CO(\poly(\nqubits))$ of these basic gate operations.\cite{KitaevBook, NielsenBook, KayeBook} 
Key resources are the ability to create superpositions and entanglement and  to use interference to one's advantage.
This is often called \emph{quantum parallelism.}
The second fundamental discovery is that any \emph{reversible classical computation} can also be performed. \cite{KitaevBook,NielsenBook,KayeBook}
This requires first computing and storing the result and then \emph{uncomputing} or reversing the order of operations on the original memory register. \cite{Bennett1982ijtp}

However, emulating a classical computer usually comes with non-negligible overhead, and so, one might prefer to use a classical computer to perform certain operations such as classical logic.
For this reason, the designers of quantum computer platforms usually consider a hybrid architecture where classical processing units (CPUs) sit beside as well as  drive quantum processing units (QPUs) much like today's heterogenous CPU and graphical processing unit (GPU) architectures.
Like CPU/GPU platforms, there will be a large overhead in input/output (I/O) operations that exchange data between CPUs and QPUs.
In fact, the I/O will be exponentially large if one attempts to exchange the entire quantum random access memory (QRAM). 
In today's world, QPUs are also far more expensive than CPU's so hybrid architectures are an  obvious choice. 
While it may be difficult to envision today, perhaps at some point in the future, quantum computers will become useful enough and cheap enough that they will become ubiquitous.

\subsection{Limitations of the Quantum Computing Model \label{sec:qcomp:limits}}
Now it is very important to discuss some key limitations of the universal quantum computing paradigm.\cite{KitaevBook,NielsenBook,KayeBook}

\paragraph*{Approximating an arbitrary unitary is exponentially hard.}
This implies that only certain classes of unitaries, precisely those that are composed of $\CO(\poly)$ basic gate operations, can be constructed ``easily.''

\paragraph*{ Initializing all quantum information is exponentially hard. \cite{NielsenBook}}
Initializing quantum information requires applying an arbitrary unitary operation (usually starting from the ground state). 
So, this also implies that initializing all information in quantum memory is exponentially hard.

\paragraph*{ Measuring all quantum information is exponentially hard. \cite{NielsenBook}}
Measuring all quantum information requires making an exponential number of measurements.  
Moreover, each measurement generically requires a different unitary transformation to be applied.
 
\paragraph*{ Directly sampling a quantum state to estimate the probability amplitude costs the same number of measurements as it does classically. }
As explained previously, the  probability distribution for measurements of the state of a qubit is given by the \emph{classical} binomial distribution.
For probabilities far from 0 or 1, there is no escape from the central limit theorem: if the measurement outcome has a finite mean and finite variance, the relative accuracy with which it can be measured converges as $1/\sqrt{\Nsamples}$ where $\Nsamples$ is the number of samples.

\paragraph*{ Directly measuring exponentially small probability amplitudes is exponentially hard. \cite{NielsenBook} }
Since these values would only be observed with exponentially small probability, one must make an exponentially large number of measurements to observe the values with any accuracy.

\paragraph*{ Nonzero failure probability. \cite{KayeBook} }
Quantum algorithms are probabilistic and, as such, they have a certain probability of failure.  This failure probability can be made quite small by batching many runs together (see the clear explanation of the \emph{Powering Lemma} in Ref.~\onlinecite{Montanaro15prsa}). However, generically, the failure probability does not vanish and this feature is quite different from   classical deterministic algorithms.

Thus, it is misleading to say that the universal quantum computation model offers a method of efficiently simulating any quantum system.
In fact, in order to remain efficient, this model can only approximately simulate certain quantum systems  and it can never actually exploit the entire Hilbert space.
Although we began with the notion that there is an exponential amount of information that can be stored and manipulated, we now see that the actual amount of such information that can be manipulated and measured is much smaller than one might have hoped. 

In fact, it is useful to compare the power of the quantum computing model to classical probabilistic models of computation, such as Monte Carlo methods.
For probabilistic models, one never manipulates all of the information required by a direct calculation, rather, one relies on the central limit theorem to provide a slow but steady increase in relative accuracy as the number of individual samples is increased.

The quantum computational model is certainly a probabilistic model and many runs of the same algorithm are needed to yield a final answer.
After the required measurements are made for each run, one can associate the outcome with a \emph{quantum trajectory} that is consistent with the outcome. \cite{Plenio98rmp}
It is only by assembling a large number of quantum trajectories  that an accurate answer is achieved. 
In a certain sense, the key difference is the fact that quantum trajectories can exploit superposition states and interference effects rather than classical trajectories which are limited to computational basis states alone.

Nonetheless, there are many elegant mathematical theorems that prove that the quantum computational model can exceed even the classical probabilistic counterpart. \cite{KitaevBook, NielsenBook, KayeBook}
Generically speaking, the ability of the quantum density matrix to store quadratically more information than the classical PDF enables up to a quadratic speedup over classical probabilistic methods. \cite{Kempe2003quantum}
However, there are certain classes of problems for which super-polynomial and even exponential speedup has been proven to hold.\cite{Childs2002example}

\subsection{  Quantum Error Correction: Improving the Information Confinement Time \label{sec:qcomp:error}}

From the inception of quantum computing, it was recognized that the ability to correct quantum errors is indispensable for building large-scale quantum computers. 
\emph{Quantum error correction} (QEC) protocols \cite{NielsenBook, KayeBook, KitaevBook} seek to undo errors that are induced by the inability to enact the desired unitary operations with perfect precision as well as the uncontrollable interactions with the environment that cause  decoherence. 
The general strategy is to use an encoding that repeats information so that it can be checked against various types of error \emph{syndromes}. 
This implies that there is a distinction between the number of \emph{physical qubits} that comprise the quantum hardware platform and the number of \emph{logical qubits} that can encode quantum information for a given time period with negligible error rates.
The subject of QEC is vast and we cannot do justice to the wide range of ideas necessary to understand the contemporary approaches to the field. 
The interested reader can begin by consulting standard textbooks,   \cite{NielsenBook, KayeBook, KitaevBook} 
books on quantum error correction, \cite{Gaitan2008quantum, Djordjevic2012quantum, Lidar2013quantum}
 and relatively recent review articles. \cite{Fowler12pra, Devitt13rpp, Terhal15rmp, Campbell17nature, Roffe19cp}
 
Generally speaking, quantum error correction protocols seek to undo errors that are induced by the inability to enact the desired unitary operations with perfect precision as well as the uncontrollable interactions with the environment that cause  decoherence. 
As discussed in Sec.~\ref{sec:qinfo:open-quantum}, if the system of interest has $\Nstates$ states, then, each time one attempts to enact a quantum gate, there are $\Nstates^4-\Nstates^2$ ways, including both decoherence operations and accidental unitary operations, that the intended gate can go awry.
If one is attempting to obtain an exponential speedup by utilizing a large number of qubits, so that $\Nstates=2^\nqubits$, then there are an exponential number, $\sim\Nstates^4=2^{4\nqubits}$. of  gate features that must be controlled.
Just obtaining enough measurements to fully characterize the operation of the gate then becomes an intractable problem.
Now we can understand the second part of Feynman's quote: \emph{by golly it's a wonderful problem because it doesn't look so easy}.

For example, from the point of view of gate design, it is very efficient to have all-to-all connectivity between qubits, as is the case for ion trap platforms. \cite{Bruzewicz2019apr}
This eliminates the need for many extra $\Operator{SWAP}$ gates that would otherwise be needed to shuttle quantum information between different parts of the memory register. 
Because these  $\Operator{SWAP}$ gates must often be performed in serial, eliminating them also increases the potential for parallelization of multiple gate operations. 

However, from the point of view of error correction, it is far superior for qubits to be well-separated both in terms of their resonant frequencies and physical locations.
For example, this can be accomplished by carefully choosing and tuning the resonant frequencies and by physically arranging qubits into a lattice, as is the case for superconducting platforms. \cite{Krantz19apr}
Another example is the \emph{quantum charge coupled device} (QCCD) ion-trap architecture,\cite{Pino2021nature} which manages its error budget by moving inactive ions to distant locations from active gate regions.
If one can limit the amount of significant crosstalk between qubits to $m$ nearest neighbors alone,  then the number of gate features that need to be controlled is reduced to $2^{4m}$ per qubit.
The key point is that now $m$ is independent of the size of the memory register.

Known quantum error correction algorithms can be proven to be successful, allowing one to store and manipulate quantum information indefinitely, as long as the error rates are below a specific limit depending on the protocol, typically at $10^{-3}-10^{-5}$ per operation.
For the best performing approaches known today, e.g. the \emph{topological surface code}\cite{Dennis02jmp} with \emph{magic state distillation},\cite{Bravyi05pra} the number of physical qubits necessary to represent a logical qubit is estimated \cite{Fowler12pra} to be on the order of $10^3-10^4$. 
Hence, the number of qubits required to perform a useful calculation is estimated to be on the order of $10^5-10^6$ or more. \cite{Campbell17nature}
Unfortunately, this large overhead cost has been investigated and interpreted \cite{Babbush2021prxq} as implying that the first successful error-corrected quantum applications should offer speedups that are beyond quadratic, e.g. at least quartic.

After decades of theoretical development of error correction codes and experimental demonstrations of primitive steps, a surface code was recently demonstrated in a superconducting circuit, \cite{Krinner2022realizing} and a \emph{color code} was recently demonstrated on a QCCD ion-trap quantum computer. \cite{Ryan-Anderson21prx} 
These recent demonstrations are starting to approach the threshold where logical qubits outperform physical qubits.  

Since each operation takes a certain physical time period, the limit on the error rate can also be translated to a limit on the norm of the effective decoherence rate matrix, $\norm{\rateMatrix}$.
Thus, the basic idea of quantum error correction is to increase the \emph{information confinement time} until this limit can be achieved. 
Perhaps surprisingly, this goal is very similar to the Lawson criterion, which requires a net energy-producing fusion reactor to achieve a specific energy confinement time.
Within the field of fusion energy, in addition to inherent physical limitations of the reactor design, it is anomalous and often turbulent transport processes that typically control the energy confinement time.
Similarly, wthin the field of quantum computing,  in addition to inherent physical limitations of the hardware platform, it is decoherence due to anomalous interactions with the environment that controls the information confinement time.
Thus, it will be interesting to gauge the relative progress in the two fields by measuring the rate of improvement of the respective confinement times.

\subsection{A Brief History of Quantum Algorithms \label{sec:qcomp:qalgo}} 
The field of quantum algorithms has seen a rapid development in interest over the last two decades and it is not possible for a short introduction such as this to do justice to the wide range of brilliant ideas that have begun to bear fruit.
Nonetheless, when compared to classical algorithms, the number of essentially quantum subroutines are actually relatively few.
In part, it is the vast number of ways that these subroutines can be configured as well as the number of applications that these quantum subroutines can speed up that leads to the considerable complexity in the field. 
\emph{Moreover, since the field is still undergoing rapid development, there may be many new concepts waiting to be discovered.}

The challenge of performing computations using an exponentially large quantum Hilbert space was reconcieved as a potential  opportunity by Richard Feynman \cite{Feynman82ijtp, Feynman86fp}  and Yuri Manin \cite{ManinBook} in the early 1980's. 
Feynman reported that his interest in the subject was piqued by Charles Bennett, who had worked to understand reversible classical computation \cite{Bennett1982ijtp} and who wanted to understand how the physics would  work  when using the laws of quantum mechanics.  \cite{Feynman86fp}
In fact, it was Charles Bennett and Giles Brassard that made the first inroads into proving rigorous theorems that demonstrated that quantum communication channels had more information processing capacity than classical communication channels. \cite{NielsenBook,KayeBook} 

In the 1980's, Paul Benioff \cite{Benioff80jsp} and David Deutsch \cite{Deutsch85prsa} defined \emph{quantum Turing machines} and showed that that they were as powerful as classical Turing machines. 
In 1992, the \emph{Deutsch-Josza algorithm} \cite{Deutsch92prsa} was discovered, a concrete example of a toy problem that could be solved faster with a quantum computer than a classical one. 
Soon after,  the \emph{Bernstein-Vazirani algorithm} \cite{Bernstein97sjc} and \emph{Simon's algorithm}, \cite{Simon97sjc}
were discovered as well.
It turns out that these simple algorithms rely on the the ability to perform the Fourier transform more efficiently on a quantum computer. 

In 1994, Peter Shor's  algorithm \cite{Shor97jc} for factoring integers and taking discrete logarithms exponentially faster than a classical computer surprised the world with the realization that quantum computers can break classical security protocols relatively easily. 
The key step is estimating the periodicity of a sequence with unknown period and the key discovery is that there is an exponentially efficient factorization of the discrete Fourier transform on quantum memory registers -- today called the \emph{quantum Fourier transform} (QFT).
After using the QFT, the period can be estimated using \emph{ phase estimation} protocol to estimate the eigenvalue of a Hamiltonian, given an eigenvector.
Shor then went on to proving that there were \emph{quantum error correction protocols} that could be used to protect quantum information, \cite{Shor95pra} which gave the abstract theory a sense of physical realizability, perhaps for the first time.

The next breakthrough was Lov Grover's search algorithm \cite{Grover97prl, Grover98prl}  which speeds up unstructured search by the square root of the number of items.
Key elements of the search algorithm were then separated into algorithms for \emph{amplitude amplification}, \emph{quantum counting}, and \emph{amplitude estimation}. \cite{Grover98arxiv, Brassard00arxiv, Brassard02qcqi}
These subroutines form the basis for quantum algorithms for computing sums and integrals \cite{Abrams99arxiv, Heinrich01jc, HeinrichNovak01arxiv, Brassard11arxiv}
and forms the basis for quadratically speeding up classical Monte-Carlo algorithms. \cite{ Montanaro15prsa}

In the late 1990's, Seth Lloyd and Daniel Abrams discovered efficient algorithms for simulating \emph{$k-$local Hamiltonian} interactions. \cite{Lloyd96sci, Abrams99prl} 
The basic idea is to split the Hamiltonian into a relatively low number of terms that do not commute with each other and to then use the Lie-Trotter-Suzuki decomposition, aka \emph{Trotterization}, to approximate the exact unitary as a product of terms.  
For example, for Hamilonians defined on a 1D lattice with nearest neighbor interactions, one can split the Hamiltonian into a sum of three pieces, wherein all terms within each piece commute: a diagonal piece, an even off-diagonal piece, and an odd off-diagonal piece.
As another example, the single particle Schr\"odinger equation can be simulated efficiently \cite{Benenti08ajp} in a manner that mimics the phase space path integral by splitting the kinetic and potential energy terms into two pieces. Then one can repeatedly use the QFT and inverse QFT to switch back and forth between the momentum and position basis, so that each term is diagonal in the respective basis.
While operator splitting methods, such as Strang splitting and higher order Suzuki methods, \cite{Suzuki91jmp} are familiar to applied mathematicians, recent work \cite{Childs21prx} has led to new understanding of error convergence for these methods.

\begin{figure*}
\centering
\includegraphics[width=6in]{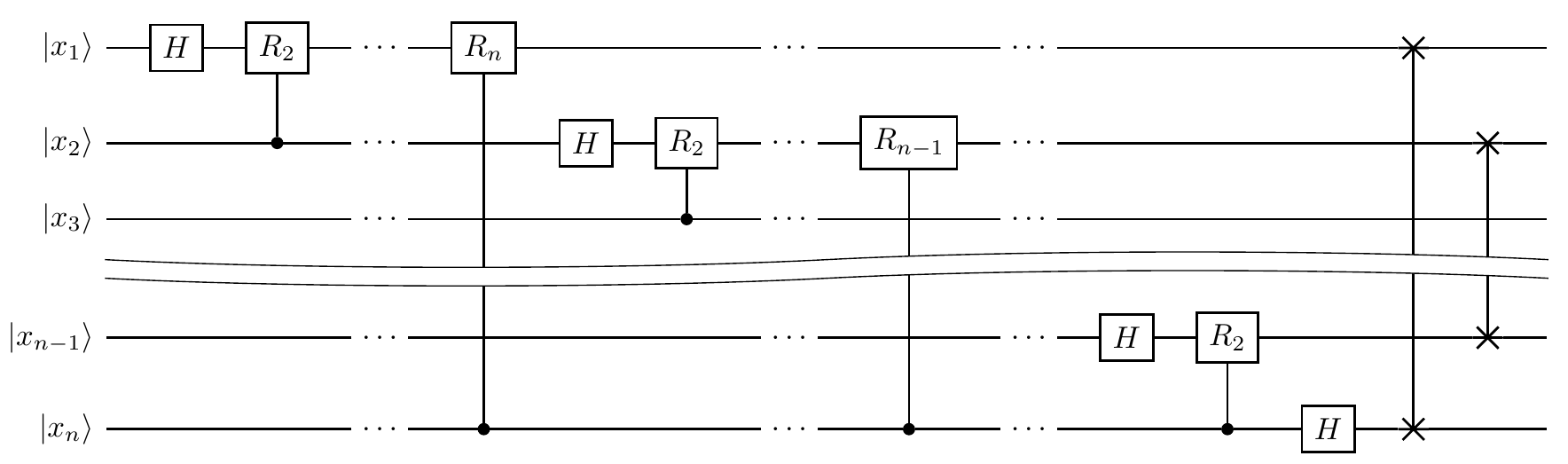}
\caption{Circuit diagram for the quantum Fourier transform $\QFTOp$.
Here, $\ROp_\kindex:= \ROp(2^{\nqubits-\kindex} \theta)=\ROp(2\pi /2^\kindex)$, where the phase gate $\ROp(\phi)$ is defined in Eq.~\ref{eq:Rphase}.
}
\label{fig:qft}
\end{figure*}

\emph{Quantum walks} \cite{Aharonov02arxiv}  are a vast generalization of the amplitude amplification paradigm that allow discrete state changes based upon the outcome of flipping a  \emph{quantum coin} or rolling \emph{quantum dice}; i.e. on the essential use of ancillary qubit registers. 
Quantum walks were originally constructed for the purpose of understanding simple models of quantum physics. 
Notably, Richard Feynman developed a quantum walk model for how Dirac fermions traverse a discrete lattice. \cite{FeynmanHibbsBook}
By the early 2000's, the essential ideas behind quantum walks were codified into a powerful set of algorithms \cite{Aharonov02arxiv} and were then proven to offer a universal computing model of their own. \cite{ Childs09prl, Childs13sci} 

Eventually, general-purpose Hamiltonian simulation algorithms that exploit sparsity in the Hamiltonian itself were developed. \cite{Aharonov03arxiv}
These algorithms were then improved by using 
quantum walks, \cite{Childs10cmp, Berry15focs}
new representations of Hamiltonians as a \emph{linear combination of unitaries} \cite{Childs12arxiv, Berry14stoc} (LCoU),
and as unitary approximations of the\emph{ truncated Taylor series} of the propagator. \cite{Berry15prl}
The latest development of these algorithms eventually achieved nearly optimal dependance on all parameters.  \cite{Berry16qic, Novo16qic, Low17prl, Martyn21prxq}

Hamiltonian simulation  powers the Harrow, Hassidim, \& Lloyd (HHL) sparse \emph{quantum linear solver algorithm} \cite{Harrow09prl} (QLSA) as well as improved versions with superior performance. \cite{Ambainis12arxiv, Childs17sjc, Gilyen19stoc, Martyn21arxiv}
Next, sparse \emph{quantum linear differential equation solver algorithms} (QLDSA) were developed that use linear solvers in combination with sparse Hamiltonian simulation algorithms. \cite{Berry14jpa, Berry17cmp} 
 
Today, there are new classes of optimal Hamiltonian simulation methods powered by
\emph{quantum signal processing} \cite{Low17prl} 
and \emph{qubitization}. \cite{Low19quantum} 
The term qubitization refers to the use of an ancillary qubit, as in quantum walks, while the term quantum signal processing refers to the way in which this ancillary qubit is manipulated to perform useful calculations.
The combination of these methods can generate nonlinear \emph{quantum eigenvalue transformations} (QEVT) of a block-encoded operator that is a normal matrix, $\Commutator{\HamiltonianOp}{\HamiltonianOp^\dagger}=0$. 
This approach was then generalized to generate nonlinear \emph{quantum singular value transformations} (QSVT) for an arbitrary matrix. \cite{Gilyen19stoc}
The use of QEVT and QSVT was shown to provide a grand unification of many quantum algorithms, \cite{Gilyen19stoc, Martyn21prxq}
including search, amplitude amplification. matrix multiplication, matrix inversion, phase estimation, and QFT. 
QSP has already been put to good use within quantum algorithms designed to solve plasma physics problems. \cite{Engel19pra, Novikau22arxiv}

\subsection{Key Quantum Subroutines  \label{sec:qcomp:qsub}}
There are a number of useful textbooks \cite{KitaevBook, NielsenBook, KayeBook} and review articles \cite{Childs10rmp, Montanaro16npjqi, Cerezo21nrp} on quantum algorithms.
Although the quantum computational model might take some time to become familiar with and new quantum algorithms with improved complexity are under rapid development, it is helpful to realize that there are actually only a few \emph{essentially quantum} subroutines that power the vast majority of quantum algorithms.

\subsubsection{Quantum Fourier Transform: Cost of $O(\log^2{\Nstates})$ rather than classical $\Nstates \log{\Nstates}$
\label{sec:qalgo:qft} }

In the 1990's it was discovered that the Fourier transform can be factorized in an extremely efficient manner on a quantum memory register.
Many Hamiltonians of interest can be simplified and, hence simulated efficiently, if the Fourier transform is exponentially cheap and if one does not need to extract intermediate information.
In fact, the key subroutine that powers Shor's algorithm for factoring integers and taking discrete logarithms is the Fourier transform.

There are many variations of the Fourier transform that arise from the character theory of commutative groups. 
The most familiar of these is the \emph{Walsh-Hadamard transform}, which arises from the group $(\Integers/2\Integers)^\nqubits$.
The Walsh-Hadamard transform on multiple qubits, $\HadamardOp_\nqubits$, is simply the tensor product of Hadamard gates
\begin{align}
\HadamardOp_\nqubits = \otimes_{\kindex=1}^\nqubits \HadamardOp .
\end{align} 
The subscript $\nqubits$ is not necessary, because the number of qubits to be acted upon should be clear from the context.
Evidently, this transform factorizes so well that it can be performed with single-qubit rotations alone.
It is simple to show by induction that this generates a uniform superposition of states from vacuum
\begin{align}
\superpositionState  :=  \HadamardOp \ket{0} =\sum_{\xcoord=1}^\Nstates \ket{\xcoord} /\Nstates^{1/2}
.
\end{align}
Then a quantum subroutine can work on the entire superposition at once: $\waveFunc =\UnitaryOp \superpositionState$.
This simple initialization step is used by many quantum algorithms.

\newcommand\floor[1]{{\lfloor #1 \rfloor}}
The discrete Fourier transform arises from the modular group $\Integers/\Nstates \Integers$ and is  defined via
\begin{align}
\bra{\kindex} \QFTOp \ket{\psi} = \Nstates^{-1/2}  \sum_{\jindex} \psi_\jindex e^ {2\pi i \jindex \kindex/\Nstates }
\end{align} 
For the binary  version of the algorithm,  $\Nstates=2^\nqubits$, and, one can express the index  in binary notation via
\begin{align}
\xcoord := \jindex/\Nstates= \sum_{\jindex=1}^\nqubits \xcoord_\jindex 2^{-\jindex}=0.\xcoord_1\xcoord_2\xcoord_3\dots \xcoord_\nqubits.
\end{align}

The \emph{quantum Fourier transform} (QFT), which is nothing but the usual Fourier transform, factorizes into the following form
\begin{align}
\QFTOp &= \Nstates^{-1/2}\sum_{\jindex\kindex} e^{2\pi i \jindex \kindex /\Nstates} \ketbra{\kindex}{\jindex}
\\
& =\Nstates^{-1/2} \sum_{\jindex\kindex} \left[ \otimes_{\ell=1}^{\nqubits}  e^{2 \pi i \jindex \kindex_\ell 2^{-\ell}  } \ket{\kindex_\ell}\right]\bra{\jindex}
\\
& = \sum_{\jindex} \left[ \otimes_{\ell=1}^{\nqubits} 2^{-1/2}\sum_{\kindex_\ell=0}^1  e^{2 \pi i \jindex \kindex_\ell 2^{-\ell}  } \ket{\kindex_\ell}\right]\bra{\jindex}
\\
& = \sum_{\jindex} \left[ \otimes_{\ell=1}^{\nqubits} 2^{-1/2}\left(\ket{0}+  e^{2 \pi i \jindex   2^{-\ell}  } \ket{1}\right) \right]\bra{\jindex}
.
\end{align} 
Now the state repeated within the tensor product is simply $\ROp(2\pi \jindex 2^{-\ell})\HadamardOp\ket{0} =\ROp( 2^{\nqubits-\ell}\xcoord \theta )\HadamardOp\ket{0} $ where  $\theta=2\pi/\Nstates=2\pi/2^\nqubits$ and  the single qubit phase rotation operator is defined as
\begin{align} \label{eq:Rphase}
\ROp(\phi)= \left( \begin{array}{cc} 1 & 0 \\ 0 & e^{i\phi } \end{array} \right)
.
\end{align}
 Thus, one can also write
\begin{align}
\QFTOp &=   \left[\sum_\jindex \otimes_{\kindex=1}^ \nqubits \left[\Operator{R}(  2^{\nqubits- \kindex}\xcoord   \theta ) \HadamardOp\right]  \ket{0}\bra{\jindex} \right]\\
&=\sum_\jindex  \left[\Operator{R}(  2^{\nqubits-1}\xcoord   \theta )  \HadamardOp \right] 
\otimes  \left[\Operator{R}(  2^{\nqubits-2}\xcoord   \theta ) \HadamardOp \right]  \dots  
\otimes	\left[  \HadamardOp \right] \ket{0}\bra{\jindex} .
\end{align} 
The effect of the factor $ 2^{\nqubits- \kindex}$ is to truncate the phase to the last $\nqubits-\kindex$ binary digits of $\xcoord$, i.e.
\begin{align}
\floor{\xcoord}_\kindex : =\floor {2^{\nqubits- \kindex}\xcoord }
=\sum_{\jindex=0}^{\nqubits-\kindex} \xcoord_{\kindex+\jindex}  2^{-\jindex-1} =0.\xcoord_{\kindex} \dots \xcoord_{\nqubits} 
\end{align}
so that
\begin{align}
\QFTOp &=\sum_\jindex   \otimes_{\kindex=1}^ \nqubits \left[\ROp( \floor{\xcoord}_\kindex   \theta ) \HadamardOp\right]\ket{0} \bra{\jindex} 
 \\
&=  \sum_\jindex \left[\Operator{R}( \floor{\xcoord}_1   \theta )  \HadamardOp \right] 
\otimes  \left[\Operator{R}(  \floor{\xcoord}_2   \theta ) \HadamardOp \right]  \dots  
\otimes	\left[  \HadamardOp \right] \ket{0} \bra{\jindex} .
\end{align} 
Now, it is easy to see that each rotation should itself factorize as the product of $\nqubits-k-1$ controlled rotations, 
$\CROp_{\ell,\kindex}$, between the $\kindex$'th qubit and all higher $\ell$'th qubits.
The operator
\begin{align}
\widehat{\ROp}( \floor{\xcoord}_{ \kindex}   \theta ) = \circ_{\ell=\kindex+1}^\nqubits \CROp_{\ell,\kindex}(  \xcoord_\ell \theta)
\end{align}
is almost correct, but it applies the phase to the $\kindex$'th qubit rather than the $\nqubits-\kindex$'th qubit and the desired rotations were supposed to be implemented in the opposite order within the tensor product.
This is easy to correct by first computing the result for both $\kindex$ and $\nqubits-\kindex$ and then swapping the results.
In other words, simply add a final stage of $\SWAP$ gates at the end defined by
\begin{align}
\Operator{SWAP}^{\floor{\nqubits/2}}:=\circ_{\ell=1}^{\floor{\nqubits/2}} \Operator{SWAP}_{\ell,\nqubits-\ell} .
\end{align} 
The final factorized expression is
\begin{align}
\QFTOp &= 
\Operator{SWAP}^{\floor{\nqubits/2}}
\sum_{\jindex=1}^\nqubits  \otimes_{\kindex=1}^\nqubits \left[\widehat{ \ROp}(\floor{\xcoord}_\kindex  \theta) \HadamardOp \right]\ket{0}\bra{\jindex}  
.
\end{align} 
The main QFT stage has $\nqubits$ single-qubit Hadamard gates and $\nqubits (\nqubits-1)/2$ controlled two-qubit rotations, for a total of $\nqubits (\nqubits+1)/2$ operations.
The final SWAP stage consists of $\floor{\nqubits/2}$ $\SWAP$ operations.
The swap stage can often be eliminated with a logical reordering of qubits in the next subroutine or by reordering the measurements that read out the ancillary register.
As shown in Fig.~\ref{fig:qft}, this circuit has a simple graphical structure.

\subsubsection{\label{sec:qalgo:phase_est}  Phase Estimation }

\begin{figure}
\centering
\includegraphics[width=3.4in]{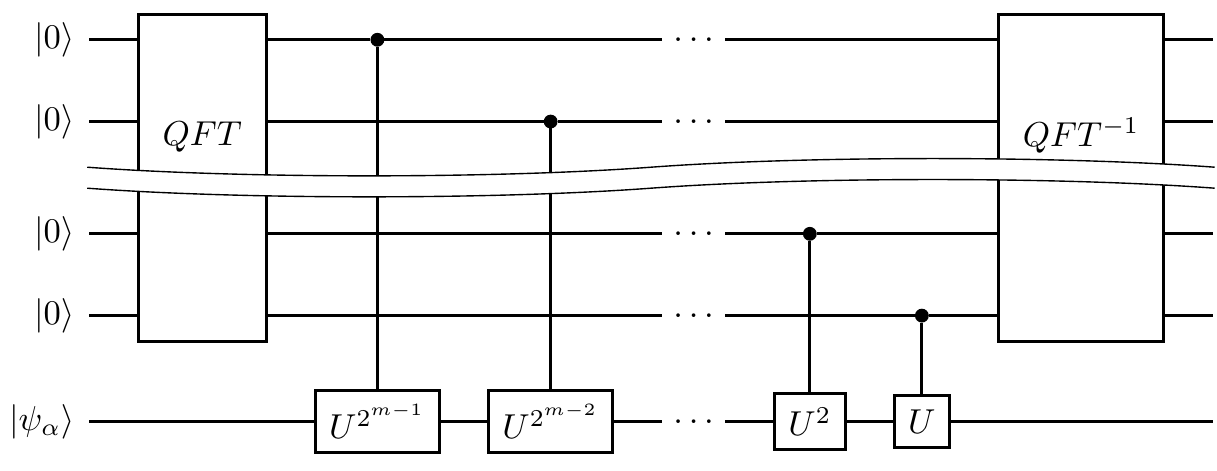}
\caption{Circuit diagram for quantum phase estimation (QPE), using QFT to read out the phase to an ancillary memory register. 
If the number of ancillary states is a power of 2, as in the case of an ancillary qubit register, then the initial $\Operator{QFT}$ can be replaced by $\HadamardOp$. 
}
\label{fig:qpe}
\end{figure}

\emph{Phase estimation} allows one to determine the eigenvalue of a unitary operator, given the ability to prepare an eigenvector.
If the application of the unitary, e.g. through Hamiltonian simulation, is inexpensive, then phase estimation is also inexpensive.
Assume that $\ket{\eigenvector}$ is an eigenvector of the unitary operator $\UnitaryOp$, so that $\UnitaryOp\ket{\eigenvector}=e^{-i\eigenphase}\ket{\eigenvector}$.
Then repeated applications of the unitary yields information about the less significant digits of $\alpha$, i.e. $\UnitaryOp^{2^\kindex}=e^{-2^\kindex i\eigenphase}\ket{\eigenvector}$.
This information can be stored in an ancillary quantum memory register by performing the operations
\begin{align}
\ket{\psi } =\sum_{\jindex=1}^\mqubits \ket{\jindex} \otimes \CUOp^{2^{\jindex-1}} \ket{\eigenvector} = \sum_{\jindex=1}^\mqubits e^{-i\, 2^{\jindex-1} \eigenphase}  \ket{\jindex} \otimes \ket{\eigenvector} .
\end{align}
Now performing the QFT on the ancillary register yields the binary digits of the eigenphase
\begin{align}
\QFTOp \ket{\psi } =  \sum_{\kindex=1}^\mqubits  \eigenphase_\kindex  \ket{\kindex} \otimes \ket{\eigenvector} .
\end{align}
The circuit diagram for quantum phase estimation (QPE) is shown in Fig.~\ref{fig:qpe}.

Thus, if the eigenvector can be prepared efficiently and the unitary can be simulated efficiently, then the QFT can be used to extract the eigenphase efficiently.
Note that there is nothing intrinsically quantum about this algorithm other than these assumptions.
In fact, Kitaev, et al.,  \cite{KitaevBook} presented a version of this algorithm that simply uses direct measurements of the phase $\cos{(2^\kindex \eigenphase)}$ and $\sin{(2^\kindex \eigenphase)}$ and dispenses with the ancillary qubit register altogether.

Phase estimation for certain classes of generalized eigenvalue problems, of the form commonly found in St\"urm-Liouville problems, MHD, and kinetic theory, was developed in Ref.~\onlinecite{Parker20pra}.

\subsubsection{\label{sec:amp_amp}
Amplitude Amplification: Cost of $\sqrt{\Nstates}$ rather than classical $\Nstates$ }

Understanding the key components of Grover's search algorithm led to the discovery of amplitude amplification and amplitude estimation.
While the utility of Grover's search is often debated because the cost of data entry (for an exponentially large phone book) must be amortized over an exponentially large number of searches, the utility of amplitude amplification and estimation is not in question.
Amplitude amplification can be used to amplify the overlap between two wavefunctions quadratically faster than this can be performed classically.
This is the key part of Grover's search that allows it to obtain the correct answer with high probability with quadratically fewer queries that the classical counterpart.

Assume that one has a \emph{function oracle operator} that marks chosen states by flipping their sign, i.e.
\begin{align}
\OracleOp_\OracleFunc \ket{\xcoord} = (-1)^{\OracleFunc(\xcoord)} \ket{\xcoord}
\end{align}
where $\OracleFunc{(\xcoord)}$ is a binary function that returns the values 0 or 1.
In particular, the oracle that marks a single particular state $\ket{\ycoord}$ will be denoted $\OracleOp_{\ket{\ycoord}}$ if $\OracleFunc(\xcoord)=-\delta_{\xcoord, \ycoord}$. 
(Note the minus sign).

The idea behind amplitude amplification is to generate a rotation that amplifies the amplitude of the subspace of interest, called the \emph{good subspace}.
Now, the product of two reflections is a rotation around the point of intersection of the two lines of reflection. 
The rotation angle  is twice the size of the angle between the two lines of reflection.
Clearly, the oracle is a  reflection operator, so, if one can find another suitable reflection, then it might be possible to generate a useful rotation.
In fact, there are an infinite number of suitable choices.

Let $\superpositionOp$ denote a unitary transformation that creates the state $\superpositionState=\superpositionOp\ket{0}$ from the vacuum state $\ket{0}$.
This state should be a superposition of all states to be searched over; for example, the choice $\superpositionOp=\HadamardOp$, which generates the uniform superposition state $\superpositionState = \HadamardOp\ket{0}$, is often a useful choice.
The amplitudes and phases of the superposition may be chosen almost arbitrarily, but only amplitudes that are not exponentially small will have a reasonable probability of being searched.
The \emph{search oracle operator} based on this state is
\begin{align} \label{eq:search_oracle}
\OracleOp_{\superpositionState}= 2\ketbra{s}{s}-\UnitOp=\superpositionOp \OracleOp_{\ket{0}} \superpositionOp^\dagger
\end{align}
where $ \OracleOp_{\ket{0}}=2\ketbra{0}{0}-\UnitOp$ marks the vacuum state $\ket{0}$.
The generalized \emph{Grover walk operator}, $\GroverOp $, alternates between the search oracle and the function oracle
\begin{align} \label{eq:grover_walk}
\GroverOp =\OracleOp_{\superpositionState} \OracleOp_\OracleFunc .
\end{align}

\begin{figure}
\centering
\includegraphics[width=3in]{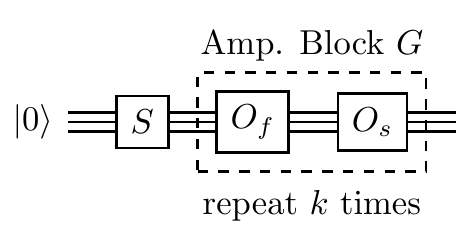}
\caption{Circuit diagram for quantum amplitude amplification.
To amplify to order unity, the amplification block is to be repeated a number of times, $\kindex$,  of order 1/$\abs{\rm amplitude}$.
}
\label{fig:qaa}
\end{figure}

Now assume one is given an initial state $\ket{\psi_0}$. 
Clearly, this can be decomposed into the sum of two parts: a part in the marked subspace, denoted $\ket{\good}$, and a part in the unmarked subspace, denoted $\ket{\bad}$, i.e.
\begin{align} \label{eq:psi_init}
\ket{\psi_0}=\cos{(\theta/2)} \ket{\bad} + \sin{(\theta/2)}\ket{\good}.
\end{align}
The action of the generalized Grover walk operator is to rotate the amplitude of the good and bad states via
\begin{align} \label{eq:psi_grover}
\ket{\waveFunc_\kindex}&=\GroverOp^\kindex \ket{\waveFunc_0} 
\\
&=\cos{((\kindex+\tfrac{1}{2}) \theta)} \ket{\bad} + \sin{(((\kindex+\tfrac{1}{2})\theta)}\ket{\good}
.
\end{align}
The amplitude amplification process is illustrated in Fig.~\ref{fig:qaa}.

For example, for Grover's search, choose the walk operator to be defined by the uniform superposition state generated by $\superpositionOp=\HadamardOp$ acting on the vacuum.
Assume that one is guaranteed that the function oracle $\OracleOp_\OracleFunc $ marks $\Mstates$ out of $\Nstates$  states.
If one begins with an initial uniform superposition, then $\sin{(\theta/2)}=\Mstates/\Nstates$.
By choosing the appropriate value of $\kindex$, i.e.
\begin{align}
\kindex &=  \round{ \left( \pi/2\theta\right)} -1/2 \\
\lim_{\Mstates/\Nstates\rightarrow 0} \kindex &=   (\pi/4) \sqrt{\Nstates/\Mstates} 
\end{align}
 one will generate a state that has the probability of measuring good states amplified to a high level.
 Then, simply performing measurements will yield these marked states with high probability.
 The final limiting form occurs in the limit that the dimension of the good subspace  is much smaller than the overall dimension $\Mstates /\Nstates  \rightarrow 0$.

\subsubsection{ \label{sec:amp_est}
Amplitude Estimation: Cost of 1/accuracy instead of classical 1/accuracy$^2$  }
\begin{figure}
\centering
\includegraphics[width=3in]{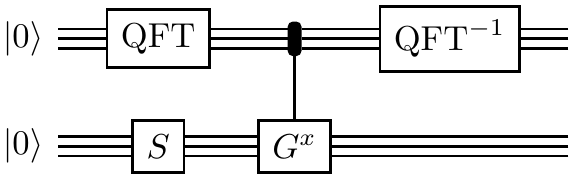}
\caption{Circuit diagram for quantum amplitude estimation (QAE), using QPE to read out the amplitude.
 The Grover walk operator operator,  $\GroverOp$ in Eq.~\ref{eq:grover_walk}, is defined by the choice of oracles and the choice of initial superposition operator $\superpositionOp$. 
 In this diagram, the multiply-controlled $\GroverOp^\xcoord$ block is defined as the block of controlled-$\UnitaryOp$ operations sandwiched between QFT's within the QPE diagram, Fig.~\ref{fig:qpe}, where $\GroverOp$ is used in place of $\UnitaryOp$.
If the number of ancillary states is a power of 2, as in the case of an ancillary qubit memory register, then the initial $\Operator{QFT}$ can be replaced by $\HadamardOp$. 
}
\label{fig:qae}
\end{figure}

Once amplified, the less significant digits of the amplitude can be measured efficiently. 
Hence, the amplitude can be measured with quadratically fewer steps than it can by directly sampling the probability. \cite{Brassard02qcqi}
In fact, whenever the answer that one seeks in encoded in a probability amplitude, then the only way to extract this answer more efficiently than by direct sampling is by using amplitude estimation.
Algorithms for quantum counting,\cite{Brassard02qcqi} which can be used to compute sums and integrals up to quadratically faster than classical probabilistic algorithms\cite{HeinrichNovak01arxiv, Montanaro15prsa} fall into this class.
Clearly, these algorithms are essential for many scientific computing applications.

Let the unitary $\superpositionOp$ generate the state of interest $\superpositionState=\superpositionOp \ket{0}$ whose overlap with the good subspace, marked by $\OracleOp_\OracleFunc$, is to be measured.
Assuming that both $\superpositionOp$  and $\superpositionOp^\dagger$ can be performed efficiently, the search oracle $\OracleOp_{\superpositionState}$ and the Grover walk operator, $\GroverOp$, can be constructed efficiently.

The essential idea of \emph{amplitude estimation} is to determine the least significant bits of $\theta$ through the action of powers of the Grover walk operator $\GroverOp^\kindex$.
For example, using phase estimation on an eigenvector of $\GroverOp$ would allow one to directly read out the value of the amplitude, $\cos{(\theta)}$.
In this case, we cannot construct an exact eigenvector, but we can construct the initial state $\ket{\psi_0}=\superpositionState$, which is a sum of two eigenvectors, as seen from Eq.~\ref{eq:psi_init} and \ref{eq:psi_grover}.
The result of applying phase estimation to $\superpositionState$, as shown in Fig.~\ref{fig:qae}, yields an equal superposition of the states $\ket{\theta}$ and $\ket{\pi-\theta}$ in the ancillary register. \cite{Brassard02qcqi}
Clearly, both states yield the same value for the amplitude.

\subsubsection{Quantum Walks:  Quadratic speedup}
Quantum walks on graphs are a vast generalization of the amplitude amplification paradigm. 
In its discrete-time version,  quantum walks allow state changes based upon the outcome of flipping a \emph{quantum coin}\cite{Aharonov02arxiv}  
(or by rolling \emph{quantum dice}).
Quantum walks accelerate the solution of various problems, e.g. the search hitting time, typically quadratically.
A heuristic explanation is that the the ability to explore superposition states increases the effective speed of the walker.

If one includes Grover's search, amplitude amplification, and estimation within this group, then the majority of quantum algorithms that achieve a quadratic speedup today use quantum walks in an essential manner.\cite{Childs10rmp}
Quantum walks have been used to speed up search on graphs, \cite{Shenvi03pra, Childs04pra}  accelerate simulation of Markov chains \cite{Szegedy04focs}, and  Chebeyshev cooling schedules for quantum annealing, \cite{ Somma2008prl, Wocjan08arxiv,Wocjan09pra}  as well as to solve the element distinctness problem. \cite{Ambainis07sjc}
Notably, they are useful for Hamiltonian simulation \cite{Childs10cmp,Berry2012qic} as well as for quantum signal processing. \cite{Low17prl}

Although quantum walks are deterministic, they share similarities with classical random walks which are used as the basis for classical probabilistic Monte-Carlo algorithms that simulate the evolution of a classical PDF in time.
Due to the Feynman-Kac formula, this evolution can often be ascribed to Langevin dynamics defined by a stochastic forcing of a deterministic system, i.e. a stochastic differential equation (SDE). 
This temporal evolution of the PDF can be used as the basis of search algorithms that seek to find marked states within a certain hitting time.  
Another important goal is to evolve the PDF towards a desired statistical distribution which may physically represent a canonical ensemble such as the Boltzmann distribution, the ground state of a wavefunction, or the solution to an optimization problem. 
After a few mixing times, the desired PDF is achieved with high accuracy and one can generate samples from the PDF and/or calculate the partition function. 
As the temperature is reduced, the PDF corresponds more and more closely to the ground state, which leads to the \emph{simulated annealing} approach to optimization and ground state preparation.

Szegedy introduced \cite{Szegedy04focs} a quantum walk analogue for any Markov process defined by a stochastic transition matrix. 
Like Grover's walk, it is composed of two reflections, but now on the tensor product of two Hilbert spaces.
He proved that the hitting and mixing times of the quantum walk are generically improved quadratically because the spectral gap for the quantum walk is the square root of that for the random walk.  
This quadratically improves the complexity of using simulated annealing to find the solution to combinatorial optimization problems in addition to the quadratic speedup in accuracy obtained through amplitude estimation. \cite{Somma2008prl, Wocjan08arxiv, Wocjan09pra}  
The improvement in convergence also improves the Chebeyshev cooling schedule. \cite{ Montanaro15prsa}
The complexity of general multi-level Monte-Carlo methods for SDEs can also generically be improved quadratically. \cite{An2021quantum}
Quantum algorithms based on these concepts have recently been explored for simulating turbulent mixing and for computing the reaction rate within turbulent flows. \cite{Xu2018aiaa, Xu2019ctm}

\subsubsection{Hamiltonian Simulation: Polynomial to super-polynomial speedup}
Simulation of many-body quantum systems was the inspiration for the development of quantum algorithms for quantum computers.
The first efficient algorithms proposed were based on the realization that many Hamiltonians of interest have a sparse decomposition into a sum of non-commuting parts, where each part is composed of commuting terms, and, hence, each part can be computed individually as a product of terms. \cite{Lloyd96sci} 
Then, one can use operator splitting techniques, such as the Trotter-Suzuki decomposition and Lie group decomposition formulas, \cite{HairerBook}  to approximate the unitary corresponding to the full Hamiltonian as a product of terms with sufficiently small time steps. 
In fact, while these techniques are well known to the operator splitting community, a rigorous formulation of the approximation theory was only recently developed. \cite{Childs21prx}

For example, for the Schr\"odinger equation for particles interacting with few-body potentials, the Hamiltonian breaks into a sum of the kinetic energy of all particles and the potential energy of all particles.
One can then efficiently switch back and forth between the momentum and position basis in order to evaluate each as a simple phase shift. \cite{Benenti01prl, Benenti08ajp}
There is also a similar decomposition for lattice Hamiltonians with nearest neighbor coupling.
These techniques can also be applied to both bosonic and fermionic systems as well as to systems of distinguishable particles. \cite{Abrams99prl} Lattice simulations may be direcly applied to model relativistic quantum plasmas. \cite{Shi21pre,Shi2021plasma}

Quantum algorithms for general sparse Hamiltonians were then developed by a number of authors. \cite{Aharonov03arxiv, Berry07cmp}
Today, optimal strategies are based on Childs' generalization of Szegedy's approach to define a quantum walk for an arbitrary Hamiltonian. \cite{Childs10cmp, Berry2012qic}
Using strategies such as the linear combination of unitaries \cite{Childs12arxiv} and Taylor series approximations, \cite{Berry15prl}
the algorithms  were improved to have nearly optimal dependance on all key parameters, including condition number, precision, and failure probability, in Refs.~\onlinecite{Berry14stoc, Berry15focs, Berry15prl} and, eventually, to optimal dependance in Refs.~\onlinecite{Berry16qic, Novo16qic}.
Optimal dependance on all parameters was actually first achieved for a large class of Hamiltonians using the methods of qubitization and quantum signal processing (QSP). \cite{Low17prl, Low19quantum}
This method works by using QSP to correct the eigenvalue distribution of the Childs walk so that it generates a more accurate approximation of the underlying Hamiltonian.

\subsubsection{Linear Equation Solvers, Linear Differential Equation Solvers, and Matrix Operations \label{sec:linear}}
Harrow, Hassidim, and Lloyd (HHL) \cite{Harrow09prl} invented the first general quantum algorithm for solving sparse linear equations, $\generalOp x = b$.
The HHL algorithm estimates the eigenvalues of $\generalOp$ and replaces them by their inverse using an ancillary qubit to as well as a technique called \emph{post-selection}, i.e. measurement of the state of the ancillary qubit.
The method was subsequently improved using the methods of \emph{variable time amplitude amplification}\cite{Ambainis12arxiv} and linear combinations of unitaries. \cite{Childs17sjc}
General matrix operations can be performed using the QSP and qubitization techniques. \cite{Martyn21prxq}

After the introduction of linear solvers and advanced Hamiltonian simulation techniques, methods for solving linear differential equations were developed.  \cite{ Berry14jpa, Berry17cmp} 
For hyperbolic partial differential equations, such as general wave equations, Ref.~\onlinecite{Costa19pra} found similar results.  
More recent linear differential equation solvers have much improved dependance on precision by using a linear combination of operations \cite{Berry17cmp} and by using spectral  methods. \cite{ Childs20cmp, Childs21quantum} 

The quantum linear solver algorithm (QLSA) and quantum linear differential equation algorithm (QLDSA) have an exponential improvement in the complexity of solving for a quantum encoded version of the solution. 
Note, however, if the amplitudes of the entire solution are required, then the speedup is reduced to quadratic.
For example, Ref.~\onlinecite{Montanaro15pra} found that, after properly accounting for precision, the improvement for solving elliptic PDEs using finite elements is only quadratic.

\subsubsection{Nonlinear Solvers \label{sec:nonlinear:solvers}}
\paragraph*{Nonlinear Operations.}
The QM Postulates require linear evolution of the state, and, if the evolution is ideal, the linear operation must be unitary. 
Hence, nonlinear operations of the entire state cannot be performed with a speedup over classical algorithms.
As discussed previously, nonlinear quantum eigenvalue transformations (QEVT) of a normal matrix \cite{Low19quantum} and nonlinear quantum singular value transformations (QSVT) \cite{Gilyen19stoc} can be performed efficiently for a block-encoded matrix. 
This approach has begun to yield efficient methods for performing nonlinear operations on blocks of the wavefunction. \cite{Guo22arxiv, Holmes22arxiv} 
An alternate approach, based on considering a nonlinear operation on a given set to be a linear operation on the Hilbert space of functions on the set was explored by Ref.~\onlinecite{Dodin20arxiv}.  
This method is the analogue of the Koopman-von Neumann approach, discussed below, for finite time intervals.

There is good potential for these methods to develop into the ability to efficiently perform arbitrary nonlinear operations.
If this is the case, then the ability to perform a nonlinear operation, combined with the ability solve linear equations, allows one to develop methods for solving nonlinear equations using either  fixed-point Picard iteration or Newton's method.

\paragraph*{Nonlinear Differential Equation Solvers.}
The first concrete quantum algorithm to simulate nonlinear differential equations  was proposed in Ref.~\onlinecite{Leyton08arxiv}, but had an exponential complexity in time.
This is because the entire wavefunction needed to be stored for each time step in order to compute the required nonlinear functions.
After this result, the quantum algorithms community began to focus on solvers for linear differential equations, as described in the previous section.

The next conceptual breakthrough was provided by Ref.~\onlinecite{Joseph20prr}, which proposed an algorithmic approach to solve nonlinear differential equations by mapping the nonlinear problem to the linear Koopman-von Neumann representation of classical dynamics.
After this step, Hamiltonian simulation could be directly applied to the system.
Then, Liu, et al.,   \cite{Liu21pnas} derived an explicit algorithm for differential equations with quadratic nonlinearities based on the closely-related technique of Carleman linearization.
The connections between these techniques will be elucidated in detail in Sec.~\ref{sec:nonlinear}.

Another algorithmic approach for solving the dissipative nonlinear differential equations that arise from many-body physics was also recently advocated by Ref.~\onlinecite{Lloyd20arxiv}.
The basic idea, which goes back to initial considerations of quantum computing, \cite{Boghosian97ijmpc, Yepez02cpc} is that such a description is an approximation of many-body quantum physics.
Hence, a simulation of the many-body quantum physics Hamiltonian should lead to an approximate solution of the classical system in the appropriate limit.

Yet another approach for solving the Navier-Stokes equations of fluid dynamics was explored by Gaitan \cite{Gaitan20nqi} based on a method for integrating differential equations developed by Kacewicz. \cite{Kacewicz06jc} 
In turn, the time-integration method developed by Kacewicz relies on quantum algorithms developed by Heinrich and Novak for computing sums and integrals. \cite{HeinrichNovak01arxiv, Montanaro15prsa}
As explained previously, these methods can obtain up to a quadratic quantum speedup over Monte Carlo methods for high-dimensional systems with non-smooth right hand sides. 
However, a key point is that this speedup occurs when only partial information is given to specify the right hand side, \cite{Kacewicz87jc} i.e. only a certain number of derivatives exist and \emph{ the position of  possible discontinuities in the last derivative are not known in advance.}
Thus, such gains can only achieved if there are non-smooth initial conditions whose evolution is too difficult to track in time or if there is a stochastic forcing of the differential equations.
Stochastic forcing is in fact useful for studies of fluid and plasma turbulence, as well as for stochastic differential equations in general.
However, it is important to keep in mind the fact that this method would not accelerate the solution of ODEs and PDEs with analytic coefficients and initial conditions.

\subsubsection{Variational Quantum Algorithms}

{Variational quantum algorithms} \cite{Moll2018quantum, Cerezo21nrp, Tilly21arxiv, Bharti22rmp}  are typically formulated as hybrid quantum-classical algorithms for solving optimization problems.
All of the quantum subroutines discussed above require large-scale fault-tolerant quantum computers which will not be available in the near term. 
As a middle ground for the NISQ-era, hybrid quantum-classical algorithms have been proposed where quantum devices are envisioned as co-processors that accelerate the otherwise classical calculations. 
It is hoped that these will serve as NISQ algorithms for difficult many-body problems in quantum chemistry and quantum materials science, typically involving fermions.
They may even be useful as methods for solving nonlinear problems. \cite{Lubasch20pra}
For example, variational simulation of the Navier-Stokes equations has recently been nvestigated.\cite{Steijl2020aqci,Steijl2020qcc,Griffin2019turb}
Variational simulation of nonlinear stochastic differential equations has also been explored. \cite{Kubo2021pra}
 
The idea is that there is a cost function of high complexity that is evaluated by a quantum computer which depends on a number of parameters that is minimized by using a classical optimization algorithm.
For example, the parameters may be in the form of phase angles associated with the circuits that are to be optimized to minimize the cost of the objective function.
Quantum resources are useful because they allow the use of Hamiltonians with many degrees of freedom and of ansatzes that are classically expensive. 
Thus, they have a few key independent parts: (i) constructing the cost function via Hamiltonian simulation, (ii) constructing a suitable ansatz for the initial wavefunction, (iii) developing efficient measurement protocols to determine the cost function and, potentially, its derivatives with respect to optimization parameters, and (iv) the classical optimization algorithm.

While it is often hoped that these algorithms will be useful for achieving at least quadratic quantum advantage on near-term NISQ hardware platforms, there are a number of outstanding challenges that must be considered and these challenges can make it difficult to obtain rigorous proofs of their complexity. 
Notable issues are associated with developing efficient ansatzes for the initial wavefunction, developing efficient measurement techniques, avoiding barren plateaus with little information on the optimization problem, and the presence of many local minima, which make the problem intrinsically difficult.
In fact, the classical-quantum hybrid scheme may not have any advantage when the classical optimization problem is NP-hard. \cite{Bittel2021prl}

\section{  Quantum Simulation \label{sec:qsim} }
 
\subsection{General Framework}
The quantum simulation application space is closest to the typical applications used within FES.  
Several authors \cite{Joseph20prr, Dodin21pop, Liu21pnas, Engel2021pop, Giannakis22pra} have outlined a framework for the quantum simulation of nonlinear dynamics.  
As discussed in the previous section, there are a number of subroutines to be employed that have steadily improved in computational complexity and accuracy.
A summary of the key steps is discussed in this section.

The key conceptual steps  are:
\begin{enumerate}
\item Determine an embedding of the dynamics into a linear differential system of equations. 
\item Numerically discretize the linear system.
\end{enumerate}
If the system is intrinsically linear, like a quantum Hamiltonian, then the first step is relatively straightforward, but the choice of representation can still have an impact on complexity, which is especially important when resources are limited.
If the system is nonlinear, then a linearization must be determined.
In either case, it is important to choose the basis so that the resulting equations have a sparse structure or an efficient Trotterization that can be utilized by the quantum simulation algorithms.

The key computational steps are:
\begin{enumerate}
\item Prepare the initial state.
\item Evolve the system in time using an efficient linear differential equation solver, or if the system is Hamiltonian, an efficient Hamiltonian simulation algorithm.
\item Estimate the observable to be measured efficiently, typically by using an ancillary quantum memory register to encode the observable and performing amplitude estimation. 
\end{enumerate}
If the linearized system is sparse, then using the sparse linear differential equation solver ensures that the key steps are efficient. \cite{Liu21pnas}
If the linearized system is both sparse and Hamiltonian, then using the sparse Hamiltonian simulation algorithm ensures that the simulation step  is efficient. \cite{Joseph20prr}
State preparation is required for  
the initialization step and is often required for the final measurement stage. 
For an efficient algorithm, one must choose initial conditions that can be prepared using an efficient method such as  Hamiltonian simulation or a quantum walk. \cite{Joseph20prr}

Many works on quantum computing algorithms consider the output of a subroutine to be a \emph{quantum encoding} of the solution, i.e. wavefunction or density matrix that can then be used as input for another subroutine.  
While this is quite reasonable for intermediate stages of a quantum program, for almost all scientific calculations of interest, the output usually consists a set of real floating point numbers that quantify the system of interest. 
Whenever this data is encoded in the amplitude of the wavefunction, then one requires a final stage where amplitude amplification and estimation is used to extract the quantities of interest.

\subsection{Sparse Representation \& Numerical Discretization}
Choosing a finite representation of the wavefunction is clearly required for numerical simulation.
Choosing a basis that keeps the operations sparse is clearly an important consideration.
In general, a complete basis allows one to represent the wavefunction, $\waveFunc(\xcoord)$, as a linear combination of basis elements, $\set{\basisFunc_\kindex(\xcoord)}$ so that
\begin{align}
\waveFunc(\xcoord, \time)=\sum_{\kindex=1}^\Nstates \psi_\kindex(\time) \basisFunc_\kindex(\xcoord).
\end{align}
The variables $\xcoord$ are indices representing the computational basis.
If the computational basis represents position, then it might be convenient to use the position basis $\basisFunc_\kindex(\xcoord) = \delta_{\kindex  \xcoord}$ or the momentum basis, $\basisFunc_\kindex(\xcoord)= \exp{(2\pi i\kindex \xcoord)}/ \Nstates^{1/2}$.
More generally, the discretized Hilbert space could  span  any finite collection of linearly independent functions, such as orthogonal polynomials,  Bessel functions, etc.
One could also use  finite difference or finite element basis functions, i.e. orthogonal polynomials (or, more generally, orthogonal functions) over  subdomains with finite support.
If the differential equation to be solved has an \emph{efficiently-computable} sparse representation in terms of these basis functions, i.e. $\sparsity$ terms, then the resulting system of equations will be $\sparsity$-sparse.

One can also use an overcomplete basis, such as the coherent states.
Coherent states are an important example of a {\bf Reproducing Kernel Hilbert Space  (RKHS)}, which is defined by the fact that the  pointwise evaluation of functions is continuous. \cite{Aronszajn50ams, PaulsenBook}
In this case, the \emph{Riesz representation theorem} implies that there is a continuous linear \emph{kernel function}, $\kernelFunc_\ycoord(\zcoord)$, that allows one to evaluate a function at the point $\ycoord$ via 
\begin{align}
\waveFunc(\ycoord)=\braket{\kernelFunc_\ycoord}{\waveFunc} =\int \kernelFunc_\ycoord(\zcoord)\waveFunc(\zcoord) d\zcoord.
\end{align}
This is called the \emph{reproducing property} of the kernel.
The definition of the bilinear kernel function in terms of the Hilbert space inner product, $\kernelFunc(\ycoord,\zcoord)=\braket{\kernelFunc_\ycoord}{\kernelFunc_\zcoord}$, demonstrates that the kernel function must be symmetric and positive definite.
The \emph{Moore-Aronszajn theorem} states that every symmetric positive definite kernel defines an RKHS.

There are actually many types of RKHS' that are used within the machine learning community for efficient representations of data in higher-dimensional spaces.
One may expect that RKHS' should play a similar role in developing efficient representations for quantum computing, and, should certainly be of interest for \emph{quantum machine learning} (QML).
Examples of RKHS' are bandwidth-limited functions, finite collections of orthogonal polynomials,  and complex analytic function spaces, such as Segal-Bargmann coherent states,  Bergman spaces, and Hardy  
 spaces.
For example, as proven in Appendix~\ref{sec:RKHS}, the Carleman representation is based on the standard Hardy space, a complex analytic RKHS that uses a Sz\"ego kernel. 
This RKHS is based on the collection of monomials, $\zcoord^\kindex$, where evaluation is given by the Sz\"ego kernel for the boundary of the unit disk in the complex plane.
Determination of which representations are more or less efficient for representing different applications is an active research area.

The Stone-von Neumann theorem implies that all representations that satisfy the (Weyl exponential form of the) \emph{canonical commutation relations} (CCR) are unitarily equivalent. \cite{HallBook}
While there are important exceptions,  the vast majority of the representations used in practice are unitarily equivalent to the position and momentum operators; i.e. there is a unitary transformation that relates the two different representations.
For example, the number basis defined by normalized Hermite polynomials, $H_\kindex(\xcoord)$ are related to the position representation by a unitary transformation.
Similarly, the coherent states are related to the number states via a unitary transformation.
The combination of these two transformations generates another unitary transformation from position space to coherent state space known as the Segal-Bargmann transform.

There have already been some interesting uses of RKHS' within quantum computing for nonlinear dynamical systems. 
Ref.~\onlinecite{Giannakis22pra} used a real RKHS that used smoothed basis functions to develop a well-defined Hamiltonian simulation algorithm for integrable systems that displays smooth pointwise convergence in the limit $\Nstates\rightarrow \infty$.  
Ref.~\onlinecite{Liu21pnas} used the  Carleman embedding, which is based on the standard Hardy 
space,  in developing a quantum algorithm for dissipative systems such as the logistic equation and the Burgers equation.
Ref.~\onlinecite{Engel2021pop} considered the possibility of exploring other representations of the CCR.
As proven in Appendix \ref{sec:RKHS}, all of the concrete examples  discussed in Ref.~\onlinecite{Engel2021pop} are well-known examples of complex analytic RKHS'.

\subsection{State Preparation: Initialization \& Measurement \label{sec:qsim:state_preparation} }
In the typical models of quantum computing, the initial wavefunction must start in a standard state, such as the ground state $\ket{0}$.
This is useful from the point of view of a physical device, because, if the temperature is much lower than the energy difference between the first excited state and the ground state, then there is an exponentially small likelihood of finding the initial wavefunction in an excited state.
Due to dissipative decoherence processes such as relaxation, simply waiting long enough allows one to re-initialize the quantum program from $\ket{0}$.
\emph{Active restart} techniques intentionally use decoherence to speed up the process of resetting to vacuum.

Then, to actively begin the program,  one must typically initialize the wavefunction in a specific state. 
For simple programs, such as search, one might use the Walsh-Hadamard transform to prepare a uniform superposition of a quantum memory register.
For a simulation, one would typically initialize the memory register with a state corresponding to the initial condition of interest.
As discussed previously, this initial condition should be efficiently computable, perhaps a smooth function with some random noise which might be useful for triggering the dynamics of interest, e.g. as in simulations of turbulence.
As discussed previously, many useful functions can be computed efficiently by using quantum walks or Hamiltonian simulation.

In order to obtain output from a given memory register, a projection step is usually required to obtain the data of interest. 
In practical models of quantum computing, there is typically only a possibility of performing measurements in the computational basis. 
Hence, in order to extract other properties of the wavefunction, e.g. measurements along another axis of the Bloch sphere, one must perform judiciously chosen unitary transformations before the measurement is taken.
This final projection step can either be thought of as a transformation of the data by some unitary operation $\UnitaryOp$ or as a transformation of the state to be measured against with the inverse operation $\UnitaryOp^\dagger$.

\subsection{ Simulation \label{sec:qsim:simulation} }
After defining an efficient representation, the simulation step has now reduced the problem to solving for the evolution of a linear set of ODEs.
In components, these ODEs can be written as 
\begin{align}
i\hbar d\waveFunc_\jindex / d\time = \sum_\kindex \Ham_{\jindex \kindex} \waveFunc_{\kindex}  = \braOpket{\basisFunc_\jindex}{\HamiltonianOp}{\waveFunc}.
\end{align}

If the evolution is unitary, then one can directly use the best available sparse Hamiltonian simulation algorithm (HSA) \cite{Lloyd96sci, Abrams99prl, Berry15focs}. 
This constructs an approximation of the form
\begin{align}
 \waveFunc(\time)=\UnitaryOp_{approx} \waveFunc_0 \simeq \TimeOrder e^{ -i\int  \HamiltonianOp d\time/\hbar} \waveFunc_0  
\end{align}
where $\TimeOrder$ is the time ordering operator. 
The result will typically be approximate due to the need to simulate the Hamiltonian efficiently. 
If the Hamiltonian can be broken into a small number of pieces where each piece is easy to construct, e.g. because it is sum of terms that commute with each other, then one can use a Trotter-Suzuki decomposition to generate an accurate approximation.
Otherwise, one can choose one of the sparse Hamiltonian solvers described previously.

If the evolution is not unitary, then one can use the best available quantum linear differential equation solver algorithm (QLDSA) \cite{Berry14jpa, Berry17cmp}
Simple statements of these algorithms might often an explicit forward Euler timestep; however, this can clearly be improved to use higher order methods as well as implicit timestepping, using the QLSA. 
For example, for each time step, one can consider performing a time splitting of the form
\begin{align}
\left[1+i\theta \HamiltonianOp \Delta \time/\hbar\right] \waveFunc(\time+\Delta \time) 
\simeq 
\left[1-i(1-\theta) \HamiltonianOp \Delta \time/\hbar\right] \waveFunc (\time)
.
\end{align}
Here, $\theta$ represents a time splitting parameter that determines the amount implicitness to be used; i.e. $\theta=0$ for fully explicit forward Euler, $\theta=1/2$ for implicit midpoint Crank-Nicolson, and $\theta = 1$ for fully implict backward Euler.

In order to handle a non-unitary evolution, the QLDSA must use post-selection.
 Care must be taken to be able to extract the information of interest and this depends on the properties of the solution itself.  
 If the linear system is unstable and the magnitude of $\waveFunc$ is exponentially unbounded, then the output will likely result in noise.
 If the linear system is stable and the magnitude of $\waveFunc$ shrinks exponentially towards zero, then the output will also eventually result in noise.

For realistic applications of interest, the \emph{no-fast-forwarding theorem} holds.\cite{Berry07cmp, Berry15focs}
The no-fast forwarding theorem implies that, in the generic situation where the solution cannot be solved for semi-analytically, then the solution must be tracked on the shortest relevant time steps.
This implies that the simulation will be restricted to short time steps, of the order of  $\delta \time \sim \hbar / \norm{\HamiltonianOp}$.
Hence, the total number of time steps, $N_t=\Delta \time / \delta \time \sim  \norm{\HamiltonianOp} \Delta \time  /\hbar$, will typically be large.
For example, to observe temporal variations on the longest relevant time scale would require $\Delta \time \sim\hbar/ \min{(\rm eigenvalues}(\HamiltonianOp))$, so the number of steps will typically scale as  the condition number   $N_t\sim \conditionNumber(\HamiltonianOp)$.
Hence, while the QLDSA is more costly than Hamiltonian simulation,  particularly with regard to condition number, the QLDSA is similar to that of Hamiltonian simulation in terms of complexity in time steps. 
Thus, which choice is  optimal may depend on other desired features of the solution.
 
\subsection{ Estimation of Physical Observables \label{sec:qsim:observables} }
In fact, there are some relatively simple methods for extracting observables from a set of measurements.\cite{Joseph20prr}
To simplify the discussion,  here we consider the case where all observables commute with each other (are in involution). 
The expectation value of a local observable, 
\begin{align}
\avg{\Observable}=\sum_\xcoord \Observable(\xcoord) \pdf(\xcoord)
\end{align}
 can be found by performing a reversible classical computation of the functions
\begin{align}
\phi(x) &= \Observable^{1/2}(\xcoord) \waveFunc(\xcoord)
\\
\phi'(\xcoord) &=\sqrt{ 1-\abs{\phi(\xcoord)}^2 }
.
\end{align}
First, form  the states
\begin{align}
\ket{\phi}  &= \sum_\xcoord \phi(\xcoord) \ket{\xcoord} /\sqrt{\sum_\xcoord \abs{\phi(x)}^2}
.
\end{align}
Then, to compute the amplitude, add an ancillary qubit to the system, $\ket{\phi} \otimes \ket{0}$, and perform a specific choice of rotation, $\RotOp_\phi$, defined via
\begin{align}
\RotOp_\phi \ket{\phi} \otimes \ket{0} &= \sum_{\xcoord=1}^\Nstates \ket{\xcoord} \otimes \left( \phi'(\xcoord) \ket{0} + \phi(\xcoord) \ket{1}\right) /\sqrt{N}\\
&=\cos{(\theta)}  \ket{ \phi' }\otimes \ket{0}+ \sin{(\theta)} \ket{ \phi }\otimes \ket{1}  
.
\end{align}
Finally, using the amplitude estimation subroutine, discussed in Sec.~\ref{sec:amp_est}, allows one to directly calculate the amplitude of the $\ket{1} $ state to determine $\sin^2{(\theta)}=\avg{\Observable} /\Nstates$ with an amount of work that scales as 1/accuracy rather than 1/accuracy$^2$.

\subsection{Quantum Speedup}
For both the intrinsically quantum and the intrinsically classical approaches, 
quantum Hamiltonian simulation of classical dynamics leads to large gains in memory and computational complexity.
The Hamiltonian simulation algorithms are efficient as long as the Hamiltonian is sparse.
It is a simple exercise to prove that the Hamiltonians of interest are sparse for $N$-body interactions and for particles interacting with electromagnetic fields.
This implies up to exponential savings relative to an Eulerian discretization of the Liouville equation.

However, the best classical algorithms of this sort are time-dependent Monte-Carlo methods. \cite{KalosMCBook} 
For example,  particle-in-cell algorithms invented in the plasma physics community 
fall into this class, as well as molecular dynamics algorithms invented in the condensed matter community. 
These probabilistic classical algorithms can also provide an exponential speedup over Eulerian methods when the same restriction on the observables of interest are applied.
In this case, one does not need to accurately sample the entire PDF, rather, one only needs to maintain good accuracy for the most probable regions of phase space. \cite{KalosMCBook}
In this case, the equivalent quantum algorithm can achieve up to a quadratic speedup over time-dependent Monte-Carlo (e.g. PIC) methods. \cite{HockneyBook, BirdsallBook}

It is important to stress that it is the use of amplitude estimation of physical observables and the use of quantum walks for state preparation and measurement transformations that enable the quadratic speedup.
 Due to the central limit theorem, when classically sampling an observable with a finite mean and variance, the relative accuracy decreases as $\accuracy \sim 1/\Nsamples^{1/2}$ where $\Nsamples$ is the number of samples.
 This implies that the amount of computational work required to achieve a certain accuracy grows as $1/\accuracy^2$.
 The utility of the amplitude estimation algorithm is that it only requires an amount of computational work that grows as $1/\accuracy$.

\section{ Quantum Representation of \\
\hspace{18pt}  Nonlinear Dynamics \label{sec:nonlinear} }
The majority of simulations used within FES today aim to solve initial-value and boundary-value problems: typically time-dependent simulations of nonlinear dynamics.
While the system may often represent partial differential equations (PDEs), such as the Navier-Stokes or magnetohydrodynamics (MHD) equations of fluid mechanics, or the kinetic equation for the probability distribution function, ultimately, numerical discretization of the set of PDEs typically reduces the system to time integration of a set of ODEs. 
The question then arises: \emph{How should quantum computers be used to efficiently simulate nonlinear dynamics?}

The past few years have seen intensive research activity on the ability to perform quantum simulations of nonlinear dynamics.
As described above, these approaches generally rely on embedding the nonlinear system into a large, 
in fact, infinite-dimensional linear system of equations. \cite{Joseph20prr,Engel2021pop,Dodin21pop,Lin22arxiv}
In this section, we simultaneously unify these approaches and systematically categorize their differences.

Classical dynamics is distinguished by the fact that the solution to a well-posed set of nonlinear differential equations typically has both existence and uniqueness theorems that apply to almost all points in phase space, aside from a set of measure zero.
The uniqueness of the individual trajectories then allows one to associate a unique evolution of other objects, such as scalars, vectors, and tensors on phase space. 
In particular, the Liouville equation expresses the fact that a volume form, which can be interpreted as a classical probability density, is advected by the flow in a specific manner that ensures that probability is conserved in time.
The action of the evolution on phase space functions presents a linear but infinite-dimensional system of equations.
Due to the Stone-von Neumann theorem, the vast majority of all possible representations are equivalent to this one, simply defined by translations in phase space.

From an algorithmic point of view, the next question that arises is: \emph{How well does a finite truncation of the linear representation approximate the underlying nonlinear dynamics? }

\subsection{    Quantized Dynamics \label{sec:nonlinear:qrep} }
\begin{figure}
	\centering
	\includegraphics[width=3in]{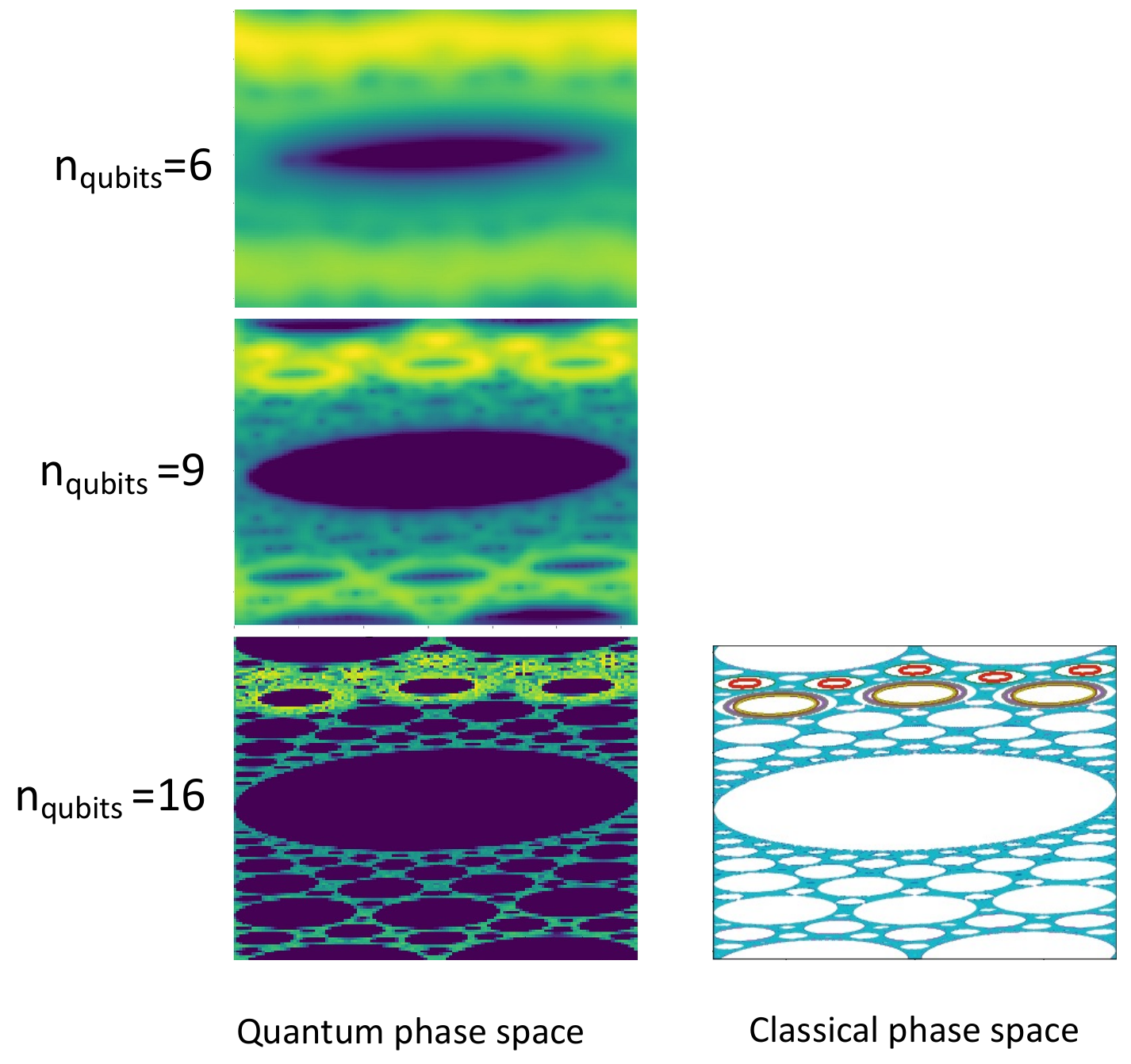}
	\caption{A sufficiently large number of qubits captures the semiclassical limit of the quantum sawtooth map corresponding to the Hamiltonian in Eq~\ref{eq:H_sawtooth}.
	Here, the parameter $K=-0.1$, so the system is in the anomalous diffusion regime.
		(Left) The wavefunction is initially in the state $\pcoord/\pi =3/4$ and evolved for 500 timesteps. 
		The colored plot shows the Husimi-Q quasiprobability function averaged over the last 50 time steps for different qubit numbers $\nqubits=$6, 9, 16.
		(Right) Poincare plot of the classical sawtooth map  for 500 time steps starting from the initial condition $\pcoord/\pi =3/4$ and various angles.
	}
	\label{fig:qsm}
\end{figure}

If one begins with a Hamiltonian system, then a natural approach is to use a quantized version of the classical physics model.
There are many potential advantages to this approach.
First of all, the quantized version may be the more accurate physical model, so important effects such as interference, tunneling, and diffraction would be correctly described. 
Second of all, there are already many quantum algorithms for the simulation of sparse Hamiltonians, and, this is especially true for quantum Hamiltonians.

Two examples of this approach, as well as their implementation on present-day quantum hardware, will be discussed in detail in Sec.~\ref{sec:apps}.
In particular, Sec.~\ref{sec:apps:wave-particle} explores the physics of the sawtooth map,\cite{Benenti04qip}  which can be viewed as a poor-man's model for wave-particle interactions that uses minimal numerical resources.
Figure~\ref{fig:qsm} compares the dynamics of the classical sawtooth map to that of the quantum sawtooth map as the number of qubits is increased from 6 to 9 to 16.
To visualize the quantum dynamics in phase space, the \emph{Husimi-Q quasi-probability distribution} is used to visualize the density matrix.
In both cases, the initial state has uniform amplitude along the $p/\pi=3/4$ momentum surface, and, as time evolves, both the quantum wavefunction and the classical PDF explore similar regions of phase space.
However, this comparison only works well once the number of quantum states becomes large enough. 
At $\nqubits=9$ and  $\Nstates=2^9 = 512$ states, for a total phase space resolution of $512\times 512\simeq 2.6\times 10^4$, the quantum Husimi-Q distribution still appears quite ``fuzzy'' to the human eye.
At $\nqubits=16$ and $\Nstates=2^{16} \simeq 6.6\times 10^4$, for a total phase space resolution of $2^{32}\simeq 4.3\times 10^9$, quantum artifacts are no longer visible at this scale.

This example makes it clear that, for the approach of quantization, it is not likely that one will have enough quantum computational resources to solve the exact quantum system underlying the classical problem. 
Instead, the basic quanta of action used to compute particle numbers, i.e. the effective Planck's constant would be larger, effectively grouping large numbers of particles together in a manner that is similar to the way that particle-in-cell codes evolve  \emph{macroparticles} that each represents large groups of physical particles.
If the model is formulated correctly, parameters of the problem could follow an appropriate \emph{renormaization group flow}, such that simulating a smaller number of groups still produces the correct results for a much larger system. 
Interestingly enough, this will generate numerical discretization error of a type not commonly explored in the traditional applied mathematics community.

Moreover, depending on the application, there can also be a number of disadvantages.
First, taking the semiclassical limit requires very large quantum numbers for many parts of the system.
As mentioned above, accuracy of the physical model must be balanced against the numerical discretization errors, e.g. due to using a value for Planck's constant that is likely to be too large.
Second, the quantum system is not equal to the classical system.
Quantum dynamics shows many well-known non-classical effects, such as interference, tunneling, and diffraction.
If the application requires understanding the behavior of a single classical trajectory, then these quantum effects may be highly undesirable.
If important decisions are to be made on the basis of classical ODE calculations, then unwanted quantum effects could have catastrophic consequences.

The method of quantization is also not necessarily directly applicable to non-Hamiltonian systems, e.g. those that include dissipation.
Yet, reduced physical models that include dissipative processes, such as friction, diffusion, and collisions, are the de facto standard used within high-performance FES simulations today.

In principle, dissipative dynamics can be  simulated using an open quantum system that interacts with its environment.
Perhaps one route could be to use measurements to generate effective dissipative interactions with the physical environment.
However, in order for such interactions to be controlled, one must embed the dissipative system into a larger ideal system that is simulated by the quantum computer. 
In fact, many of the dissipative dynamical systems of interest to FES are naturally realized as reduced order models of large ideal systems.
Ideas such as simulating a many-body quantum fluid to approximate a classical viscous fluid were explored early on in the quantum computing community. \cite{Boghosian97ijmpc, Yepez02cpc}
In principle, simulating a modest-sized ideal system yet only measuring a subset of this system -- while controlling discretization errors -- could be an efficient approach.

Due to the \emph{Schmidt decomposition}, one can generate an arbitrary mixed state of dimension $\Nstates\times\Nstates$ from the reduction of a pure state in a \emph{bipartite} quantum system of dimension $\Nstates^2$.
Thus, for a given system of interest, the size of the Hilbert space that is needed for the embedding may need to be only be as small as the dimension of the system squared, $\Nstates^2$. 
The benefit is that would only require an embedding with twice as many qubits, $2\nqubits$, rather than a macroscopically large system.
Recently, there has been renewed interest in these methods \cite{Lloyd20arxiv} precisely for the benefit of classical applications of the type that are of interest to FES.

\subsection{ Semiclassical Dynamics \label{sec:nonlinear:screp}  }

\subsubsection{Remember What to Hold Fixed }

Consider a set of ODEs for coordinates $\zcoord=\{\zcoord^i\}$ on a manifold, $\Manifold$.
In local coordinates, the ODEs are presented in the form
\begin{align}
 d\zcoord^i/d\time = \Velocity^i(\time,\zcoord).
\end{align}
In order to understand this equation, it is crucial to understand that the initial conditions $\zcoord_0:=\zcoord(\time=0)$ are being held fixed during the evolution of the trajectory. 
Thus, the solution is actually a multi-variable function, $ \zcoord(\time,\zcoord_0)$, that depends on both time and on the set of initial conditions, $\zcoord_0$.
This simple fact implies that one is actually solving a set of PDEs of the form
\begin{align} \label{eq:dzdt}
\left. \partial_\time \zcoord^i  \right|_{\zcoord_0}= \Velocity(\time,\zcoord )\cdot \nabla \zcoord^i
\end{align}
where here $\nabla:=\set{\partial_{\zcoord^i}}$. 

It is also possible to derive the evolution of the initial conditions,  $\zcoord_0(\time, \zcoord)$, as a function of the final coordinates $\zcoord $.
Using the chain rule, one finds that
\begin{align}
\left. \partial_\time   \right|_{\zcoord_0 } = \left. \partial_\time  \right|_{\zcoord  } +\left. \partial_\time  \zcoord^\jindex \right|_{\zcoord_0} \partial_{\zcoord^\jindex} 
= \left. \partial_\time  \right|_{\zcoord  } +\Velocity\cdot \nabla
.
\end{align} 
Thus, the initial conditions satisfy the equation of motion
\begin{align}  \label{eq:dz0dt}
\left. \partial_\time \zcoord^i_0  \right|_{\zcoord }=
-\Velocity(\time, \zcoord(\time, \zcoord_0))\cdot \nabla \zcoord^i_0
.
\end{align}
Notice that this equation has the opposite sign of the velocity so it appears to go ``backwards in time.''
However, this is absolutely consistent with the definition of initial and final coordinates.

Consider a classical  observable, $\Observable(\time,\zcoord)$, that is a scalar function which is constant along the trajectory, so that $\Observable(\time, \zcoord(\time,\zcoord_0))=\Observable(\time=0,\zcoord_0)$. 
Applying the chain rule again implies that
\begin{align}  \label{eq:dsdt}
\left. \partial_\time \Observable  \right|_{\zcoord}=
-\Velocity(\time, \zcoord )\cdot \nabla \Observable
.
\end{align}
If, instead, one would like to express the observable as a function of the initial coordinates via $\Observable_0(\time,\zcoord_0):= \Observable(\time,\zcoord(\time,\zcoord_0))$, then again using the chain rule yields
\begin{align}  \label{eq:ds0dt}
\left. \partial_\time \Observable_0  \right|_{\zcoord_0}=
\Velocity(\time, \zcoord )\cdot \nabla \Observable_0
.
\end{align}
In the dynamical systems literature, the evolution equation of scalar observables is the infinitesimal generator of the {\bf Koopman evolution operator}.

The adjoint of the scalar equation (over the standard phase space measure $d\zcoord$) yields the evolution of a volume-form, $\pdf(\time,\zcoord)$, on phase space.
The Liouville equation for volume forms is 
\begin{align} \label{eq:dfdt}
\left. \partial_\time \pdf \right|_\zcoord = 
-\nabla \cdot (\Velocity \pdf) 
.
\end{align}
In the dynamical systems literature, this is the infinitesimal generator of the {\bf Perron-Frobenius evolution operator}.
Again using the chain rule, this equation can be expressed in the completely equivalent form
\begin{align} \label{eq:df0dt}
\left. \partial_\time \pdf_0 \right|_{\zcoord_0} =  
\nabla \cdot (\Velocity \pdf_0) 
.
\end{align}

The representations above converts ODEs that are finite dimensional but nonlinear into PDEs that are linear.
The trade-off is that the PDE should be viewed as an infinite-dimensional set of ODEs.
The representation that focuses on observables is analogous to the Heisenberg picture in quantum mechanics, and the representation that focuses on the volume form is analogous to the Schr\"odinger picture. 
These two pictures are complementary and equivalent. 
The advection equation satisfied by observables in Eq.~\ref{eq:dsdt} is linear in $\Observable$, which can be discretized, for example using upwinding or discontinuous Galerkin schemes, to a set of linear ODEs and solved using quantum algorithms described in Sec.~\ref{sec:linear}. 
Similarly, the continuity equation satisfied by the volume form in Eq.~\ref{eq:dfdt} can be directly solved using quantum algorithms. 

On classical computers, one may not want to solve nonlinear ODEs using the detour of linear embedding, because discretizing the PDEs could introduce numerical artifacts and the resultant linear ODEs are of much higher dimensions than the original problem.
Nevertheless, for quantum computers, this detour offers a compromise that enables nonlinear ODEs to be solved in the first place. 
The cost of increasing the dimensionality and introducing discretization artifacts may eventually be offset by quantum speedups that are not available on classical computers. 
For example, Koopman methods are quite popular method for using machine learning to develop reduced-order models of dynamical systems from both simulated and experimental data.\cite{Mezic04physd, Mezic05nld, Brunton16pnas, Kaiser18rspa}

In this paper, we primarily focus on the embedding of nonlinear ODEs, but it is worth pointing out that similar methods can be applied to nonlinear PDEs, which are are ubiquitous in FES. 
One approach is to first discretize the PDEs to become nonlinear ODEs, and to then use the ODE embedding schemes discussed below. 
However, this approach yields a linear problem of high dimensionality and it remains to quantify the quantum advantage expected for this case. 
An alternative approach has been developed for nonlinear hyperbolic PDEs.
In this case, they can be directly embedded in linear PDEs of higher dimension using the \emph{ level-set embedding}. \cite{Jin2003level, Liu2006multi, Jin2022quantum} 
It still remains to proven whether these embedding schemes will offer quantum advantage for FES-relevant problems.

\subsubsection{Prequantization\label{Sec:Prequantization}}
The Liouville equation, which already linearizes the ODEs, can be further mapped to the Schr\"odinger equation using the approach developed by Koopman and von Neumann \cite{Koopman31pnas,Koopman32am,vonNeumann32am1,vonNeumann32am2} almost a century ago.
By definition, the norm of the wavefunction squared  represents the classical PDF, i.e. a volume form.
Thus, the wavefunction itself must be treated as the square-root of a volume form, also known as a \emph{half-form} in the language of \emph{geometric quantization}.
The  law for the evolution of a form of this type can be obtained by expressing the wavefunction as 
\begin{align}
\waveFunc=\pdf^{1/2} \exp{(i\phase)},
\end{align}
 where $\phase$ is a scalar function and $\pdf$ is a volume form.
If one assumes that both $\phase$ and $\pdf$ are simply advected along the flow, then this yields the standard {\bf Koopman-von Neumann (KvN) equation}
\begin{align} \label{eq:dpsidt}
\left. \partial_\time \waveFunc\right|_\zcoord = 
-\tfrac{1}{2}\left( \Velocity \cdot \nabla + \nabla \cdot  \Velocity \right) \waveFunc 
.
\end{align}
Crucially, this form is both linear and unitary.
Once again, using the chain rule, this can also be expressed as 
\begin{align} \label{eq:dpsi0dt}
\left. \partial_\time \waveFunc_0 \right|_{\zcoord_0} =  
\tfrac{1}{2}\left( \Velocity \cdot \nabla + \nabla \cdot  \Velocity \right) \waveFunc_0 
.
\end{align}

More generally, let us assume that the complex phase of the wavefunction evolves locally along the trajectory, with a source function, $\Lag(\time,\zcoord)$ that is a function of the coordinates of the trajectory.
Since the phase is a scalar, the evolution law must be of the form
\begin{align}
d\phase/d\time = \left. \partial_\time \phase \right|_\zcoord+
\Velocity \cdot \nabla  \phase
=  \Lag /\hbar
.
\end{align}
This assumption then yields a generalized form of the KvN equation
\begin{align} \label{eq:KvH} 
\left. \partial_\time \waveFunc  \right|_{\zcoord_0} =
-\tfrac{1}{2}\left( \Velocity \cdot \nabla + \nabla \cdot  \Velocity \right) \waveFunc 
-i\Lag \waveFunc /\hbar
.
\end{align}
In this case, time reversal requires using the chain rule  and taking the complex conjugate which leads to 
\begin{align} \label{eq:KvH0} 
\left. \partial_\time \waveFunc_0 \right|_{\zcoord_0} =  \left. \tfrac{1}{2}\left( \Velocity \cdot \nabla + \nabla \cdot  \Velocity \right) \waveFunc_0 \right|_\time + i\Lag \waveFunc_0 /\hbar
.
\end{align}

If the equations of motion are Hamiltonian, with Hamiltonian $\Ham(\zcoord)$ and canonical coordinates $\zcoord=\set{\qcoord,\pcoord}$, then there is a special choice of  the source function that matches Feynman's prescription for the phase space path integral.
This choice is simply the classical Lagrangian
\begin{align} \label{eq:Lag} 
\Lag(\zcoord) = \Ham- \pcoord\cdot \partial_\pcoord\Ham.
\end{align}
Differential operators of the form 
\begin{align}
i\hbar \CL_\Ham \waveFunc =  \Ham- \pcoord\cdot \partial_\pcoord\Ham +  i\hbar \PoissonBracket{\Ham}{\waveFunc}  
\end{align}
were first studied by Leon van Hove in his Ph.D. thesis. \cite{vanHovePhD}
Van Hove noted that these operators have the special Lie group property  
\begin{align} \label{eq:vanHove} 
\PoissonBracket{i\hbar \CL_A}{i\hbar \CL_B} =i\hbar \CL_{\PoissonBracket{A}{B}}
\end{align}
that was emphasized by Dirac in his study of canonical quantization.
In fact, this reproduces the classical Poisson algebra of observables on a Poisson manifold.
Because these operators also represent the first order asymptotic expansion of the Schr\"odinger equation, Bertram Kostant referred  to the construction of these operators as the process of \emph{pre-quantization}.
Pre-quantization is the first step in the program of \emph{geometric quantization} which seeks to determine a mathematically consistent framework for defining quantization on arbitrary symplectic manifolds. \cite{HallBook}
Using the terminology introduced by a number of authors, \cite{Bondar19prsa, Tronci21jpp} this evolution equation
\begin{align}
\CL_\Ham\waveFunc=0.
\end{align}
is now called the {\bf Koopman-van Hove (KvH) equation.} 

Notice that although KvN and KvH equations may resemble the Schr\"odinger equations for quantum systems, they are completely equivalent to the Liouville equation, which describes classical dynamical systems.
 A crucial difference is that the KvN Hamiltonian only involves first order derivatives  while the non-relativistic Schr\"odinger equation involves second-order derivatives. 
 Although the Dirac equation, which describes relativistic quantum systems, also only involves first order derivatives, each component of the \emph{Dirac spinor} satisfies a second-order hyperbolic PDE.   
The mathematical behavior of first-order and second-order hyperbolic PDEs is rather different and often requires different types of numerical techniques for their solution.

\subsubsection{ Forwards or Backwards?}

As we have shown in the previous sections, classically, it does not matter whether the forward or backward form is used because both equations are completely equivalent. 
The real question is whether the wavefunction is considered to be a function of the final coordinates, $\zcoord$, i.e. the Eulerian picture, or of the initial coordinates, $\zcoord_0$, i.e. the Lagrangian picture. 
From the point of view of the coordinates with fixed initial conditions, $\zcoord_0$, this evolution is forward, but from the point of view of scalar functions, with fixed coordinates, $\zcoord$, this evolution is backwards.
This answers the questions posed by Ref.~\onlinecite{Lin22arxiv} on the conventions used for backward vs. forward evolution.

The use of Koopman evolution operators in the dynamical systems community \cite{Mezic04physd, Mezic05nld} follows from the desire to use data, collected either from experiment or simulation, to deduce the underlying equations of motion, and, then, given a data-driven model of the dynamics, to control the dynamics of the system.
Thus, this approach is usually applied  to the evolution of the coordinates themselves, using the Lagrangian framework, rather than to the evolution of scalar functions.

Since the dynamical systems community often uses Carleman linearization,  recent work on quantum algorithms for nonlinear dynamics \cite{Liu21pnas, Engel2021pop, Krovi22arxiv} has used the Lagrangian framework. 
However, the Carleman representation can be applied to the Eulerian formulation equally well. 
In fact, the Eulerian framework might be more natural from the point of view of quantum mechanics.

Once one understands that there is, in fact, a natural geometric setting in which to view the evolution of the wavefunction, it is natural to try to compare this theory with quantum mechanics  itself.
Of course, a typical quantum Hamiltonian, such as the Schr\"odinger equation, is a PDE on configuration space, $\set{\qcoord}$, that is 
at least second order in derivatives. 
As is clear from the path integral, the quantum evolution operator is a weighted sum over all possible trajectories.
Thus, unlike the classical system, one cannot unambiguously invert the Eulerian coordinates to become a function of initial Lagrangian coordinates.
This singles out the Eulerian picture as the approach that is most natural for comparing classical, semi-classical, and quantum dynamics.

\subsubsection{Integrable Systems}

The case of integrable classical dynamics is discussed in some detail in Ref.~\onlinecite{Joseph20prr} and an explicit quantum algorithm for simulation of Hamiltonian systems was derived in Giannakis, et al. \cite{Giannakis22pra}.
In this case, a set of action-angle variables, $\{\action_\jindex,\angle^\jindex\}$  can be determined in which the action variables are constant in time and the angle variables have a constant rate of change
\begin{align}
\dot \action_\jindex = 0 && \dot \theta^\jindex = \omega^\jindex(\action).
\end{align} 
The total phase space is the tensor product of the domain of the action variables and the phase angle variables.
An integrable system can always be recast as a linear system for the variables $\zcoord^\jindex_\pm = \action_\jindex ^{1/2}e^{\pm i\theta^\jindex}$, since 
\begin{align}
\dot \zcoord^\jindex =\pm \omega^\jindex(\action)  \zcoord^\jindex.
\end{align}
Hence, the quantum linear differential equation solver (QLDSA) would certainly work. 

It is clear that the evolution of the angle variables can be encoded in the evolution of the phase of a single qubit. 
Hence, it would be simple to associate each trajectory in phase space with a single qubit. 
Then,   phase space evolution could be accomplished by single-qubit $\RZ$ rotations.
However, this would be an exponentially inefficient use of resources.

In fact, an $\nqubits$-qubit register represents a linear system with $\Nstates=2^\nqubits$ states.
Thus, it is actually possible to associate each trajectory in phase space with a single state in Hilbert space.
If the trajectories are labeled by a finite set of action variables, $\action_\jindex$, that are held fixed during the evolution, then each trajectory experiences its own evolution.
The KvN Hamiltonian is simply
\begin{align} \label{eq:Ham_integrable}
\Ham =  \sum_\jindex \omega^\jindex(\actionOp ) \angularMomOp_\jindex .
\end{align}
where the operators $\actionOp_\jindex=\action_\jindex$ are simply the action coordinates that index each trajectory and the $ \angularMomOp_\jindex=-i\hbar \partial/\partial\angle^\jindex $ are a set of angular momentum operators that are conjugate to the phase angles $\angle^\jindex$.

In order for an efficient quantum algorithm to exist for this system, the phase space must first be transformed to action-angle variables.
In general, this is an intractable problem.
However, let us proceed by assuming that this generically nonlinear coordinate transformation can be accomplished efficiently.
Even in this case, it may not be easy to simulate the Hamiltonian in Eq.~\ref{eq:Ham_integrable}.   
The reason is that performing an arbitrary phase shift of each state can be an exponentially difficult problem.
One of the most efficient algorithms for performing such a phase shift is given by Ref.~\onlinecite{Welch2014efficient}, but has exponential complexity, $2\Nstates$, in the worst case scenario.
Alternatively, the method of qubitization and QSP allows one to form an efficient polynomial approximation of smooth functions, \cite{Low17prl}
so this method may be preferable for many Hamiltonians of interest. 
Hence, it is clearly  important to continue improving these algorithms.

Giannakis, et al., \cite{Giannakis22pra}  improves this basic algorithm through the use of an RKHS approach that acts to smooths the function space of interest. \cite{Das20arxiv}
Their algorithm achieves an exponential advantage, in agreement with Ref. \onlinecite{Joseph20prr}, even when considering output.
Note, however, that the cost of the full output of the dependance of an observable over the entire phase space would presumably reduce the speedup to quadratic at best.
Moreover, they discarded the dependance on the $\set{\action_\jindex}$, which occurs generically for Hamiltonian systems and, as discussed in the previous paragraph, can potentially make the problem much more expensive.
Quite impressively, they actually demonstrated this method for simulating integrable dynamics on the 2-torus, i.e. a surface of constant action with two phase angles, on the IBM Quantum System One with 4 qubits per dimension, i.e. 8 qubits for 256 states.  

\subsection{ Semiclassical Representations  }

\subsubsection{Koopman-von Neumann, Koopman-van Hove, \& RKHS}

The natural setting for linearization is the Liouville equation for the evolution of a classical PDF on the classical phase space. \cite{Joseph20prr}
For Hamiltonian dynamics, the phase space flow is divergence-free and the evolution operator, called the Perron-Frobenius operator, that is generated by the Liouville equation is unitary. \cite{Koopman31pnas,Koopman32am}
For phase-space flows that are not divergence-free, one must be careful to distinguish between the evolution of a scalar function and the evolution of a volume-form or density.
In order to determine a unitary evolution operator, one must consider the evolution of the square-root of a volume form, called half-forms by the geometric quantization community.
If there is no evolution of the phase of the half-form, then this evolution equation is called the Koopman-von Neumann (KvN) equation. 
Ref.~\onlinecite{Joseph20prr} referred to generalizations of the KvN equation that include the evolution of the complex phase as generalized KvN equations. 
In fact, for Hamiltonian dynamics, there is one particular evolution of the phase that corresponds to semiclassical dynamics called the Koopman-van Hove (KvH) equation, \cite{Bondar19prsa, Tronci21jpp} as discussed in Sec.~\ref{Sec:Prequantization}.

Integrating the KvN or KvH equations in time leads to the generation of a nonlinear map.
Ref.~\onlinecite{Dodin21arxiv} explained that the technique described above also applies to such maps in a similar sense that the mapping between phase space functions is linear and unitary.

Concrete algorithms that fully exploit this method are still in their infancy. 
First, Engel et al. \cite{Engel19pra} determined that there is a speedup for solving the linearized Vlasov-Poisson system. 
 They used the method of qubitization \cite{Low17prl} to determine an efficient encoding of the Hamiltonian.

Next, Ref.~\onlinecite{Joseph20prr} explained the general approach of simulating the KvN equation.
However, Ref.~\onlinecite{Joseph20prr}  did not advocate a specific  discretization for the PDF nor did it explicitly specify algorithms for the state preparation required for the initialization and output subroutines.  
While it is clear that finite difference, finite volume, and finite element approaches to discretization all lead to sparse representations of the KvN Hamiltonian, there are issues associated with the fact that the Hilbert space only exactly reproduces phase space in the limit that an infinite number of basis functions is retained.
For this reason, Ref.~\onlinecite{Joseph20prr} discussed the potential need to smooth the phase space basis functions and initial conditions, etc.
In fact, a particular choice of smoothing operation was advocated in Ref.~\onlinecite{Giannakis22pra} through their choice of RKHS.\cite{Das20arxiv}

\subsubsection{Carleman, Coherent States, \& Complex Analytic RKHS}
 
Recently, explicit quantum algorithms for quadratic differential equations were developed using Carleman linearization, \cite{Liu21pnas,Xue21njp} 
which was able to obtain an exponential speedup over the method of Ref.~\onlinecite{Leyton08arxiv} when dissipation dominates nonlinearity.
The algorithms are applied to simulate the logistic equation, the Burger's equation, 
and reaction-diffusion equations, \cite{an2022efficient} none of which is Hamiltonian. 
Note that their algorithm produced a quantum encoding; hence, they did not consider the complexity of measuring the output state nor did they compare this complexity to that of Monte Carlo algorithms.
 Ref.~\onlinecite{Krovi22arxiv} has worked to improve this approach.

Carleman developed a linearization approach for embedding classical dynamics \cite{Carleman1932am} just after Koopman's seminal publication, which is today known as  \emph{Carleman embedding} or \emph{Carleman linearization}. 
Carleman linearization is used today by many in the dynamical systems and control theory community for finding sparse approximations of nonlinear dynamical systems.
The Carleman method has the advantage that, for linear systems of equations, the embedded linear system exactly reproduces the eigenvalue spectrum of the Koopman evolution operator.
If the location of a fixed point of interest is known, it is useful to linearize a nonlinear dynamical system around a given fixed point. Then, applying the Carleman approach ensures that the linear dynamics near the fixed point is faithfully reproduced.

However, unless the original system is exactly linear, a truncation of the linearized system will not generically converge to the evolution operator in other regions of phase space nor will the linearized dynamics capture all of the topological features of the original system.
For example, nonlinear systems often have multiple fixed points, but any finite-dimensional linearized system can only have a single fixed point.
Thus, the linear system can never display truly chaotic dynamics, such as trajectories that are attracted to multiple fixed points. 
These issues are discussed further in Sec.~\ref{sec:limitations}.

The Carleman approach was connected to Hilbert space methods and further generalized by W.-H. Steeb \cite{Steeb87hj} and K. Kowalski \cite{Kowalski87physica} 
and their approach is clearly described in Refs.~\onlinecite{Kowalski91Book, Kowalski94Book}. 
 Carleman embedding  is closely related to the use of number states and coherent states, i.e. the Segal-Bargmann space,  which represent a type of complex-analytic RKHS, defined over the entire complex plane.
For this technique, assume there is a destruction operator $\destroyOp$ and a creation operator $\createOp$ that satisfy the CCR $\Commutator{\destroyOp}{\createOp} =1$. 
In addition, assume there is a vacuum state $\ket{0}$ annihilated by the destruction operator $\destroyOp$ and that there is no subspace that is invariant under the action of $\createOp$. 
Then, there is a countable infinity of orthogonal states $\ket{\kindex}=(\createOp)^\kindex\ket{0}$ that are unnormalized versions of the usual number states, defined by the number operator  $\numberOp=\createOp\destroyOp$.
The \emph{unnormalized coherent states}  are defined via $\ket{z}=e^{ z \cdot   \createOp} \ket{0}$, and, can easily be shown to be eigenstates of the destruction operator,  $\destroyOp\ket{z} =z\ket{z}$. 
If the eigenvalues  satisfy the nonlinear equations of motion,   $d\zcoord/d\time = \Velocity(\zcoord,\time)$, then the coherent states satisfy the  general linear equations of motion  \cite{Kowalski87physica, Kowalski91Book, Kowalski94Book}
\begin{align}
\partial_\time \ket{\zcoord} &=\generalOp  \ket{\zcoord} & \generalOp&:=\createOp\cdot  \Velocity(\destroyOp,\time)
.
\end{align}
In fact, due to the completeness of the coherent states, any complex analytic function, $\waveFunc$, can be written as a linear combination of these states.
Due to the Stone-von Neumann theorem, one can make the identifications $\destroyOp=\zcoord$ and  $\createOp= -\partial_\zcoord$.
Hence, complex analytic functions of $\zcoord$ satisfy the equations of motion
\begin{align}
\left. \partial_\time \psi \right|_{\zcoord_0}&=-\nabla \cdot (\Velocity(\zcoord,\time) \psi ) 
\end{align}
In other words, for this approach, the states appear to evolve as a PDF rather than as a wavefunction, as assumed for the KvN approach.
As explained in Appendix~\ref{sec:complex-KvN}, this is in fact the natural evolution law for a complex analytic wavefunction on a holomorphic RKHS.

Another important difference is that the evolution occurs on the overcomplete space of coherent state eigenvalues.
\emph{More generally, one could use a potentially overcomplete multi-dimensional \emph{feature map} and assume that the dynamics takes place on the feature map rather than evolving directly on the space of observables.}

In order to generate a discrete set of equations, the next step is to project onto a set of complex analytic functions.
For example, one could directly use the coherent states, which are overcomplete.
This leads to many different potential choices of bases.
Another natural choice is to use the number states generated by $\numberOp=\createOp\destroyOp$, which leads to the monomials, $\zcoord^\kindex/\sqrt{\kindex!}$.
In the position representation, $\qOp=(\destroyOp+\createOp)/\sqrt{2}$, these basis functions become the Hermite polynomials, $H_\kindex(\qcoord)$.

Another interesting RKHS is the standard Bergman space of complex analytic functions inside the unit disk in the complex plane.
This naturally generalizes to the unit disk within an $\nqubits$-dimensional complex space.
The Bergman basis  uses the monomials $\zcoord^\kindex \sqrt{\kindex+1}$.
Using a standard conformal transformation called the Cayley transform, this naturally maps to complex analytic functions on the upper half plane (where the unit disk maps to the real line) and as well as to the functions with positive real part.
The generalized Bergman spaces are also defined by a weight function on the unit disk, and this generalization is described in Appendix \ref{sec:RKHS}.

The case where the weight function only has support on a curve in the complex plane, is called a Sz\"ego kernel. 
In fact, the Carleman method uses the Hardy 
space of complex analytic functions inside the boundary of the unit disk, where the weight function lies entirely on the unit circle along the boundary.  
The Carleman basis  uses the monomials $\zcoord^\kindex $.
Again using the Cayley transform, this can be conformally mapped to a region of the upper half plane where the weight function lies entirely along the real line.

Engel, et al., \cite{Engel2021pop} explored alternate approaches for generating new bases of functions for Hilbert space by exploring alternate choices of operators that satisfy the CCR. 
As proven in Appendix \ref{sec:RKHS}, this freedom corresponds to exploring other choices of complex analytic RKHS.
Exactly which bases are more convenient likely depends on the application of interest. 

\subsubsection{ Which Form is Best?  }

The approach advocated by Ref.~\onlinecite{Joseph20prr} directly uses one of the generalized KvN representations, e.g. standard KvN or KvH.
This  treats the wavefunction as a geometric object that represents the square root of a real volume form, and, because this object corresponds to a quantum pure state, the evolution is  unitary.
This implies that a quantum simulation can directly apply the techniques of Hamiltonian simulation.

In contrast, when working with a complex analytic RKHS, it is desirable to ensure that the evolution remains complex analytic so that it stays within the chosen RKHS.
Thus, as shown in Appendix~\ref{sec:complex-KvN}, the Carleman embedding considers the evolution of a complex analytic wavefunction that undergoes holomorphic dynamics.
The direct projection of the complex-analytic KvN law to real space corresponds to the evolution of a PDF rather than as a density.
If the velocity, $\Velocity$, is not divergence-free, then the evolution law for scalars and for volume forms clearly differs from that of a half-form by terms proportional to $\nabla\cdot \Velocity$.
Hence, after projection, the resulting linear differential equations may not be Hermitian and the resulting evolution operator may not be unitary.
In such cases, a quantum simulation  must use the quantum linear differential equation solver.

\subsection{Limitations of Linear Embedding \label{sec:limitations} }

\subsubsection{Finite-Dimensional Truncation \label{sec:truncation} }
The dynamical systems community has put much effort into understanding how to efficiently estimate the underlying dynamics experienced by a system from measurements along.
For example, the Sparse Identification of Nonlinear Dynamics (SINDy) algorithm\cite{Brunton16pnas, Kaiser18rspa} has become quite popular.
The goal of these efforts is typically cast in the following form: embed the observed data into a dynamical system that is almost linear.
While the majority of the variables evolve linearly in response to the other variables, in order to exactly represent the nonlinear dynamics, a subset of variables must still experience nonlinear equations of motion.
When the SINDy algorithm is successful, the nonlinear terms are typically small for the majority of the evolution, but they always eventually kick in within certain regions of phase space to provide the nonlinearities that are necessary for truly chaotic dynamics.\cite{Brunton16pnas}

The present version of quantum algorithms for nonlinear dynamics is actually of a different character, where only exactly linearized systems are evolved.
Unfortunately, a finite linear system can never display truly chaotic dynamics, such as trajectories that are attracted to multiple fixed points. 
For the Carleman approach, there is the benefit that the exact linearized dynamics near a fixed point can be followed exactly.
However, because a linear system can only have a single fixed point, trajectories cannot be attracted to other fixed points, nor can there be changes in which basin of attraction a chaotic trajectory is orbiting.  
Clearly, this is qualitatively quite different than truly nonlinear dynamics, in which chaotic dynamics and bifurcations are ubiquitous.

\subsubsection{ Domain of Convergence \label{sec:domain} }
For all numerical methods, which are intrinsically finite dimensional, only a finite region of phase space can  be described  with high accuracy.
Clearly, both the initial conditions and the subsequent evolution must fit within the domain.
For example, for the discrete Fourier representation, the domain can be considered periodic.
Note, however, that if the actual region of interest is not periodic, then undesirable interference effects will occur near the boundary.

In contrast, for the case of the complex analytic Hilbert spaces, e.g. the standard Bergman and Hardy spaces described above, the domain is the inside of the unit disk in the complex plane. 
In fact, the solution will blow up if it attempts to cross outside the boundary of the domain, i.e. the unit disk.
Thus, in order for the computation to remain valid, it is important for the solution to avoid traveling outside the boundary.
Clearly, such a rescaling needs to be applied so that initial conditions lie within the domain. \cite{Engel2021pop}
Moreover, one must estimate the maximum excursion of the trajectory from the origin and  rescale the equations in a manner that ensures that the maximum excursion lies within the boundary. 
See the remarks at the end of Appendix \ref{sec:RKHS} for more discussion of these issues.

\subsubsection{Dissipation vs. Information Scrambling \label{sec:scrambling} }
The KvN/KvH approaches present the dynamics as explicitly occurring in phase space. 
Because the dynamics is unitary, there are no true Lyapunov exponents for the wavefunction itself.
The basis functions of the KvN/KvH approaches are oscillatory functions are analogous to the Fourier basis. 
Consequently, any truncation of the basis at finite order can potentially introduce Gibbs phenomena and  spurious oscillations. 
Yet, there is still a possibility of observing truly chaotic behavior due to the fact that the dynamics does follow the classical trajectories.

However, there are limitations for modeling dissipative systems: if information cannot be dissipated, then there are situations where the information must eventually become scrambled. 
In fact, Ref.~\onlinecite{Dodin21arxiv} noted a certain scrambling of information that appears to generically occur for finite difference approximations of the KvN embedding for nonlinear maps. 
A similar phenomenon was seen in numerical studies of Ref.~\onlinecite{Lin22arxiv} that targeted the dynamics near a stable sink.
These authors introduced another approach that they called the \emph{chemical master equation} approach, but this method had similar issues.

The scrambling effect appears to be related to the structure of the KvN/KvH eigenfunctions for standard finite difference approximations.
Perhaps there are alternate numerical discretizations that would solve this problem, but it seems likely that such discretizations would need to trade scrambling for dissipation, e.g. in the form of upwinding the finite difference approximation of the advection operator.
If dissipation is introduced at the numerical discretization stage, then one would need to use the linear differential equation solver rather than Hamiltonian simulation.

\subsection{New Hybrid Algorithms: Nonlinear Classical Dynamics Coupled to Linear Quantum Dynamics}
Perhaps another concept in hybrid classical-quantum dynamics would be interesting to study.
Separate the nonlinear system into a nonlinear classical reduced-order model with relatively few degrees of freedom and a linearized quantum high-order model.
The high-order model represents an effective bath that the reduced order model experiences.
For example, the SINDy type algorithms discussed above can sometimes represent the system as being completely linearized except for a single variable that depends nonlinearly on the overall system.

A classical computer can efficiently evolve the nonlinear reduced dynamics, while a quantum computer can efficiently compute the  linear bath dynamics.
Assuming that the high-order model has exponentially more degrees of freedom than the reduced-order model, then the amount of information that needs to be passed between quantum-classical interface, namely, the classical inputs to the quantum algorithm and the output measured from the quantum simulation, is relatively small.

An important example of this is the ideal $N$-body problem in astrophysics, plasma physics, and molecular dynamics. 
In this case, the electromagnetic and/or gravitational fields can be considered the reduced order model since it only has the three dimensions of configuration space.
The linear evolution of the classical PDF / quantum wavefunction can be considered to be the linear model for evolution on the very high $6N$ dimensional phase space / $3N$ dimensional configuration space of the particles.

In fact, this is precisely the way that quantum co-processors are expected to be used to accelerate calculations in chemistry, materials science, and quantum hydrodynamics. 
The macroscopic degrees of freedom could be simulated by a classical computer, while the microscopic degrees of freedom could be simulated by a quantum computer. 
If the microscopic degrees of freedom are assumed to respond linearly to
the macroscopic degrees of freedom, as in an analogue of the Born-Oppenheimer approximation, then the quantum computer could efficiently simulate their dynamics.
Assuming the combined system is sparse, the hybrid simulation will remain efficient even if there are exponentially more microscopic degrees of freedom in the high-order model.

\section { Quantum Computing Applications \label{sec:apps} }

There are a number of quantum software simulators and actual hardware platforms available today for use through both commercial and government funding sources.
This has led to a number of impressive demonstrations that now claim to have achieved quantum advantage or even quantum supremacy. \cite{Arute19nat, Wu21prl, Zhong21prl}

In the literature, the phrase \emph{quantum simulator} and \emph{quantum emulator} are often used to denote hardware platforms that can emulate the behavior of a given quantum mechanical system. 
Here, we shall use the phrase  \emph{classical simulator} to denote the use of classical computer software to simulate a quantum mechanical system.
Clearly, such classical simulators are less efficient than a future error-corrected quantum computer should be.
However, they are also much more efficient than today's quantum computing platforms. 
Today, using a classical simulator it is relatively easy to simulate up to $\sim 15$ qubits on a laptop and up to $\sim$50 qubits on a supercomputer.
Yet, in practice, it is difficult to perform high-precision calculations with more than a handful of physical qubits on present-day quantum hardware. 

For the classical dynamics application space, there have been a number of notable recent quantum software simulations including: 
the linearized Vlasov equation, \cite{Engel19pra}  
the diffusion equation, \cite{Gaitan20nqi}  
the Burgers equation, \cite{Liu21pnas} and 
the linear mode conversion problem. \cite{Novikau22arxiv}
In contrast, there are few demonstration of realistic algorithms on experimental quantum computing platforms.  
Notable experimental results include simulations of 
integrable classical dynamics, \cite{Giannakis22pra}  
non-integrable quantum maps, \cite{Pizzamiglio21entropy, Porter21arxiv, Porter22arxiv} which can be considered models of wave-particle interactions, 
simulations of quantum wave-wave interactions, \cite{Shi21pra} and 
simulations of the Dirac quantum walk for quantum hydrodynamics. \cite{Zylberman2022hybrid}

\subsection{Quantum Hardware Platforms \label{sec:apps:hardware} }
Anyone can get started exploring laws of quantum mechanics by simulating their first quantum algorithms  using the open-access IBM-Quantum (IBM-Q)  cloud-based superconducting quantum computing platform. 
Today, this gives free but limited access to up to 5 qubit devices.
There are also many other companies developing quantum hardware platforms including D-Wave, Google, Intel, IonQ, Microsoft, Quantinuum (Honeywell), Rigetti, Xanadu, etc.

From a scientific perspective, having good experimental access to the hardware platform and  enough runtime to characterize the performance of the system is crucial for improving performance.
For this reason, the U.S. Department of Energy's (DOE) Advanced Scientific Computing Research (ASCR) program supports the activities of several research groups, called \emph{Quantum Testbeds for Science}, for providing open science, quantum hardware platforms for  research on experimental quantum computing architectures.

For superconducting qubits today,\cite{Krantz19apr} the typical gate times are  $\sim 10-100$ \unit{ns}  while the typical relaxation and dephasing times are $\sim 10-100$ $\unit{\mu s}$. 
This implies a minimum effective error rate on the order of 0.1\% per gate application.
Measured two-qubit entangling gate error rates are typically ten times higher than this, on the order of 1\%.
Clearly, this implies the existence of additional sources of error, such as imperfect gate design and cross-talk between neighboring qubits.
For ion-trap platforms, \cite{Bruzewicz2019apr}  the the gate times on on the order of 1-100 \unit{$\mu$s. Although} relaxation and dephasing times are quite long when qubits are in isolation, 10-100 \unit{s}, errors are introduced when gates are enacted and typical gate fidelities are better than but still comparable to superconducting platforms.

 
\begin{figure*}
\centering
\includegraphics[width=6in]{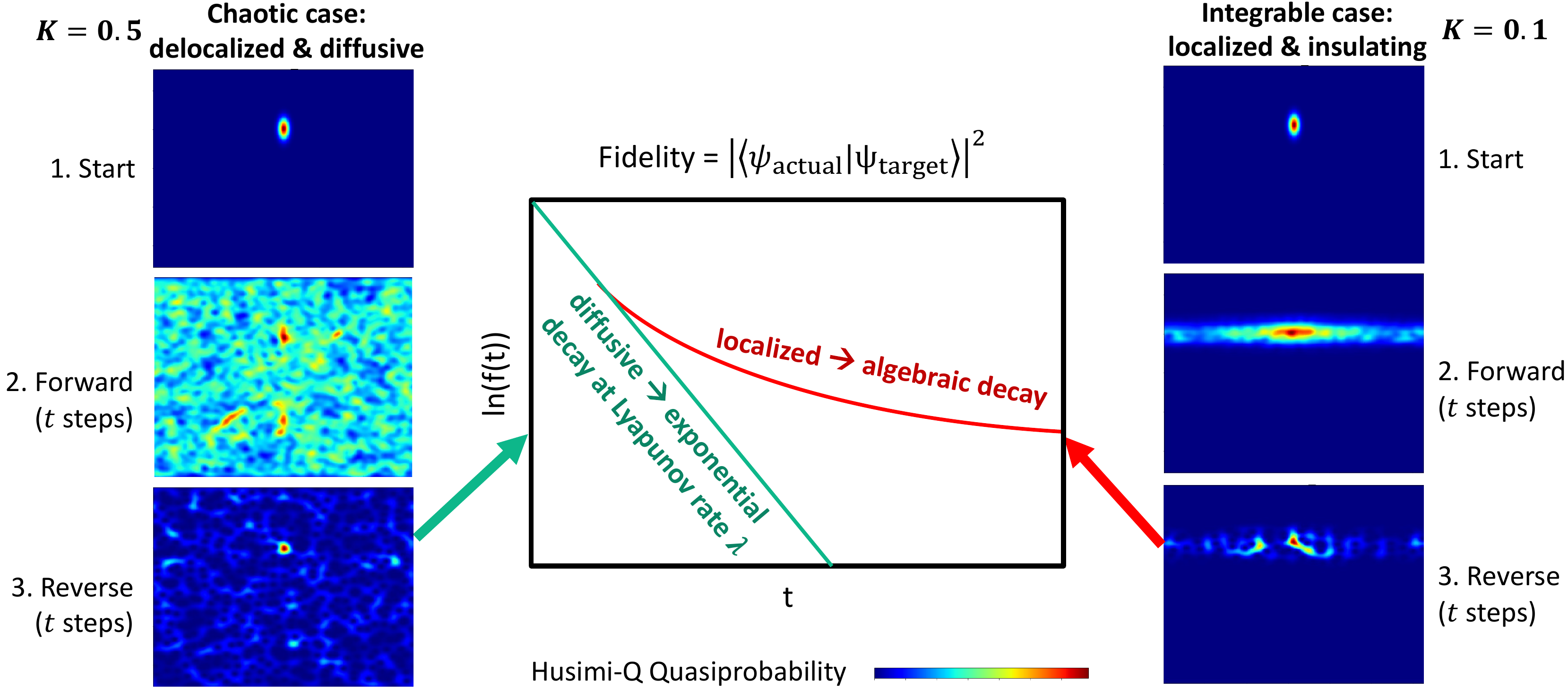}
\caption{
Here, the Loschmidt echo protocol measures the fidelity of the quantum sawtooth map (Eq.~\ref{eq:H_sawtooth}) after simulating the dynamics for $t$ steps forward in time and $t$ steps backward in time.
(Left) If the dynamics is chaotic / diffusive / conducting, as in the sawtooth map at $K=0.5$, then the wavefunction spreads out over a large region of phase space and the fidelity decays exponentially at the Lyapunov rate (Center).
(Right) If the dynamics is integrable / localized / insulating, as in the sawtooth map at $K=0.1$, then the wavefunction only spreads out over the energy (action) surface and the fidelity decays algebraically (Center). \cite{Jaquod09aip, Porter21arxiv}
}
\label{fig:fidelity}
\end{figure*}

 \subsection{ Wave-Particle Interactions \label{sec:apps:wave-particle} } 
Some of the first quantitative studies of classical chaos within plasma physics were focused on wave-particle interactions.
This led to highly simplified mathematical dynamical systems called nonlinear maps that displayed all of the features of chaos in a form that was numerically precise and independent of additional approximations, e.g. introduced by numerical integration and/or nonlinear solvers. 
Quantized versions of these classical nonlinear maps, called \emph{quantum maps} or \emph{Floquet systems}, were developed for similar purposes: understanding quantum chaos in highly simplified yet precisely defined systems that were relatively easy to simulate.

In the 1970's and 80's Giulio Casati, Boris Chirikov, and Dima Shepelyansky made some of the first fundamental discoveries about quantum chaos by studying quantum maps.\cite{Casati79book, Chirikov81ssr, Chirikov88pd}
They found that, if the quantum map had chaotic dynamics, the wavefunction would rapidly spread out along the unstable manifolds of the semiclassical trajectories with a growth rate set by the fastest Lypanov exponent, $\LyapunovRate$.
This leads to a rapid spreading of the wavefunction or density matrix on the Ehrenfest time, $\tau_E=  \log{(\Nchaotic)}/\LyapunovRate$, where $\Nchaotic$ is the number of eigenstates accessible within the chaotic sea.
An effective chaotic diffusion in momentum or action space proceeds until the Heisenberg time, $\tau_H=\hPlanck/\Delta\Energy$, set by the average difference between energy levels; i.e.  $2\pi/\Delta \Energy$ is the mean density of states at the energy scale of interest.
Since the quantum dynamics is linear, for time scales much longer than the Heisenberg time, $\tau_H$, one can observe discrete eigenfrequencies, and the actual Lyapunov exponent of the quantum dynamics must vanish.

In the  chaotic regime, where the dynamics is diffusive/conductive, the typical wavefunction extends over the entire chaotic sea.
In contrast, in the integrable regime, where the dynamics is localized/insulating, the wavefunction can only tunnel slowly through KAM surfaces.
Due to the effects of quantum interference, the insulating regime persists even after the classical transition to chaos, namely, the quantized map will still exhibit \emph{dynamical localization}, even when the corresponding classical map is chaotic.
Localization occurs when the number of quantum states in the chaotic sea is too small. Since the number of states roughly scales as the classical action in chaotic sea divided by $\hbar$, the localization effect occurs when Planck's constant is too large relative to the physical value.

Dynamical localization was eventually realized to be closely related to Anderson localization in condensed matter theory. \cite{Fishman82prl}
Thus, in the generic case, the wavefunction is localized with exponential tails.
Chirikov, et al.,  \cite{Chirikov88pd} realized that the effective localization length can be estimated from a simple scaling argument:
the number of momentum states included in the wavefunction is roughly the number that can be reached by chaotic diffusion up to the Heisenberg time: 
$(\hbar \localizationLength)^2 = \diffusion \tau_H$.
Estimating $\tau_H/\tau\simeq \Energy_{max} /\Delta \Energy \simeq  \localizationLength$, then yields the estimate $\localizationLength\simeq \diffusion \tau /\hbar^2$.
 
The lack of a true Lyapunov exponent in quantum dynamics has made the regime of quantum chaos more subtle to define than its classical counterpart.
While the \emph{quantum-classical correspondence principle} is known to hold for integrable systems, the meaning for chaotic systems is more subtle and questions still remain.
Clearly, for chaotic systems, the correspondence principle only holds up until the Heisenberg time.
In fact, in the presence of noise, chaotic quantum dynamics is much easier to reverse than the classical counterpart.
After reversal, classical trajectories will eventually begins to separate at the Lyapunov rate if there is any uncertainty present, but the quantized version will never do so indefinitely.
In other words, the quantum version often has a much stronger \emph{Loschmidt echo}, i.e. overlap between the initial and final states, than the classical version.
Although the relevant Heisenberg time might be longer than the lifetime of the universe for realistic values of Planck's constant, this qualitative difference may be somewhat surprising.

Asher Peres realized that there is still a kind of quantum-classical correspondence in the dependance of the Loschmidt echo on the dynamics.\cite{Peres84pra} 
For both the classical and quantum cases, the coarse-graining introduced by lack of a precise knowledge of the exact system Hamiltonian is responsible for increasing entropy with time.
Eventually, it was discovered \cite{Jalabert01prl, Jaquod01pre} that the Loschmidt echo has three different regimes: 
a perturbative regime of slow quadratic decay, \cite{Peres84pra} 
a Fermi golden rule regime often observed in the microscopic quantum world, \cite{Jaquod01pre}
and a semiclassical regime where it is controlled by the classical Lyapunov exponent. \cite{Jalabert01prl, Jaquod01pre, Jaquod09aip}
 
In fact, the decay of the fidelity of a noisy quantum simulation yields information about the nature of the dynamics being simulated. \cite{Benenti04qip} 
 The protocol for measuring the so-called Loschmidt fidelity echo is to simulate the dynamics for a certain number of time steps and then to simulate the dynamics backwards in time for the same number of steps.
 In the presence of noise, the final result will not exactly match the initial conditions.
 When the dynamics is chaotic and diffusive (conducting), as shown on the left of Fig.~\ref{fig:fidelity}, then the wavefunction spreads out over a large region of phase space. In this case, the fidelity decays exponentially, with a decay rate set by the Lyapunov exponents of the system. \cite{Jalabert01prl, Jaquod01pre, Benenti02pre}
 When the dynamics is approximately integrable and localized (insulating), as shown on the right of Fig.~\ref{fig:fidelity}, then the wavefunction only spreads out over a surface of constant energy (action). 
 In this case, the fidelity decays algebraically as $ \time^{-\ndim/2}$ where $\ndim$ is the dimension of phase space. \cite{Jaquod09aip}
 
For Hamiltonians with separable kinetic and potential energies, a quantum computer can use QFT's to go back and forth between position and momentum space exponentially faster than a classical computer. 
The only remaining question is whether the remaining phase shifts associated with the kinetic and potential energies can be simulated efficiently.
In fact, for many energy functions, such as polynomials or trigonometric functions, the phase shifts can be performed efficiently. \cite{Georgeot01prl}
Hence, Benenti, et al.,   \cite{Benenti01prl, Benenti04qip} and Georgeot and Shepelyansky \cite{Georgeot01prl} proposed that this quantum factorization can be used to calculate many properties of the dynamical systems more quickly on a quantum computer.
For example, in the localized regime, this algorithm can speed up the calculation of the localization length, and, in the chaotic regime, the chaotic diffusion coefficient (or, when transport has anomalous scaling, the anomalous transport coefficients) can also be measured with polynomial speedup.  \cite{Benenti01prl, Benenti04qip}
They also proposed that simply measuring the fidelity decay can be used to calculate the Lyapunov exponent with polynomial or even super-polynomial speedup.  \cite{Benenti02pre, Benenti04qip}

The idea that one can use a noisy quantum computer to measure the classical Lyapunov exponent is of profound importance to the classical computations of interest to FES. 
Following in their footsteps, we studied the sawtooth map, which is one of the easiest maps to simulate on quantum hardware and developed efficient implementations of the map for the IBM-Q platform. \cite{Porter21arxiv, Porter22arxiv}
The sawtooth map is relatively easy to simulate because it uses a quadratic form for both the kinetic and \emph{potential} energies:
\begin{align} \label{eq:H_sawtooth}
\Ham_{SM}(\pcoord,\qcoord,\time) =\tfrac{1}{2} \pcoord^2 - \tfrac{1}{2} \Kick \qcoord^2  \sum_\jindex \delta(\time-\jindex \tau) 
.
\end{align} 
 with timestep $\tau$ and kicking strength $K$.  
The potential energy is defined as above over the interval $-\pi \leq \qcoord <  \pi$ and is assumed to be periodic outside this interval, so that it is continuous but not smooth.

The classical sawtooth map (CSM) is defined by the Lagrangian 
\begin{align}
\Lag_{CSM}(\pcoord_\jindex,\qcoord_\jindex) = \sum_\jindex \pcoord_{\jindex+1} \qcoord_\jindex - \Ham_{SM}(\pcoord_\jindex,\qcoord_\jindex).
\end{align}
where $\qcoord_\jindex$ and $\pcoord_\jindex$ are the coordinates at time $\time=\jindex \tau$. 
Hence, the CSM is explicitly given by
\begin{align}
\qcoord_\jindex-\qcoord_{\jindex-1} &= \pcoord_\jindex \tau
\\
\pcoord_{\jindex+1}-\pcoord_\jindex &= \Kick \qcoord \tau .
\end{align}
Despite the simplicity of the CSM, it still has very rich dynamics.
It is chaotic for $\Kick<-4$ and $\Kick>0$, integrable for $\Kick=-4,-3,-2,-1,0$, and generically displays anomalous diffusion without a positive Lyapunov exponent for $-4<\Kick<0$.
For any Hamiltonian that has a quadratic kinetic energy and a periodic potential energy, the dynamics is periodic on $\pcoord$ over the interval $-\pi \leq \pcoord \tau  < \pi$. 

The quantum version, the QSM, is defined by promoting the coordinates $\{\qcoord,\pcoord\}$ to operators $\{\qOp,\pOp\}$ that satisfy the canonical commutation relation: $\Commutator{\qOp}{\pOp}=i\hbar$. 
Since there are no terms that mix the $\qOp$'s and $\pOp$'s, there is no operator ordering issue and there is a natural quantization of the sawtooth Hamiltonian given by $\HamiltonianOp_{QSM}:=\Ham_{SM}(\qOp,\pOp)$.
Because the quantum dynamics must be unitary, the quantum sawtooth map evolution operator is then defined to be $\UnitaryOp_{QSM} = \exp{(-i\HamiltonianOp_{QSM} \tau/\hbar)} $.

A finite discretization must have a finite number of degrees of freedom; e.g. $\Nstates$ discrete points along $\qcoord$.
For this choice, the momentum operator has discrete eigenvalues, and is periodic with a range of twice the Nyquist frequency $\hbar \Nstates$.
Interestingly enough, for any map periodic in $\pcoord$, in order for the classical periodicity of $\pcoord$ over the range of $2\pi/\tau$ to consistently match the quantum periodicity of $\pcoord$ on $\hbar \Nstates$, one must then choose the specific value for Planck's constant $\hbar=2\pi/\Nstates \tau$.
Because the Hamiltonian can be considered to consist of separate applications of the kinetic and potential energy operators, each time step naturally divides into four steps: QFT, phase shift, inverse QFT, and another phase shift:
\begin{align}
\UnitaryOp_{QSM} =\PhaseShiftOp_{kinetic}  \QFTOp^{\dagger} \PhaseShiftOp_{potential} \QFTOp
.
\end{align} 
Again, in this case, there are efficient algorithms for performing the phase shift operations because each term is only quadratic in momentum/position operators. 

Figure \ref{fig:qsm} compares the result of simulating the quantum and classical versions of the map in the regime of anomalous diffusion where $K=-0.1$.
This choice is interesting because there is a lot of structure in phase space that is not quickly destroyed by chaotic diffusion. 
On the right is a classical Poincar\'e plot of the result, starting all trajectories from $\pcoord/2\pi=3/4 $ and uniformly sampling the phase.
On the left are the quantum results, shown by the Husimi-Q quasiprobability distribution function.
In this case, the wavefunction is started in the pure momentum eigenstate with $\pcoord/2\pi=3/4$.
As the number of qubits is increased from $n=6$ to 9 to 16, the Husimi-Q function begins to resemble the Poincar\'e plot more and more closely. 

Our theoretical investigations showed that the QSM requires more than 5 qubits to measure fidelity decay at the Lyapunov rate \cite{Porter21arxiv} and our experimental investigations imply that much lower error rates are needed than are observed at present. \cite{Porter22arxiv}
The operating window for observing Lyapunov decay is limited for three reasons: (i) the need for the Lyapunov rate to be slower than the decay rate initially induced by noise, so that an intermediate time asymptotic appears where semiclassical effects are observable, (ii) the need for the dynamics to be in the diffusive/conducting chaotic phase, rather than the dynamically localized/insulating phase, and (iii) the need for the fidelity decay rate to be slow enough to be able to measure multiple timesteps in the semiclassical  intermediate time window.
Based on our experimental investigations on the IBM-Q superconducting platform as well as on existing experimental data on the IonQ ion trap platform, we estimated the reduction in error that would be required to measure the Lyapunov exponent depending on potential hardware and software capabilities.
In the best case scenario of all-to-all parallelization, the gate depth is significantly reduced and then errors might only need to be reduced by a factor of a few. 
However, in the worst case scenario of linear QPU topology, errors may need to be reduced by a factor of 100 or more.\cite{Porter21arxiv} 

Hence, we were surprised to find that the experimental fidelity of quantum simulations on IBM-Q hardware using only three qubits does in fact depend on the dynamics.\cite{Porter22arxiv} 
In fact, relative to the IBM-Q reported error rate based on randomized benchmarking, the fidelity decays $3\times$ faster for integrable dynamics and $5\times$ faster for chaotic dynamics.
While it is well known by experts in the field that such effects are possible, the most commonly used methods of characterizing quantum hardware platforms, randomized benchmarking, can only be used to estimate the effects of a simple \emph{depolarizing noise} model.
Since the depolarizing noise model only generates a combination of the original density matrix, $ \densityOp$, and an incoherent mixture, $ \densityOp_{incoherent} =\mathbf{1}/\Nstates $,  it can never display more than a single decay rate.
Yet, for three qubits, this effect also does not appear  using the unitary noise models  studied by Refs. \onlinecite{Benenti02pre, Porter21arxiv}.

Instead, we discovered that a gate-based Lindblad model, in which single-qubit relaxation and dephasing rates are enhanced during gate operation, can potentially explain this effect. \cite{Porter22arxiv}
The reason is that highly localized wavefunctions are insensitive to dephasing, whereas the highly entangled wavefunctions that are naturally generated by chaotic dynamics are quite sensitive to dephasing.
The noise that generates dephasing acts to randomly alter the phases of each term in the wavefunction, and, hence, generates more fidelity loss when the wavefunction is extended over many states.
Fitting the Lindblad model to three different types of noisy quantum circuit simulations consistently found that gate operation appears to induce a $3\times$ increase in the dephasing rate relative to the rates measured while the qubits are idle.
If the gate-based Lindblad model is an accurate model of reality, then, unfortunately, this pushes the operating window for measuring the Lyapunov regime to even higher qubit numbers. \cite{Porter22arxiv}

\subsection{Wave-Wave Interactions \label{sec:apps:wave-wave} } 
Wave-wave interactions are important nonlinear processes in plasma physics, fluid dynamics, nonlinear optics, and quantum field theory.  
For a general dynamical system in a weakly-coupled regime, decomposing nonlinear dynamics into wave-wave interactions is a ubiquitous method for perturbative analysis and can also yield highly nontrivial non-perturbative predictions, as in the study of wave turbulence. \cite{ZakharovBook, NazarenkoBook}
When the system fluctuates near its fixed points with small amplitudes, the linearized system possesses a spectrum of eigenmodes, which are often called linear waves for hyperbolic systems. 
The linear waves form a convenient basis of the system, and the Hamiltonian is determined by a rank-2 Hermitian tensor, which represents a second-order coupling between waves. 
At larger wave amplitudes, nonlinearities manifest in perturbative analysis as higher order couplings between waves. 
To lowest order, the nonlinear coupling typically involves three waves, leading to what are known as three-wave interactions. 
Near stable fixed points, three-wave interactions are of decay type, where one pump wave decays to two daughter waves, or, conversely, two daughter waves merge to excite the pump wave. 
Near unstable fixed points, three-wave interactions are of explosive type, where the three waves are either spontaneous generated or destroyed. 
Couplings of higher order, such as four-wave coupling, emerge in next-to-leading order perturbative analysis, giving rise to $N$-wave interactions.

\begin{figure}
\centering
\includegraphics[width=3in]{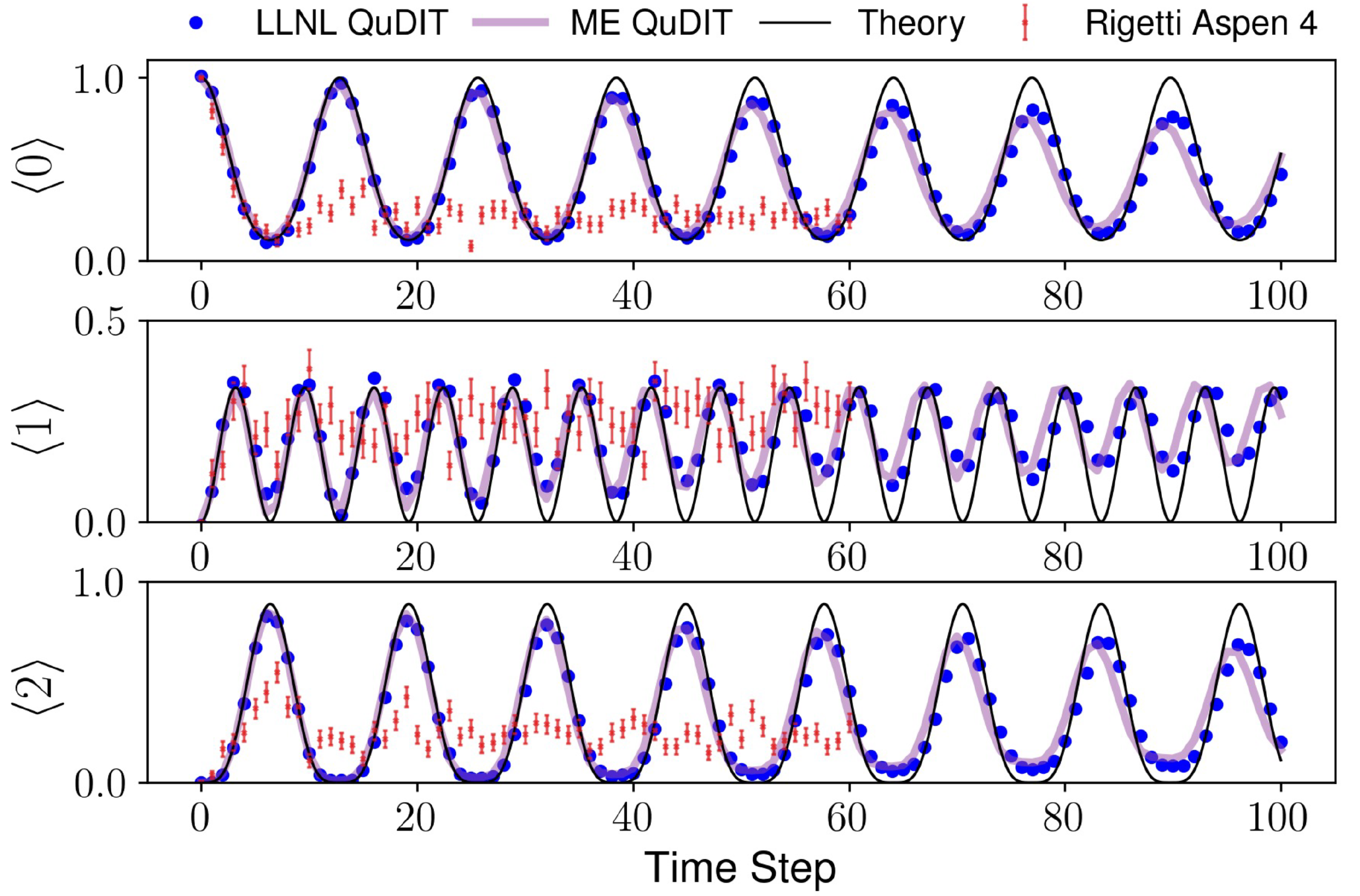}
\caption{Experimental results of simulations of a three-level three-wave interaction problem using different compilation techniques. On Rigetti Aspen-4-2Q-A (red error bars), the  unitary for a single time step is compiled into a circuit composed of $\CZ$ and Pauli gates. In contrast, on the LLNL QuDIT platform (blue dots) each time step is compiled into a single optimized customized microwave control pulse. The later approach better matches the exact results (black line), and the deviations are explained well by a phenomenological GKLS master equation (ME) model (purple line) using experimentally measured decoherence rates.
}
\label{fig:3-wave-expt}
\end{figure}
 
\begin{figure*}
	\centering
	\includegraphics[width=0.9\textwidth]{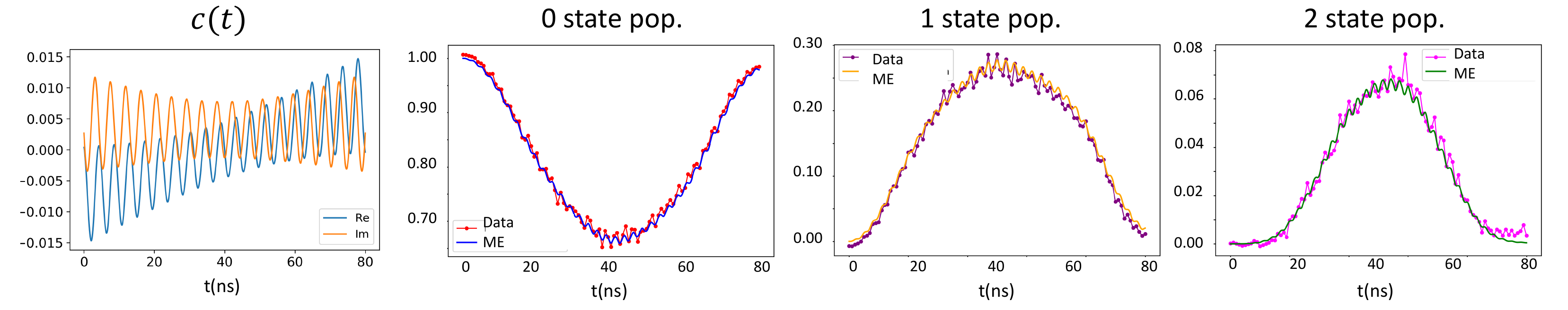}
	\caption{ The optimized control pulse that realizes the single-step unitary $\UnitaryOp(\Delta t)$ for the three-level three-wave problem on LLNL's QuDIT platform. Probabilities of the ground, first excited, and second excited states evolve during the control pulse application. The experimentally observed probabilities on the quantum hardware (dotted line) match the prediction of master equation (ME) simulations on classical computers (lines). By the end of the control pulses, the desired unitary $\UnitaryOp(\Delta t)$ is realized. 
	}
	\label{fig:control}
\end{figure*}

For our first attempt at quantum simulation, we investigated how to simulate three-wave interactions on quantum computers. \cite{Shi21pra}
The three-wave interaction is a nonlinear process, which in its classical form is difficult to map onto quantum hardware using embedding methods discussed earlier. As an alternative approach discussed in Sec.~\ref{sec:nonlinear:qrep}, we chose to simulate a quantized version of the problem, which approaches the classical interaction in the limit of large photon number, where the statistics are typically Poissonian.
This version is implementable on current quantum hardware platforms.
For decay type three-wave interactions, both classical and quantum processes are described by what are known as the three-wave equations:
\begin{eqnarray}
	\label{eq:dA1}
	d\destroyOp_1/d\time&=&+g \destroyOp_2 \destroyOp_3,\\
	\label{eq:dA2}
	d\destroyOp_2/d\time&=&-g^* \destroyOp_1 \createOp_3,\\
	\label{eq:dA3}
	d\destroyOp_3/d\time&=&-g^* \destroyOp_1 \createOp_2,
\end{eqnarray}
where $d/d\time=\partial_\time + \Velocity\cdot \nabla$ is the convective derivative of waves at the group velocity, $\Velocity$, and $g$ is a complex-valued coupling coefficient. 
In the quantum problem, the $\destroyOp_j$ are bosonic operators that satisfy canonical commutation relations $[\destroyOp_j, \createOp_k]=\delta_{jk}$, whereas in the classical problem, these operators are reduced to commuting $c$-numbers, $[a_j,a^\dagger_k]=0$.
In fact, the coupling coefficient $g$ for classical three-wave interactions may be more easily derived from quantum theory. \cite{Shi2017three}

The  action of the $j$-th wave is proportional to the number operator $\numOp_j=\createOp_j  \destroyOp_j$. 
For both the quantum and classical problems, the relative action invariants
\begin{eqnarray}
	\label{eq:S2}
	\Operator{S}_2=\numOp_1+\numOp_2,\\
	\label{eq:S3}
	\Operator{S}_3=\numOp_1+\numOp_3.
\end{eqnarray}
are conserved.
For the quantum problem, these operators commute with the Hamiltonian and each other, while, for the classical problem, these quantities are conserved by the Hamiltonian and are in involution with each other.
Because the operators $\Operator{S}_j$ commute with the number operators $\numOp_j$, they can be simultaneously diagonalized, with eigenvalues $s_j$ and $n_j$, respectively. 
In the Heisenberg picture, the $s_j$ are constants of the motion, but the $n_j$ evolve in time.

The difference between the quantum and classical versions of the problem becomes manifest when computing observables. 
The quantum problem satisfies 
\begin{multline} \label{eq:quantum_3-wave}
\partial_t^2 \langle \numOp_1\rangle=-\partial_t^2 \langle \numOp_2\rangle=-\partial_t^2 \langle \numOp_3\rangle =
\\
2\abs{g}^2\left[s_2s_3-(2s_2+2s_3+1) \langle \numOp_1\rangle+3\langle \numOp_1^2\rangle\right]
.
\end{multline}
 In contrast, the classical problem satisfies
\begin{multline} \label{eq:classical_3-wave}
\partial_t^2 \langle n_1\rangle=-\partial_t^2 \langle n_2\rangle=-\partial_t^2 \langle n_3\rangle=
\\
2\abs{g}^2 \left[s_2s_3-2(s_2+s_3) \langle n_1\rangle+3\langle n_1\rangle^2\right]
\end{multline}
where, now, the bracket denotes the expectation value averaged over a classical ensemble of initial conditions.
Hence, the quantum equation of motion is almost the same as the classical equation except for the final two terms. 
First, notice the extra $1$ in the parentheses of the middle term Eq.~\ref{eq:quantum_3-wave}. 
This term is due to \emph{spontaneous emission}, which only occurs in the quantized version. 
However, it is subdominant in the large photon number limit $s_2, s_3\gg1$. 
Second, the variance of the final term $\langle \numOp_1^2\rangle$ often differs between the quantum and the classical versions, precisely because of the difference in statistics implied by the density matrix vs. the classical PDF, as well as because of the different evolution equations that the higher moments satisfy.
For example, for a single classical trajectory with a particular value of $n_j$, then $\avg{n_j^2}=\avg{n_j}^2=n_j^2$ (and the same result holds true for any power of $n_j$). 
This relation between the first two moments is similar to the statistics generated by the Poisson distribution.

The three-wave Hamiltonian 
\begin{equation}
	\label{eq:H3}
	\HamiltonianOp_{3-wave}=ig \createOp_1  \destroyOp_2 \destroyOp_3 -ig^* \destroyOp_1 \createOp_2 \createOp_3.
\end{equation}
governs the three-wave interaction, because its Heisenberg equations of motion are precisely the three-wave Eqs.~\ref{eq:dA1}-\ref{eq:dA3}. 
The quantum problem can be readily solved in the Schr\"odinger picture once a basis is chosen. 
A convenient basis is formed by eigenstates of the relative action invariants,  $\Operator{S}_2$ and $\Operator{S}_3$, in Eqs.~\ref{eq:S2} and \ref{eq:S3}. 
The basis is convenient because these operators commute with both the Hamiltonian and the number operators.
Since  $\Operator{S}_2$ and $\Operator{S}_3$ are constants of motion, the dynamics in subspaces with different eigenvalues $s_2$ and $s_3$ are decoupled. 
Denoting the Fock states by $\ket{n_1,n_2,n_3}$, then any quantum state can be represented by 
\begin{align}
\ket{\psi}=\sum_{s_2\ge s_3, j} \psi_j \ket{s_3-j,s_2-s_3+j,j}.
\end{align} 
Therefore, it is sufficient to focus on each of the subspaces independently. 
Moreover, since $s_2, s_3\ge 0$, the values of $j$ are constrained by $0\le j \le \min(s_2,s_3)$. 
In other words, each subspace is finite dimensional. 
Notice that the total dimension of the problem is infinite, but the system is a direct sum of disjoint irreducible subspaces each of finite dimension. 
This special property makes artificial truncation unnecessary, thereby avoiding the issues discussed in Sec.~\ref{sec:truncation}. 
Notice that the special property only holds for decay type wave-wave interactions near an elliptic fixed point. For explosive type interactions near a hyperbolic fixed point, the entire Hilbert space will become connected.  

In each subspace, the nonlinear three-wave problem is mapped to a linear Hamiltonian simulation problem, where the Hamiltonian matrix is tridiagonal with zero diagonal elements. 
Explicitly, the Schr\"{o}dinger equation $i\hbar\partial_\time \ket{\psi}=\HamiltonianOp_{3-wave}\ket{\psi}$ becomes a system of linear equations for the expansion coefficients $c_j(t)$, which satisfies
\begin{eqnarray}
	i \hbar\partial_\time \psi_j
	=ig \Ham_{j+\frac{1}{2}}c_{j+1}-ig^*\Ham_{j-\frac{1}{2}}\psi_{j-1}.
\end{eqnarray}
This problem may be thought of as quantum walk on a one dimensional lattice, where the quantum state hops from one lattice site to its two nearest neighbors. What distinguishes this quantum walk from others is the specific matrix element 
\begin{align}
\Ham_{j-\frac{1}{2}}=\sqrt{j(s_3+1-j)(s_2-s_3+j)},
\end{align}
 which encodes the nonlinear three-wave interaction. Notice that the transition matrix elements satisfy $\Ham_{-\frac{1}{2}}=\Ham_{D+\frac{1}{2}}=0$, where $D=\min(s_2,s_3)+1$ is the dimension of the problem. In other words, the quantum walk has a compact support, and the quantum state is confined within a lattice of a finite number of sites. 
On ideal quantum computers, the Hamiltonian simulation problem may be solved efficiently using algorithms discussed in Sec.~\ref{sec:qsim}. 
However, on current noisy devices, we do not yet have the capability of implementing the necessary oracles. 

Instead of performing quantum Hamiltonian simulation, we tested current quantum hardware platforms by performing a simple unitary evolution that can be represented using only two qubits. 
For a time step $\Delta t$, the unitary evolution under the time-independent Hamiltonian $\HamiltonianOp$ is given by the matrix exponential $\UnitaryOp(\Delta t) = \exp(-i\HamiltonianOp\Delta t/\hbar)$. 
For small problem size, the exponential may be computed analytically, and for larger problem sizes, the exponential can be calculated using classical computers. 
Of course, the point of quantum computing is that for very large problem sizes, the unitary can be realized to arbitrary precision without already knowing the unitary matrix classically.  
However, for the few qubits that are available on present-day hardware, the problem size is not too large, so the short-time propagator $\UnitaryOp(\Delta t)$ can be computed classically. 
Then, the propagator is compiled into quantum gates, and the quantum computer is responsible for repeating the gates to carry out the time advance $\UnitaryOp(\Delta t)^N$ for $N$ time steps. 

In this problem, compounding the unitary is somewhat trivial because $\UnitaryOp(\Delta t)^N= \UnitaryOp(N\Delta t)$ can be fast forwarded. 
However, this is a good benchmark for more complicated problems where the total Hamiltonian is composed of non-commuting terms, such as $\HamiltonianOp=\HamiltonianOp_A+\HamiltonianOp_B$. 
In this more typical non-commuting case, where $\HamiltonianOp_A$ and $\HamiltonianOp_B$ are easy to simulate separately but the total $\HamiltonianOp$ is difficult to simulate, product formula approximations such as $\left(\UnitaryOp_A(\Delta t)\UnitaryOp_B(\Delta t)\right)^N\ne \UnitaryOp_A(N\Delta t)\UnitaryOp_B(N\Delta t)$ cannot be fast forwarded. 

Quantum computers realize unitary evolutions by repeating precompiled gate sequences, unlike classical computers which compute the total unitary by a series of matrix multiplications. 
We implement unitary evolution for a $D=3$ test problem, which can be embedded in two qubits, on the Rigetti superconducting hardware platform. \cite{Karalekas2020quantum,Smith2016practical} 
The results are shown in Fig.~\ref{fig:3-wave-expt} as red error bars. 
For comparison, the exact answers are shown by the black line. 
As can be seen from the figure, only about 10 time steps can be performed before results become indistinguishable from noise.
Although the dynamics of this system is integrable, results on current hardware are only able to capture about half of the nonlinear wave period.
This unfortunate result is caused by the fact that harware error rate is about 1\% per gate, so after about $10^2$ gate applications, the error becomes of order unity. 
For the two-qubit problem, the unitary matrix has $2^4=16$ elements, so a brute-force compilation to hardware native gates results in about $20$ gates for each timestep. 
Consequently, the unitary evolution can only reach a depth of about $10$, as we have observed in the test run.

In order to increase the simulation depth, we performed unitary evolution of the three-wave problem on LLNL's Quantum Design and Integration Testbed (QuDIT) platform, \cite{Wu20prl} which gives users white-box access to control pulse design for a transmon qutrit. 
Instead of preparing a universal gate set and then compiling the desired unitary in terms of these gates, we prepared a customized gate that directly realizes the unitary as a single control pulse. 
In this way, each simulation time step is realized by only one customized gate, instead of $\sim 20$ universal gates. 
At constant device performance, the compressed gate depth enables roughly $10\times$ longer effective simulation depth. 

Optimal control uses standard optimization techniques, which were first applied to manipulate quantum systems in the nuclear magnetic resonance community \cite{Khaneja05} and were later applied to many other quantum systems. \cite{Glaser2015training,Petersson2020discrete}
For a given hardware Hamiltonian, $\HamiltonianOp_0$, and control Hamiltonian, $\HamiltonianOp_c$, quantum optimal control searches for classical wave form, $c(t)$, such that by the end of the time evolution under $\HamiltonianOp=\HamiltonianOp_0+c(t)  \HamiltonianOp_c$, the final state of the quantum device is mapped from its initial state by a unitary that is $\epsilon$-close to the desired unitary. 
For the three-wave problem, control pulses for $\UnitaryOp(\Delta t)$ can be readily generated using \textsf{optimize\_pulse\_unitary} in QuTIP, \cite{Johansson2012qutip, Johansson2013qutip} using $\HamiltonianOp_0$ and $\HamiltonianOp_c$ that are calibrated immediately before the production runs on the QuDIT platform. 
Typical control pulses are shown in Fig.~\ref{fig:control}.

Repeating the single-time-step pulses for $N$ times, we achieve unitary evolution shown by blue points in Fig.~\ref{fig:3-wave-expt}. 
The experimental results match the exact solutions much better, and the deviations can be explained by a phenomenological decoherence model described through simulations of the GKLS master equation (ME).
The GKLS model is calibrated using experimentally measured decoherence rates of the quantum hardware (and was not calibrated by fitting the three-wave experimental data). 
The optimal control technique allowed us to extend the number of simulated time steps by approximately an order of magnitude, so that about five time periods could be simulated overall in this test problem. 
This result constituted the first interesting plasma simulation ever performed on quantum computing hardware.

\section{ Conclusions \& Outlook \label{sec:conclusion} }
Quantum information science holds great promise for accelerating scientific breakthroughs through the three pillars of quantum sensing, quantum communications, and quantum computing.
Fusion energy science (FES) stands to benefit, in particular, from quantum computing, due to the importance of scientific computing in driving discoveries in the field.
Many useful quantum algorithms have been developed that could be put to good use including: efficient quantum Fourier transforms, sparse linear solvers, sparse Hamiltonian simulation, and variational eigensolvers.
Intrinsically quantum simulations of interest to FES that may stand to benefit include plasma chemistry, fusion materials science, relativistic laser-plasma interactions, as well as high energy quantum processes that take place in the    most extreme plasma environments in the universe, such as black holes and neutron stars.

For a long time, it was not known whether intrinsically classical simulations, of the type usually performed within FES computing applications, would also stand to benefit in similar ways. 
Recent work has proven that, in principle, Hamiltonian simulation of classical dynamics can achieve an exponential speedup over Eulerian methods.
It is important to point out that classical probabilistic algorithms, such as Monte Carlo methods,  also have the ability to perform an effective exponential reduction in complexity relative to direct solvers. \cite{KalosMCBook}
For generic problems, it is expected that quantum algorithms can achieve up to a quadratic speedup over classical probabilistic algorithms.
The anticipated quantum speedup increases with the dimensionality of the problem and the lack of smoothness of the solution.
If the overhead, such as input and output between classical-quantum interface, and physical costs can eventually be made sufficient small, then these quantum algorithms may one day become the algorithms of choice.
This conclusion opens up the door to improving simulations of classical  $N$-body problems, molecular dynamics, and fluid and kinetic plasma models that are central to FES applications. 

At present, multiple quantum computing hardware platforms have developed high fidelity qubits that can enact multiple entangling gate operations.
Unfortunately, the lack of error correction severely limits the performance for practical applications -- especially for classical applications such as classical logic and deterministic simulations.
If supremacy is defined as one computing platform being able to perform calculations that another platform cannot, then today is certainly the era of \emph{classical supremacy}.
Hence, much of today's quantum computing applications research is focused on passive and active error reduction techniques, such as the quantum optimal control methodology discussed in this work.
Unfortunately, the estimated cost of fault-tolerant error correction is large enough that the first useful quantum applications may need to offer speedups that are beyond quadratic. \cite{Babbush2021prxq}

It is plausible that before perfect error correction becomes widely available, quantum computing hardware  could still provide a novel and useful platform for the simulation of open many-body quantum dynamics.
It may even be the case that such platforms have significant potential for near-term quantum advantage.
However, there is much work to be done in order to understand the accuracy with which such simulations can be actually performed and how this accuracy compares to classical algorithms that run efficiently on high performance supercomputers. 
Simply verifying that the calculation achieves the desired fidelity becomes quite challenging once the complexity of the calculation pushes beyond the limits of classical computers.
Finally, since it is decoherence that controls the \emph{information confinement time,} it can be hoped that the grand pursuit of a developing a  functioning quantum computer will lead physicists to gain a much deeper understanding of decoherence and, ultimately, of the fundamental workings of the universe. 

\begin{acknowledgments}
The authors would like to thank A. H. Boozer, as well as positive interactions with the 2022 International Sherwood Fusion Theory Workshop community, for encouraging the writing of this review paper. 
This work, LLNL-JRNL-839545, was performed under the auspices of the U.S. DOE by LLNL under Contract DE-AC52-07NA27344 and was supported by the DOE Office of Fusion Energy Sciences ``Quantum Leap for Fusion Energy Sciences'' project  FWP-SCW1680.
\end{acknowledgments}

\section*{Data Availability Statement}
The data that support the findings of this study are available within the article.

\appendix

\section{ Hilbert Space Primer \label{sec:HilbertSpace} }
\vspace{6pt}
\paragraph*{\bf Hilbert Space} 
\begin{itemize}
\item {{\bf Definition:} A Hilbert space (HS) is a vector	space with an inner product that is complete under the metric induced by the inner product.}
\begin{itemize} 
\item {The inner product, denoted $\braket{\cdot}{\cdot}$ must be linear, skew-symmetric, and positive-definite}
\begin{align}
\braket{\phi}{a \psi+b \chi} &= a \braket{\phi}{\psi} + b \braket{\phi}{\chi}
\\ 
\braket{\psi}{\phi} &= \braket{\phi}{\psi}^*
\\
\braket{\psi}{\psi} &\geq 0
.
\end{align}
\end{itemize}

\item {{\bf Theorem:} All finite-dimensional Hilbert spaces are equivalent.}
\begin{itemize}
\item{ A finite-dimensional Hilbert space is spanned by a finite set of basis elements $\ket{\jindex}$.}
\item { Define the complex conjugate transpose operation $\bra{\jindex}=\ket{\jindex}^\dagger$.
}
\item { In this basis, the HS metric has the representation}
\begin{align}
\braket{\jindex}{\kindex} &= \HSmetric_{\jindex\kindex}
\end{align}
where $ \HSmetric_{\jindex\kindex}= \HSmetric^*_{\kindex\jindex}$, i.e. as a matrix operator $\HSmetricOp=\HSmetricOp^\dagger$.  

\item The Hilbert space metric, $\HSmetricOp$, must be a positive definite matrix, i.e. it must have positive definite  real eigenvalues. Hence, $\HSmetricOp$ is invertible and $\HSmetricOp^{-1}$ exists.

\item {Dual vectors can be defined that satisfy $\dbraket{\jindex}{\kindex} = \delta_{\jindex\kindex}$. The solution is }
\begin{align}
\dbra{\jindex}=\left(\HSmetricOp^{-1} \ket{\jindex}\right)^\dagger =  \bra{\jindex}\HSmetricOp^{-1} = \sum_\kindex \HSmetric^{-1}_{\jindex \kindex} \bra{\kindex}.
\end{align}
\item {An orthonormal basis exists in which the Hilbert space metric is unity. The orthonormal basis vectors are defined via}
\begin{align}
 \nket{ \jindex} = \HSmetricOp^{-1/2} \ket{\jindex}
\\
\nbra{ \jindex}= \bra{\jindex} \HSmetricOp^{-1/2}
.
\end{align}
\end{itemize} 
\end{itemize}

\section{ Prequantization for Holomorphic Hilbert Spaces \label{sec:complex-KvN}}

Due to the many important uses of complex analysis, it is natural to try to understand how to apply the Koopman-von Neumann (KvN) approach to nonlinear dynamics on complex manifolds.
In order to pass from real to complex variables, one must combine two copies of phase space.
Hence, given two copies of coordinates for the original phase space, $\xcoord,\ycoord\in \Reals^\ndim$,   construct a combined list of coordinates, $\zcoord=\set{\xcoord,\ycoord}\in \Reals^\ndim\otimes \Reals^\ndim=\Reals^{2\ndim}$.
By defining the complex conjugation operator, $\conj : \set{\xcoord,\ycoord}\rightarrow \set{\xcoord,-\ycoord}$, which is clearly an involution, the doubled phase space is isomorphic to a complexified version of phase space, $\zcoord=\xcoord+i\ycoord \in \Complex^\ndim$.

Now, if the dynamics is to act \emph{naturally} on the complex space, the generator of the motion must commute with the complex conjugation operator, $\conj$. 
Hence,  the equations of motion  for $\zcoord$ and $\zconj$ must be compatible in the following sense
\begin{align}
d\zcoord/d\time&=\Velocity(\zcoord,\zcoord^*, \time)& & 
&d\zconj/d\time&= \Vconj(\zcoord,\zconj, \time)=\Velocity(\zconj,\zcoord, \time)
.
\end{align} 

Thus, complex scalar fields evolve via 
\begin{align}
\left(\partial_\time +\Velocity\cdot \partial_\zcoord +\Vconj \cdot \partial_\zconj\right)\scalar (\zcoord,\zconj,\time)=0
\end{align}
and complex volume forms evolve via
\begin{align}
\left(\partial_\time +\partial_\zcoord\cdot \Velocity + \partial_\zconj\cdot \Vconj \right)\pdf (\zcoord,\zconj,\time)=0.
\end{align}
A wavefunction that is defined as the square root of a volume form, should evolve via the complex KvN equation
\begin{multline}
\left[ \partial_\time +
 \tfrac{1}{2} 
\left( \partial_\zcoord\cdot \Velocity + \Velocity\cdot \partial_\zcoord  + \partial_\zconj\cdot \Vconj +\Vconj \cdot \partial_\zconj\right)
\right] 
\waveFunc(\zcoord,\zconj,\time)=
\\
i \Lag (\zcoord,\zconj,\time) \waveFunc(\zcoord,\zconj,\time)/\hbar
\end{multline}
Here, the function $\Lag (\zcoord,\zconj,\time)$ introduces the freedom of adding a non-trivial evolution of the complex phase of the wavefunction.

Now, consider a complex analytic flow velocity $\Velocity(\zcoord, \time)$, which satisfies $\partial_\zconj \Velocity=0$.
The complex analytic scalar field, $\scalar(\zcoord,\time)$,  satisfies the constraint, $\partial_\zconj \scalar=0$, so it evolves via
\begin{align}
\left(\partial_\time +\Velocity\cdot \partial_\zcoord  \right)\scalar (\zcoord, \time)=0.
\end{align}
However,  a volume form cannot be complex analytic because it must carry the weight of the full Jacobian, so it must still satisfy
\begin{align}
\left(\partial_\time +\partial_\zcoord\cdot \Velocity + \partial_\zconj\cdot \Vconj \right)\pdf (\zcoord,\zconj,\time)=0.
\end{align}

A complex analytic  flow keeps the solution complex analytic as a function of time, so that $\zcoord(\zcoord_0,\time)$ is independent of $\conjugate{\zcoord_0}$, i.e. $\partial \zcoord/\partial\conjugate{\zcoord_0}=0$.
Hence, the  Jacobian, $\weightFunc(\zcoord,\zconj)$, on the full space, $\set{\zcoord,\zconj}$, is the norm squared  of the complex analytic Jacobian $\Jac=\partial\zcoord_0/\partial \zcoord$; i.e. the full Jacobian is 
\begin{align}
 \weightFunc(\zcoord,\zconj)=\Jac(\zconj)\Jac(\zcoord)=\abs{\Jac}^2.
\end{align}
Because of this splitting of the Jacobian, one can still define a complex analytic wavefunction via
$\waveFunc(\zcoord,\time) = \waveFunc_0(\zcoord_0)  \Jac(\zcoord,\time)$.
Thus, the {\bf complex analytic KvN equation}  
\begin{align}
\left(\partial_\time +\partial_\zcoord\cdot \Velocity  \right)\waveFunc(\zcoord,\time)=0
\end{align}
corresponds to the evolution law for a complex analytic half-form -- with an appropriate choice for the dynamics of the phase.
Note how it appears to be advected as a volume form on a space with half of the real dimension of the full system.
Although this form of the evolution is not Hermitian, it is actually a special case of the Hermitian evolution law of a ``wavefunction'' that is a $\zcoord$ volume form and a $\zconj$ scalar field 
\begin{align}
\left(\partial_\time +\partial_\zcoord\cdot \Velocity +\Vconj \cdot \partial_\zconj  \right)\waveFunc(\zcoord,\time)=0
\end{align}
subject to the initial condition $\partial_\zconj\waveFunc=0$.

Given the complex-analytic form of the KvN evolution law, the norm of a complex analytic wavefunction is preserved in time
\begin{multline}
\braket{\waveFunc(\time) }{\waveFunc(\time)} 
= \int  \abs{ \waveFunc_0(\zcoord_0) \Jac(\zcoord)}^2  d\zcoord d\zconj
\\
=\int  \abs{\waveFunc_0(\zcoord_0)    }^2 d\zcoord_0d\conjugate{\zcoord_0} = \braket{\waveFunc_0 }{\waveFunc_0}
.
\end{multline} 

The complex analytic evolution law is still sensible when a nontrivial Hilbert space inner product is included. 
Assume the inner product is defined by  the function $\weightFunc(\zcoord,\zconj)$, which is fixed in time,  via
\begin{align}
\braket{\basisFunc  }{\waveFunc } =  \int  \basisFunc(\zconj,\time) \waveFunc(\zcoord,\time) 	\weightFunc(\zcoord,\zconj)d\zcoord d\zconj.
\end{align}
For example, if $\weightFunc(\zcoord,\zconj)$ represents a complex analytic coordinate transformation, then it has the form $\weightFunc(\zcoord,\zconj)=\abs{\Jac(\zcoord)}^2$.
In this case, the inner product simplifies to
\begin{align}
\braket{\basisFunc  }{\waveFunc } =  \int  \basisFunc(\zconj,\time) \waveFunc(\zcoord,\time) 	\abs{\Jac(\zcoord)}^2d\zcoord d\zconj.
\end{align}
The relevant evolution law can be achieved by defining the divergence operation as 
\begin{align}
\nabla \cdot \Velocity := \Jac^{-1}(\zcoord) \partial_\zcoord \left( \Jac(\zcoord)\cdot \Velocity(\zcoord, \time) \right)
\end{align}
Then, the  complex analytic KvN equation is generalized to  
\begin{align}
\left(\partial_\time +\nabla\cdot \Velocity  \right)\waveFunc(\zcoord,\time)=0.
\end{align}
Again, this is equivalent to the Hermitian evolution law for a ``wavefunction'' that is a $\zcoord$ volume form and a $\zconj$ scalar field
\begin{align}
\left(\partial_\time +\nabla\cdot \Velocity +\Vconj\cdot\nabla^* \right)\waveFunc(\zcoord,\zconj,\time)=0
\end{align}
subject to the initial condition $\partial_\zconj\waveFunc=0$.

\section{Holomorphic Reproducing Kernel Hilbert Spaces and Canonical Commutation Relations (CCR) \label{sec:RKHS}}

Motivated by Kowalksi, \cite{Kowalski94Book} Ref.~\onlinecite{Engel2021pop} recommended exploring the possibility that operators that satisfy the canonical commutation relations (CCR)  
\begin{align}
\Commutator{\zOp}{\wOp} = 1 
\end{align}
but differ from the canonical creation and destruction operators could lead to interesting representations that would be useful for representing classical dynamics.
The study of such generalized representations is the domain of the theory of Reproducing Kernel Hilbert Spaces (RKHS).

The Stone-von Neumann theorem  \cite{HallBook} states that, if these operators are sufficiently well-behaved, so that they satisfy an exponential form of the CCR, known as the Weyl commutation relations (WCR)
\begin{align}
 e^{ \zOp s}e^{\wOp t}  = e^{st}  e^{ \wOp t}e^{\zOp s} 
\end{align}
then, there is a unitary transformation that relates $\wOp$ and $\zOp$ to the usual operators $\qOp$ and $\pOp=-i\partial/\partial \qcoord$, respectively.
In fact, for the form above, they are analogous to the destruction and creation operators, $\destroyOp=(\qOp+i\pOp)/2^{1/2}$ and $\createOp=(\qOp-i\pOp)/2^{1/2}$, respectively.
In practice, the cases of interest often do satisfy the WCR, as was the case for all examples explored in Kowalski \cite{Kowalski94Book} and in Ref.~\onlinecite{Engel2021pop}.

Assume there is a \emph{vacuum state} $\ket{0}$ annihilated by the destruction operator $\zOp\ket{0}=0$, and assume there is no finite-dimensional subspace that is left invariant by the action of $\wOp$. 
Then, the infinite tower of states 
\begin{align}
\wket{\jindex} :=  \wOp^\jindex \ket{0}
\end{align}
 for $\jindex \in \Naturals$ spans the Hilbert space.
 Let us define the notation $ \wbra{\jindex} :=\wket{\jindex}^\dagger$. 
In order to complete the definition of the Hilbert space,
one must specify the metric for these states
\begin{align}
\wbraket{\jindex}{\kindex} = \HSmetric_{\jindex \kindex}.
\end{align}
Due to the fact that the Hilbert space inner product must be  Hermitian, $\HSmetricOp^\dagger=\HSmetricOp$, the dual basis elements, $\wbra{\overline \jindex} $, which are defined to satisfy $\wbraket{\overline \jindex} {\kindex} =\delta_{\jindex\kindex}$, are given by
\begin{align}
\wbra{\overline \jindex} :=\wbra{\jindex}\HSmetricOp^{-1} = \sum_\kindex \wbra{\kindex} \HSmetric^{-1}_{\jindex \kindex} .
\end{align}
Then, using the CCR, one finds the representation
\begin{align}
\wOp&=\sum_{\jindex=0}^\infty  \wket{\jindex}  \wbra{\overline{\jindex-1}}  
\\ 
\zOp&=\sum_{\jindex=1}^\infty  \wket{\jindex-1}  \wbra{\overline \jindex}   \jindex
.
\end{align}

An orthonormal set of \emph{number states} is defined via 
\begin{align}
\nket{\jindex} = \HSmetricOp^{-1/2}\wket{\jindex} = \sum_\kindex \HSmetric^{-1/2}_{\kindex \jindex} \wket{\kindex} .
\end{align}
Given the assumptions, these states are equivalent to the standard number states, with the standard destruction  $\destroyOp$, creation   $\createOp$, and number $\createOp\destroyOp$ operators.
Using the definitions above
\begin{align}
\wOp&=\sum_{\jindex,\kindex,m=0}^\infty  \nket{\jindex} \nbra{\kindex}   \HSmetric^{1/2}_{\jindex m}  \HSmetric^{-1/2}_{m-1 \kindex} 
\\ 
\zOp&=\sum_{\jindex,\kindex,m=0}^\infty  \nket{\jindex}  \nbra{\kindex}   \HSmetric^{1/2}_{\jindex m} m  \HSmetric^{-1/2}_{m-1 \kindex}
.
\end{align}

The \emph{generalized coherent states} are defined via 
\begin{align}
\wket{\zCoherent} = e^{\zCoherent \wOp} \ket{0}=\sum_\jindex \zCoherent^\jindex \wket{\jindex}/\jindex!=\sum_{\jindex\kindex}\zCoherent^\jindex \HSmetric^{-1/2}_{\jindex \kindex} \nket{\kindex}/\jindex!
.
\end{align}
These states are eigenstates of the destruction operator $\zOp$ because they satisfy $\zOp\wket{\zCoherent} = \zCoherent\wket{\zCoherent}$.
The inner product of these states is the reproducing kernel for this space
\begin{align}
\kernelFunc_\ycoord (\zcoord ) =
\wbraket{\yCoherent}{\zCoherent} = \sum_{\jindex\kindex} (\conjugate{\yCoherent})^\kindex   \zCoherent^{\jindex}  \HSmetric^{-1}_{\jindex\kindex}/\jindex! \kindex!.
\end{align}
The coherent space coordinates offer yet another equivalent parameterization of the Hilbert space.
Assume the Hilbert space inner product in coherent state space is given by the definition
\begin{align}
\wbraket{\phi}{\psi} = \int \phi^*(\zcoord)\psi(\zcoord) \weightFunc(\zcoord, \zconj) d\zcoord d\zconj/\Omega 
\end{align}
where  $\weightFunc(\wcoord, \zconj)=\conjugate{\weightFunc}(\zcoord , \conjugate{\wcoord})$ is a positive Hermitian weight function and  $\Omega=\pi^\ndim$ is a normalization constant. 
In order for the coherent state inner product to be consistent,  the Hilbert space metric must be given by the moments of the weight function 
\begin{align}
\HSmetric_{\jindex\kindex}  = \int  (\zconj)^\jindex   \zcoord^\kindex \weightFunc(\zcoord, \zconj) d\zcoord d\zconj / \jindex!\kindex!\Omega
.
\end{align}
Thus, the choice of  the coherent space weight function, $\weightFunc(\zcoord, \zconj)$, is tied to the choice of metric on Hilbert space, $\HSmetricOp$.

For example, in the simple yet important special case where the metric is diagonal, i.e. the states $\wket{\jindex}=\wOp^\jindex \ket{0}$ are orthogonal, then the states can be normalized  by the definition
\begin{align}
\ket{\widehat \jindex} = \HSmetric_{\jindex \jindex}^{-1/2} \wket{\jindex} = \HSmetric_{\jindex \jindex}^{-1/2}   \wOp^\jindex \ket{0} 
\end{align}
and one finds the representation
\begin{align}
\wOp&=\sum_{\jindex=1}^\infty  \nket{\jindex }  \nbra{\jindex-1}  ( \HSmetric_{\jindex \jindex}/ \HSmetric_{\jindex-1 \jindex-1})^{ 1/2} 
\\ 
\zOp&=\sum_{\jindex=1}^\infty  \nket{\jindex-1}  \nbra{\jindex}   \jindex (\HSmetric_{\jindex \jindex}/  \HSmetric_{\jindex-1 \jindex-1})^{- 1/2} 
.
\end{align}
For any quantity $\weightCoef_\jindex$ labeled by $\jindex$,   define
\begin{align}
\weightCoef_\jindex! := \prod_{\kindex=1}^\jindex \weightCoef_\kindex.
\end{align} 
Then, the number states are related via
\begin{align}
\wket{\jindex}:= ( \HSmetric_{\jindex \jindex}/ \HSmetric_{\jindex-1 \jindex-1})^{ 1/2} ! \nket{\jindex}
.
\end{align} 

For the standard Segal-Bargmann coherent states over the complex plane, the weight function is $\weightFunc(\norm{\wcoord}) = \exp{(-\norm{\wcoord}^2 )}$ which leads to $\HSmetric_{\jindex\kindex} =\delta_{\jindex\kindex}  / \jindex! $.   
This implies that $w_{\jindex,\jindex-1} = z_{\jindex-1,\jindex} = \jindex^{1/2} $.
The number states are normalized, $\wket{\jindex}=\nket{\jindex}$ so that $\braketw{\zcoord}{\jindex}= \braketw{\zcoord}{\widehat \jindex}= (\zconj)^\jindex/(\jindex!)^{1/2}$ and the reproducing kernel is 
\begin{align}
\kernelFunc_\ycoord(\zcoord)=\wbraket{\ycoord}{\zcoord} =e^{\conjugate{\ycoord} \zcoord}.
\end{align}

For the standard Bergman space over the complex unit disk, the weight function is 1 inside the unit disk and vanishes outside the disk.
This leads to $\rho_{\jindex\kindex}= \delta_{\jindex\kindex} /( \jindex+1)!\jindex!$ which yields $w_{\jindex+1,\jindex} = (\jindex(\jindex+1))^{1/2}$ and $z_{\jindex-1,\jindex} = (\jindex+1)^{-1/2}$.
The number states are $\braketw{\zcoord}{\widehat \jindex} = (\jindex+1)^{1/2} (\zconj)^\jindex$ and the reproducing kernel is 
\begin{align}
\kernelFunc_\ycoord(\zcoord)=\wbraket{\ycoord}{\zcoord} =1/(1-\conjugate{\ycoord} \zcoord)^2.
\end{align}

For Sz\"ego kernels,
the weight function has all of its support on  the boundary of a domain in the complex plane, e.g in the standard case, by the unit disk 
\begin{align}
\HSmetric_{\jindex\kindex}  = \int  (\zconj)^\jindex   \zcoord^\kindex \weightFunc(\zcoord, \zconj) \delta{(\norm{\zcoord}-1)}  d\zcoord d\zconj/\jindex!\kindex! \Omega 
.
\end{align}
For the standard  Hardy space, the weight function is 1, which yields $\rho_{\jindex\kindex}=  \delta_{\jindex\kindex} /\jindex!^2$.  
Hence, this leads to $w_{\jindex,\jindex-1 } = \jindex$ and $z_{\jindex-1,\jindex} =1$. 
The un-normalized number states are $\braketw{\zcoord} {\widehat \jindex}= (\zconj)^\jindex $ and the reproducing kernel is 
\begin{align}
\kernelFunc_\ycoord(\zcoord)=\wbraket{\ycoord}{\zcoord} =1/(1-\conjugate{\ycoord} \zcoord).
\end{align}
This is the basis used by Carleman linearization.

The different choices of the Hilbert space metric are also tied to the domain over which the function spaces are defined. 
For Segal-Bargmann space, the Hilbert space is composed of holomorphic functions of the complex plane that vanish sufficiently rapidly at complex infinity.
In contrast, for the Bergman and Hardy spaces, the Hilbert space is composed of functions that are holomorphic within the unit disk.
For the standard coherent states, the eigenfunctions are localized with a Gaussian form factor near the given point in phase space.
In contrast, for the Bergman and Hardy
spaces, the eigenfunctions \emph{inside} the unit disk are generated by a singularity \emph{outside} the unit disk.
In these case, the kernel has the form $\kernelFunc_\ycoord(\zcoord)= (1-\conjugate{\ycoord}\cdot \zcoord)^{-a}  $, and, so, the singularity actually resides at the point $\zcoord=1/\conjugate{\ycoord}$.  
Thus, the kernel becomes unbounded if the coordinate $\ycoord$ ever crosses over to the inside of the unit disk.

Another important point to keep in mind is that the observable state space, i.e. the corresponding number states, have not yet been specified. 
While the observable states could correspond to complex analytic functions, they could also be defined in many other ways.
For example, they could correspond to any infinite series of orthogonal functions over a given weight function such as Fourier series, Hermite polynomials, Bessel functions, etc.

The interesting idea proposed by Kowalski \cite{Kowalski94Book} is to encode the dynamics through translations on the space of coherent states rather than  through translations on the space of physical observables. While the action may be quite complicated in observable space, the Stone-von Neumann theorem guarantees that there is a unitary transformation to coherent state space where the dynamics is simply encoded in the usual manner: 
\emph{through trajectories in coherent space}. 

\bibliography{Quantum-Bibliography.bib}{}

\end{document}